\documentclass[aps,prd,preprint,tightenlines,superscriptaddress,nofootinbib]{revtex4}
\usepackage{amsmath,amssymb,url}
\usepackage{graphicx,tabularx}
\usepackage{psfrag}

\newcommand{\D}{\displaystyle}

\newcommand{\etal}{\textit{et al.}}
\newcommand{\ie}{{\sl i.e. }}
\newcommand{\eg}{{\sl e.g. }}

\newcommand{\di}[1]{\ensuremath\text{d}#1\;}
\newcommand{\om}{\ensuremath\mathcal{O}}

\newcommand{\gev}{{\ensuremath\rm GeV}}

\newcommand{\br}{{\ensuremath\rm BR}}
\newcommand{\sign}{{\ensuremath\rm sign}}

\begin{document}

\preprint{Edinburgh 2007/25}

\date{\today}

\title{Measuring Supersymmetry}

\author{Remi Lafaye}
\affiliation{LAPP, Universit\'e Savoie, IN2P3/CNRS, Annecy, France}

\author{Tilman Plehn}
\affiliation{SUPA, School of Physics, University of Edinburgh, Scotland}

\author{Michael Rauch}
\affiliation{SUPA, School of Physics, University of Edinburgh, Scotland}

\author{Dirk Zerwas}
\affiliation{LAL, Universit\'e Paris-Sud, IN2P3/CNRS, Orsay, France}

\begin{abstract}
  If new physics is found at the LHC (and the ILC) the reconstruction
  of the underlying theory should not be biased by assumptions about
  high--scale models.  For the mapping of many measurements onto
  high--dimensional parameter spaces we introduce SFitter with its new
  weighted Markov chain technique. SFitter constructs an exclusive
  likelihood map, determines the best--fitting parameter point and
  produces a ranked list of the most likely parameter points. Using
  the example of the TeV--scale supersymmetric Lagrangian we show how
  a high--dimensional likelihood map will generally include
  degeneracies and strong correlations. SFitter allows us to study
  such model--parameter spaces employing Bayesian as well as
  frequentist constructions. We illustrate in detail how it should be
  possible to analyze high--dimensional new--physics parameter spaces
  like the TeV--scale MSSM at the LHC.  A combination of LHC and ILC
  measurements might well be able to completely cover highly complex
  TeV--scale parameter spaces.
\end{abstract}

\maketitle
\tableofcontents

\newpage
\section{New Physics at the TeV Scale}

In the coming years, the major effort in high--energy physics will be
the search for a Higgs boson or an alternative to such a fundamental
Higgs scalar at the LHC. However, fundamental scalars are difficult to
accommodate in field theory --- their masses are quadratically
divergent with the cutoff scale of the theory. This problem naturally
leads to speculations about the necessary ultraviolet completion of
the Standard Model, which should remove such quadratic divergences and
allow to extrapolate our understanding to maybe even the Planck
scale. Such an ultraviolet completion can (and should) at the same
time solve the second big mystery of high--energy physics, the
existence of cold dark matter.

An overwhelming amount of data on possible ultraviolet completions of
the Standard Model has been amassed over the past decades,
consistently confirming the Standard Model.  LEP and Tevatron have put
stringent bounds on the masses of new particles, cutting into the
preferred region for example for supersymmetric dark
matter~\cite{SUSY,cdm} not only via the derived light Higgs mass, but
also via direct searches~\cite{tevatron_direct}.  The anomalous
magnetic moment of the muon may or may not seriously threaten the
Standard Model, but it will certainly disfavor many possible
interpretations of LHC signatures~\cite{g-2}. Flavor physics lead
to the postulation of additional symmetries in ultraviolet completions, 
an example being
supersymmetry~\cite{susy_flavor,hitoshi}.  And last but not least, the
measured relic density of the dark matter agent puts very stringent
constraints not only on the mass and coupling of such a candidate, but
also on other particles involved in the annihilation process or in its
(direct or indirect) detection~\cite{cdm}.\bigskip

Many new--physics scenarios do not simply predict a new narrow
resonance, such as for example a $Z'$.  Instead, a wealth of
measurements at the LHC, and later on at the ILC and other experiments
might be available, and with it the need to be combined properly.  The
situation could be similar to current fits of electroweak precision data,
but most likely it will be much more complex.  The LHC era with all its 
experiments can give a great many hints
about new--physics scenarios, it will certainly rule out large classes
of extensions of the Standard Model --- but it will definitely not give
a one--to--one map between a limited number of observables and a
well--defined small set of model parameters.

Bayesian probability distributions and frequentist profile likelihoods
are two ways to study an imperfectly measured parameter space, where
some model parameters might be very well determined, others heavily
correlated, and even others basically unconstrained. This situation is
different for example from $B$ physics, where theoretical degeneracies
and symmetries have become a major challenge~\cite{ckmfitter,
utfit}. A careful comparison of the benefits and traps of the
frequentist and the Bayesian approaches in the light of new--physics
searches is therefore necessary. SFitter follows both paths.\bigskip

If heavy strongly interacting particles can be produced at the LHC,
they will decay into lighter weakly interacting particles and finally
into the dark matter candidate~\cite{edges,cascade,per,lhc_dm}, with
decay cascades longer than the top--quark decay chain.  These cascade
measurements not only carry information on the masses of the particles
involved. The angular correlations also reflect the spins of the
particles in the cascade and allow tests for example of the SUSY
hypothesis against an extra--dimensional hypothesis~\cite{lhc_spins}.
At the ILC, detailed analyses of kinematically accessible particles
will be possible, for example using threshold scans~\cite{tesla_tdr}.
Masses, branching ratios as well as measurements of
particle spins will shed additional light on the underlying theory.
Currently, no attempt is made to measure discrete quantum numbers of
new--physics particles using SFitter. Instead the analysis is limited to
the continuous space of the parameters.\bigskip

In this paper the analysis will be restricted to the parameter point
SPS1a~\cite{sps}, only because it has been studied in detail by the
experimental communities at the LHC and ILC. After briefly reviewing
the experimental results and the treatment of the experimental and
theoretical errors in SFitter, the relevant features of these
measurements as well as the approach of SFitter~\cite{sfitter} will be illustrated 
in the MSUGRA model, before moving on to the weak--scale
MSSM.

The organization of the MSUGRA and the MSSM sections follows the
general logic of SFitter: First, a fully exclusive log-likelihood map of the
respective parameter space is constructed and a ranked list of the
best--fitting points is produced. We will see that already for the MSUGRA 
parameter space the LHC measurements will lead to strong correlations and
alternative likelihood maxima. The situation will become more complex 
in the case of the MSSM, where equally good alternative best--fitting
points are induced by the structure of the gaugino--higgsino mass parameters, 
the sign of the higgsino mass parameter, and the correlations between the
trilinear coupling in the top sector and the top--quark mass. This 
degeneracy would have to be broken by additional measurements, at the 
LHC or elsewhere.

Starting from the log-likelihood map we then use
frequentist and Bayesian constructions to study lower--dimensional
probability distributions including correlation effects. Again,
this analysis illustrates the complex structure of the MSSM parameter
space as well as the features of the statistics methods employed.
Finally, the weak--scale MSSM Lagrangian is reconstructed 
with proper experimental and theory error distributions. This 
weak--scale result should serve as a starting point to probe supersymmetry 
breaking bottom--up without theoretical bias. In
the appendices we discuss the techniques of SFitter using a simple toy
model.\bigskip

The approach of mapping measurements onto a high--dimensional
parameter space as well as the SFitter tool are completely
general\footnote{Fittino~\cite{fittino} follows a very similar logic to SFitter,
including a scan of the high--dimensional MSSM parameter space.}:
model parameters as well as measurements are included in the form of
model files and can simply be replaced.  SFitter serves as a general
tool to map typically up to 20--dimensional highly complex parameter
spaces onto a large sample of highly correlated measurements of
different quality.

\section{Collider Data}
\label{sec:measurements}

The analysis in this paper critically depends on detailed
experimental simulations of measurements and errors at the LHC and at
the ILC. Therefore the well--understood parameter point
SPS1a~\cite{sps} is used. This point has a favorable phenomenology for both
LHC and ILC. The original version SPS1a instead of the
dark--matter corrected SPA1/SPS1a' point is used, since cosmological
measurements like the relic density are not part of this
work~\cite{with_dan}.

\subsection{LHC and ILC measurements}

The parameter point SPS1a is characterized by moderately heavy squarks
and gluinos, which leads to long cascades including the neutralinos
and sleptons. Gauginos are lighter than Higgsinos, and the mass of the
lightest Higgs boson is close to the mass limit determined at LEP.
The summary of particle mass measurements is listed in
Table~\ref{tab:mass_errors}, taken from Ref.~\cite{giacomo}. The
central values are calculated by SuSpect~\cite{suspect}.  In general,
we see from the table that the LHC has the advantage of a better
coverage of the strongly interacting sparticle sector, whereas a
somewhat better coverage and precision in particular in the gaugino
sector can be obtained with the ILC~\cite{bagger,weak_corrections}. It
should be noted that the quoted LHC mass measurements are obtained
from measurements of kinematical endpoints and mass
differences~\cite{giacomo}, using the observables shown in
Table~\ref{tab:edges}. The systematic error quoted in these
measurements is essentially due to the uncertainty in the lepton and
jet energy scales, expected to be 0.1\% and 1\%, respectively, at the
LHC.  These energy--scale errors are each taken to be 99\% correlated
as discussed in Ref~\cite{giacomo}.\bigskip

Precision mass measurements at the LHC are not possible from the
measurement of production rates of certain final states, \ie
combinations of ($\sigma \cdot \br$). The reason are the sizeable QCD
uncertainties on the cross section~\cite{susy_prod}, often largely due
to gluon radiation from the initial state, but by no means restricted
to this one aspect of higher--order corrections. Generic errors on the
cross section alone of at least $20\%$, plus errors due to detector
efficiencies and coverage imply that one would only rely on ($\sigma
\cdot \br$)--type information in the absence of other useful
measurements~\cite{focuspoint,ayres_lhc}. For such cases, the
next--to-leading order production rates for strongly interacting
sparticles (based on Prospino2~\cite{prospino}) are implemented in
SFitter and can be readily included in the analysis. The same is true
for the branching ratios, where interfaces to MSMlib~\cite{msmlib} and
Sdecay/S-HIT~\cite{sdecay} are implemented. The QCD
corrections to measurements of the decay kinematics are known to be
under control: additional jet radiation is well described by shower
Monte Carlos~\cite{susy_jet} and will not lead to unexpected QCD
effects. Off-shell effects in cascade decays can of course be large
once particles become almost mass degenerate~\cite{off_shell}, but in
the standard SPS1a cascades these effects are expected to be
small.\bigskip

For the ILC, as a rule of thumb if particles are light
enough to be produced in pairs given the center-of-mass energy of the
collider, their mass can be determined with impressive accuracy. The
mass determination is possible either through direct reconstruction or
through a measurement of the cross section at production threshold
with comparable accuracy but different systematics. Precision
measurements of the branching ratios, \eg of the Higgs boson, are
possible.  Additionally discrete quantum numbers like the spin of the
particles can be determined similarly well.
\bigskip

\begin{table}[t]
\begin{small} \begin{center}
\begin{tabular}{|l|cccc||l|cccc|}
\hline
 & $m_{\rm SPS1a}$ & LHC & ILC & LHC+ILC &
 & $m_{\rm SPS1a}$ & LHC & ILC & LHC+ILC\\
\hline
\hline
$h$  & 108.99& 0.25 & 0.05 & 0.05 &
$H$  & 393.69&      & 1.5  & 1.5  \\
$A$  & 393.26&      & 1.5  & 1.5  &
$H+$ & 401.88&      & 1.5  & 1.5  \\
\hline
$\chi_1^0$ &  97.21& 4.8 & 0.05  & 0.05 &
$\chi_2^0$ & 180.50& 4.7 & 1.2   & 0.08 \\
$\chi_3^0$ & 356.01&     & 4.0   & 4.0  &
$\chi_4^0$ & 375.59& 5.1 & 4.0   & 2.3 \\
$\chi^\pm_1$ & 179.85 & & 0.55 & 0.55 &
$\chi^\pm_2$ & 375.72 & & 3.0  & 3.0 \\
\hline
$\tilde{g}$ &  607.81& 8.0 &  & 6.5 & & & & & \\
\hline
$\tilde{t}_1$ & 399.10&     &  2.0  & 2.0 & & & & & \\
$\tilde{b}_1$ & 518.87& 7.5 &       & 5.7 &
$\tilde{b}_2$ & 544.85& 7.9 &       & 6.2 \\
\hline
$\tilde{q}_L$ &  562.98&  8.7 & &  4.9 &
$\tilde{q}_R$ &  543.82&  9.5 & &  8.0 \\
\hline
$\tilde{e}_L$    & 199.66   & 5.0 & 0.2  & 0.2  &
$\tilde{e}_R$    & 142.65   & 4.8 & 0.05 & 0.05 \\
$\tilde{\mu}_L$  & 199.66   & 5.0 & 0.5  & 0.5  &
$\tilde{\mu}_R$  & 142.65   & 4.8 & 0.2  & 0.2  \\
$\tilde{\tau}_1$ & 133.35   & 6.5 & 0.3  & 0.3  &
$\tilde{\tau}_2$ & 203.69   &     & 1.1  & 1.1  \\
$\tilde{\nu}_e$  & 183.79   &     & 1.2  & 1.2  & & & & & \\
\hline
\end{tabular}
\end{center} \end{small} \vspace*{-3mm}
\caption[]{Errors for the mass determination in SPS1a, taken 
  from~\cite{giacomo}. Shown are the nominal parameter values (from SuSpect),
  the error for the LHC alone, from the LC alone, and from a combined 
  LHC+LC analysis. Empty boxes indicate that the particle cannot, to
  current knowledge, be observed 
  or is too heavy to be produced.
  All values are given in GeV.}
\label{tab:mass_errors}
\end{table}

\begin{table}[t]
\begin{small} \begin{center}
\begin{tabular}{|ll|r|rrrr|}
\hline
\multicolumn{2}{|c|}{ type of } & 
 \multicolumn{1}{c|}{ nominal } & 
 \multicolumn{1}{c|}{ stat. } & 
 \multicolumn{1}{c|}{ LES } & 
 \multicolumn{1}{c|}{ JES } & 
 \multicolumn{1}{c|}{ theo. } \\
\multicolumn{2}{|c|}{ measurement } & 
 \multicolumn{1}{c|}{ value } & 
 \multicolumn{4}{c|}{ error } \\
\hline
\hline
$m_h$ & 
 & 108.99& 0.01 & 0.25 &      & 2.0 \\
$m_t$ & 
 & 171.40& 0.01 &      & 1.0  &     \\
$m_{\tilde{l}_L}-m_{\chi_1^0}$ & 
 & 102.45& 2.3  & 0.1  &      & 2.2 \\
$m_{\tilde{g}}-m_{\chi_1^0}$ & 
 & 511.57& 2.3  &      & 6.0  & 18.3 \\
$m_{\tilde{q}_R}-m_{\chi_1^0}$ & 
 & 446.62& 10.0 &      & 4.3  & 16.3 \\
$m_{\tilde{g}}-m_{\tilde{b}_1}$ & 
 & 88.94 & 1.5  &      & 1.0  & 24.0 \\
$m_{\tilde{g}}-m_{\tilde{b}_2}$ & 
 & 62.96 & 2.5  &      & 0.7  & 24.5 \\
$m_{ll}^\mathrm{max}$: & three-particle edge($\chi_2^0$,$\tilde{l}_R$,$\chi_1^0$)  
 & 80.94 & 0.042& 0.08 &      & 2.4 \\
$m_{llq}^\mathrm{max}$: & three-particle edge($\tilde{q}_L$,$\chi_2^0$,$\chi_1^0$)  
 & 449.32& 1.4  &      & 4.3  & 15.2 \\
$m_{lq}^\mathrm{low}$: & three-particle edge($\tilde{q}_L$,$\chi_2^0$,$\tilde{l}_R$)
 & 326.72& 1.3  &      & 3.0  & 13.2 \\
$m_{ll}^\mathrm{max}(\chi_4^0)$: & three-particle edge($\chi_4^0$,$\tilde{l}_R$,$\chi_1^0$)
 & 254.29& 3.3  & 0.3  &      & 4.1 \\
$m_{\tau\tau}^\mathrm{max}$: & three-particle edge($\chi_2^0$,$\tilde{\tau}_1$,$\chi_1^0$)
 & 83.27 & 5.0  &      & 0.8  & 2.1 \\
$m_{lq}^\mathrm{high}$: & four-particle edge($\tilde{q}_L$,$\chi_2^0$,$\tilde{l}_R$,$\chi_1^0$)
 & 390.28& 1.4  &      & 3.8  & 13.9 \\
$m_{llq}^\mathrm{thres}$: & threshold($\tilde{q}_L$,$\chi_2^0$,$\tilde{l}_R$,$\chi_1^0$)
 & 216.22& 2.3  &      & 2.0  & 8.7 \\
$m_{llb}^\mathrm{thres}$: & threshold($\tilde{b}_1$,$\chi_2^0$,$\tilde{l}_R$,$\chi_1^0$)
 & 198.63& 5.1  &      & 1.8  & 8.0 \\
\hline
\end{tabular}
\end{center} \end{small} \vspace*{-3mm}
\caption[]{
LHC measurements in SPS1a, taken 
  from~\cite{giacomo}. Shown are the nominal values (from SuSpect) 
  and statistical errors, systematic errors from the lepton (LES)
  and jet energy scale (JES) and theoretical errors. 
  All values are given in GeV.}
\label{tab:edges}
\end{table}

\subsection{Error determination}
\label{sec:sugra_errors}

In order to obtain reliable error estimates for the fundamental
parameters, a proper treatment of experimental and theory errors
depending on their origin is mandatory. The CKMfitter
prescription~\cite{ckmfitter} is largely followed.  The complete set of
errors in the MSUGRA as well as in the MSSM analysis includes
statistical experimental errors, systematic experimental errors, and
theory errors. The statistical experimental errors are treated as
uncorrelated in the measured observables. In contrast, the systematic
experimental errors for example from the jet and lepton energy
scales~\cite{giacomo} are fully correlated. Hence, both are
non-trivially correlated in the masses determined from the
endpoints. Theory errors are propagated from the masses to the
measurements.  \bigskip

As there is no reason why unknown higher--order corrections should be
centered around a given value or even around zero, the theory error of
the weak--scale masses is not taken to be gaussian but flat
box--shaped: the probability assigned to any measurement does not
depend on its actual value, as long as it is within the interval
covered by the theory error. A tail could be attached to these
theory--error distributions, but higher--order corrections are
precisely not expected to become arbitrarily large. Confronted with a
perturbatively unstable observable one would instead have to rethink
the perturbative description of the underlying theory.

Taking this interval approach seriously impacts not only the
distribution of the theory error, but also its combination with the
combined (gaussian) experimental error. A simple convolution of a
box--shaped theory error with a gaussian experimental error leads to
the difference of two one--sided error functions.  Numerically, this
function will have a maximum, so the convolution still knows about the
central value of theoretical prediction. On the other hand, the
function is never flat and differentiable to arbitrarily high orders
at all points.

A better solution is a distribution which is flat as long as the
measured value is within the theoretically acceptable interval and
outside this interval drops off like a gaussian with the width of the
experimental error.  The log--likelihood $\chi^2 = -2 \log
\mathcal{L}$ given a set of measurements $\vec d$ and in 
the presence of a general correlation matrix $C$ reads
\begin{alignat}{7}
\chi^2     &= {\vec{\chi}_d}^T \; C^{-1} \; \vec{\chi}_d  \notag \\ 
\chi_{d,i} &=
  \begin{cases}
  0  
          &|d_i-\bar{d}_i | <   \sigma^{\text{(theo)}}_i \\
  \frac{\D |d_i-\bar{d}_i | - \sigma^{\text{(theo)}}_i}{\D \sigma^{\text{(exp)}}_i}
  \qquad  &|d_i-\bar{d}_i | >   \sigma^{\text{(theo)}}_i \; ,
  \end{cases}
\label{eq:flat_errors}
\end{alignat}
where $\bar{d}_i$ is the $i$-th data point predicted by the model
parameters and $d_i$ the actual measurement. This definition corresponds
to the RFit scheme described in Ref.~\cite{ckmfitter}. The
experimental errors are considered to be gaussian, so they are summed 
quadratically. The statistical error
is assumed to be uncorrelated between different measurements. The
first systematic error $\sigma^{(\ell)}$ originates from the lepton
energy scale and is taken as $99\%$ correlated between two
measurements. Correspondingly, $\sigma^{\text{(j)}}$ stems from the
jet energy scale and is also $99\%$ correlated. The correlations are
absorbed into the correlation matrix $C$
\begin{equation}
C_{i,i} = 1 \qquad \qquad \qquad
C_{i,j}  = 
C_{j,i}  = \frac{ 0.99 \; \sigma^{(\ell)}_i     \; \sigma^{(\ell)}_j
                 +0.99 \; \sigma^{\text{(j)}}_i \; \sigma^{\text{(j)}}_j }
                { \sigma^{\text{(exp)}}_i \; \sigma^{\text{(exp)}}_j} 
\; .
\end{equation}\bigskip

While box--shaped error distributions for observables are conceptually
no problem, they lead to a technical complication with hill--climbing
algorithms. All functions used to describe such a box--shaped
distribution will have a discontinuity of higher derivatives in at
least one point. The prescription above has a step in the second
derivative at $\bar{d} \pm \sigma^{\text{(theo)}}$, which leads to a
problem for example with Minuit's Migrad algorithm. Details on this
problem are given in the Appendix.

A second complication with flat distributions is that in the central
region the log--likelihood is a constant as a function of some model
parameters. In those regions these parameters vanish from the counting
of degrees of freedom. For all results shown in this paper flat
theory errors are assumed, unless stated otherwise. Results with different
theory errors are discussed in Sec.~\ref{sec:sugra_errors}.\bigskip

To determine the errors on the fundamental parameters two
techniques are used: a direct determination for the best fit using Minuit and a
statistical approach using sets of toy measurements. The advantage of
Minuit is that only one fit is necessary to determine the errors.  For
the non-gaussian error definition used above only Minos (of Minuit)
can be used, as it determines the intervals $\chi^2 \pm 1$ without
assuming gaussian errors. However, there is a complication because
of the flat region. Its algorithm computes the second derivative
of the log--likelihood for example in its convergence criterion. This
second derivative has two steps precisely in the region where one
would expect the algorithm to converge. Therefore the Minos
algorithm may not perform well with flat error distributions in the
log--likelihood.\bigskip

SFitter provides the option to smear the input measurement sets
according to their errors, taking into account the error form (flat or
gaussian) and the correlations \eg of the systematic energy scale
errors.  For each of the smeared toy--measurement sets SFitter
determines the best--fit value. The width of the distribution of the
best--fit values of a parameter gives the error on this
parameter. This option is time consuming (many fits are needed), but
necessary to be able to obtain the correct confidence level
intervals. Hence, this is the method used to determine the parameter
errors whenever flat theory errors are assumed.  For other cases this
smearing technique can be used as a cross-check.

\section{MSUGRA}
\label{sec:sugra}

No model for supersymmetry breaking should be assumed for
analyses. Instead, the breaking mechanism should be inferred from
data.  

However, the supersymmetric parameter
space can be simplified by unification assumptions, leading to an easily solvable
problem. A simple Minuit fit is sufficient to determine the
MSUGRA~\cite{msugra} parameters $m_0, m_{1/2}, A_0$ and $\tan\beta$
from the mass or endpoint measurements at the LHC and/or ILC.  The
correct sign of $\mu$ is determined by the quality of the fit which is
worse for the hypothesis with the wrong sign.  Such a fit can be an
uncorrelated gaussian $\chi^2$ fit or it can include all correlations
and correlated errors, and none of the errors have to be assumed to be
gaussian. Using SFitter a log-likelihood fit is performed, extracting
the best--fitting point in the respective MSUGRA (or later MSSM)
parameter space and determining the errors including all
correlations.\bigskip

Because of the sizeable error on the top mass (LHC target: 1~GeV; ILC
target: 0.12~GeV), the top mass or Yukawa must be included in any SUSY
fit~\cite{ben}. In a way, the running top Yukawa is defined at the
high scale as one of the MSUGRA model parameters, which through
coupled renormalization group running predicts all low--scale masses,
including the top--quark mass, all supersymmetric partner masses, and
the light Higgs mass~\cite{m_h}. In principle, this approach should be
taken for all Standard Model parameters, couplings and
masses~\cite{ben}, but at least for moderate values of $\tan\beta$ 
for example the bottom Yukawa coupling has
negligible impact on the extraction of supersymmetric parameters.

The first question to be discussed in the simplified MSUGRA context is
whether it is possible to unambiguously identify the correct
parameters from a set of observables and their errors. In other words,
which parameter point has the largest likelihood value $p(d|m)$,
evaluated as a function over the model--parameter space $m$ for a
given data set $d$. Note that discrete model assumptions (like MSUGRA
vs extra dimensions) are not included. Instead, one model with
a multi--dimensional vector of continuous parameters is scanned. The
question immediately arises if there are secondary maxima in the
likelihood map of the parameter space.\bigskip

In a one--dimensional problem the probability distribution function
(pdf) $p(d|m)$ for an observable $d$ given a vector of model
parameters $m$ can be used to compare two hypotheses for a given data
set: decide which of the two hypotheses $m_j$ with their
central values $d_j^*$ is preferred and compute the integral over the
`wrong' pdf $p(d|m_{\rm wrong})$ from $d_{\rm right}^*$ to
infinity. This integral gives the confidence level of the decision in
favor of one of the two hypotheses. Note that this extraction applies
to discrete and to continuous parameter determination, but it requires
that we start from a mathematically properly defined pdf in the
observable space.

For the procedure described above the Neyman--Pearson lemma states
that if the correct hypothesis is picked as `right', a
likelihood--ratio estimator will produce the smallest possible type-II
error, \ie the smallest error caused by mistaking a fluctuation of
`wrong' for `right'. A likelihood ratio can be extracted from
simulations~\cite{kyle}, or from data combined with
simulations~\cite{topmass} or from data alone~\cite{ckmfitter}. To
test well--defined hypotheses using powerful data, including for
example the top--mass measurement, likelihood methods can yield
impressive results. Such a likelihood method can easily be generalized
to high--dimensional observable spaces or model--parameter vectors, as
long as it is applied to properly defined probability distributions. The
crucial and highly controversial question is how to produce a pdf when
the parameter space is high--dimensional and
poorly constrained dimensions of it are ignored.\bigskip

{\sl SFitter provides the relevant frequentist or Bayesian results in
three steps:} first (1), SFitter computes a log-likelihood map of the
entire parameter space. This map is completely exclusive,
\ie it includes all dimensions in the parameter space. Then (2),
SFitter ranks the best local likelihood maxima in the map according to
their log--likelihood values. It identifies the global maximum, and a
bias towards secondary maxima (\eg SUSY breaking scenario) can be
included, without mistaking such a prior for actual likelihood.  Last
(3), SFitter computes profile likelihood or Bayesian probability maps
of lower dimensionality, down to one--dimensional distributions, by
properly removing or marginalizing unwanted parameter dimensions. Only
in this final step frequentist and Bayesian approaches need to be
distinguished. The three steps are illustrated in the Appendix for a
simple toy model.\bigskip

\subsection{Likelihood analysis}
\label{sec:sugra_likelihood}

\begin{figure}[t]
 \begin{minipage}{6cm}
    \begin{tabular}{l|rrrrrr}
     $\chi^2$&$m_0$ &$m_{1/2}$ &$\tan\beta$&$A_0$&$\mu$&$m_t$ \\ \hline
     0.09  &102.0 & 254.0 & 11.5 & -95.2  & $+$ & 172.4 \\
     1.50  &104.8 & 242.1 & 12.9 &-174.4  & $-$ & 172.3 \\
     73.2  &108.1 & 266.4 & 14.6 & 742.4  & $+$ & 173.7 \\
    139.5  &112.1 & 261.0 & 18.0 & 632.6  & $-$ & 173.0 \\
          \dots
     \end{tabular}
 \end{minipage} 
 \hspace*{2cm}
 \begin{minipage}{6cm}
 \includegraphics[width=6cm]{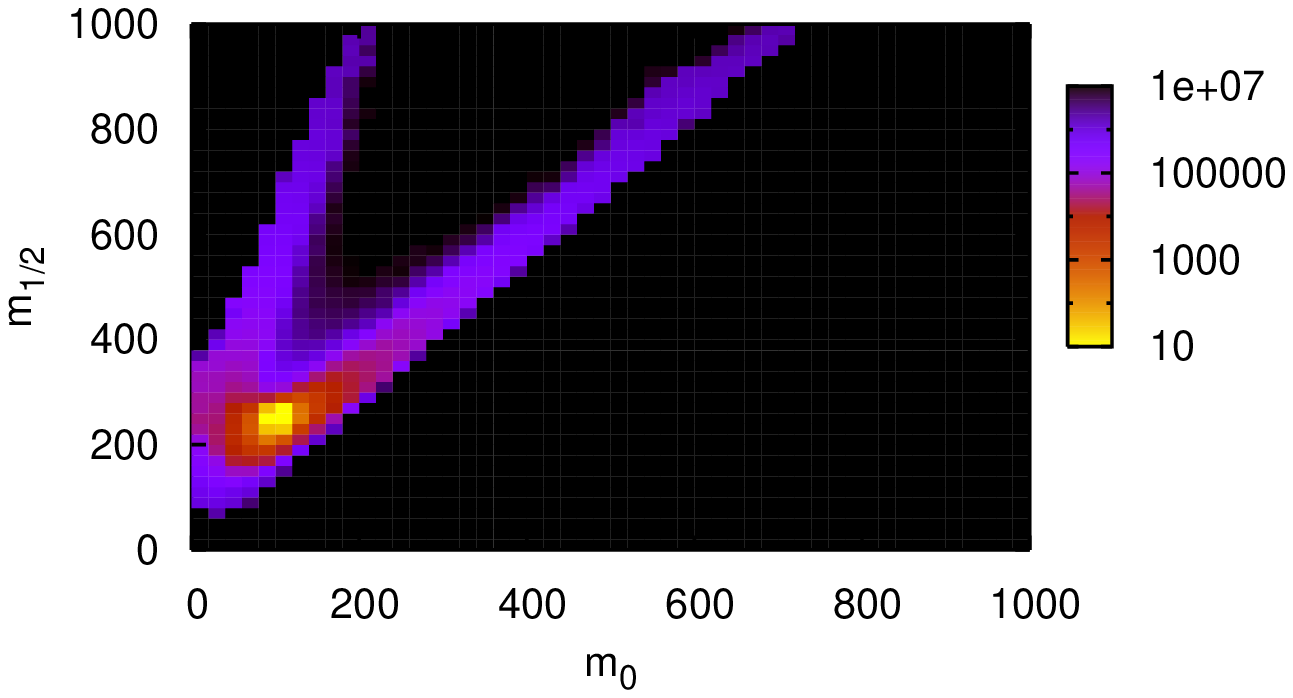}
 \end{minipage} \\
  \includegraphics[width=5cm]{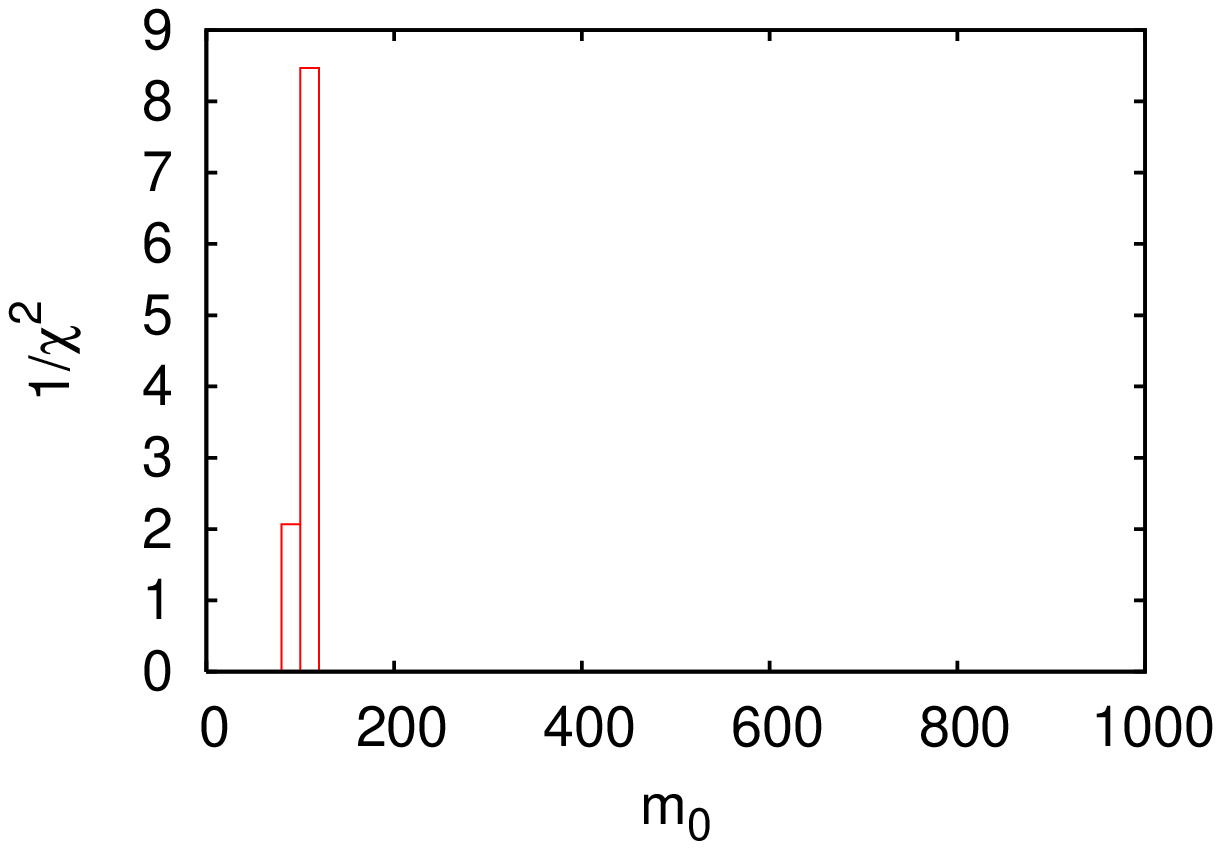} \hspace*{2cm}
  \includegraphics[width=5cm]{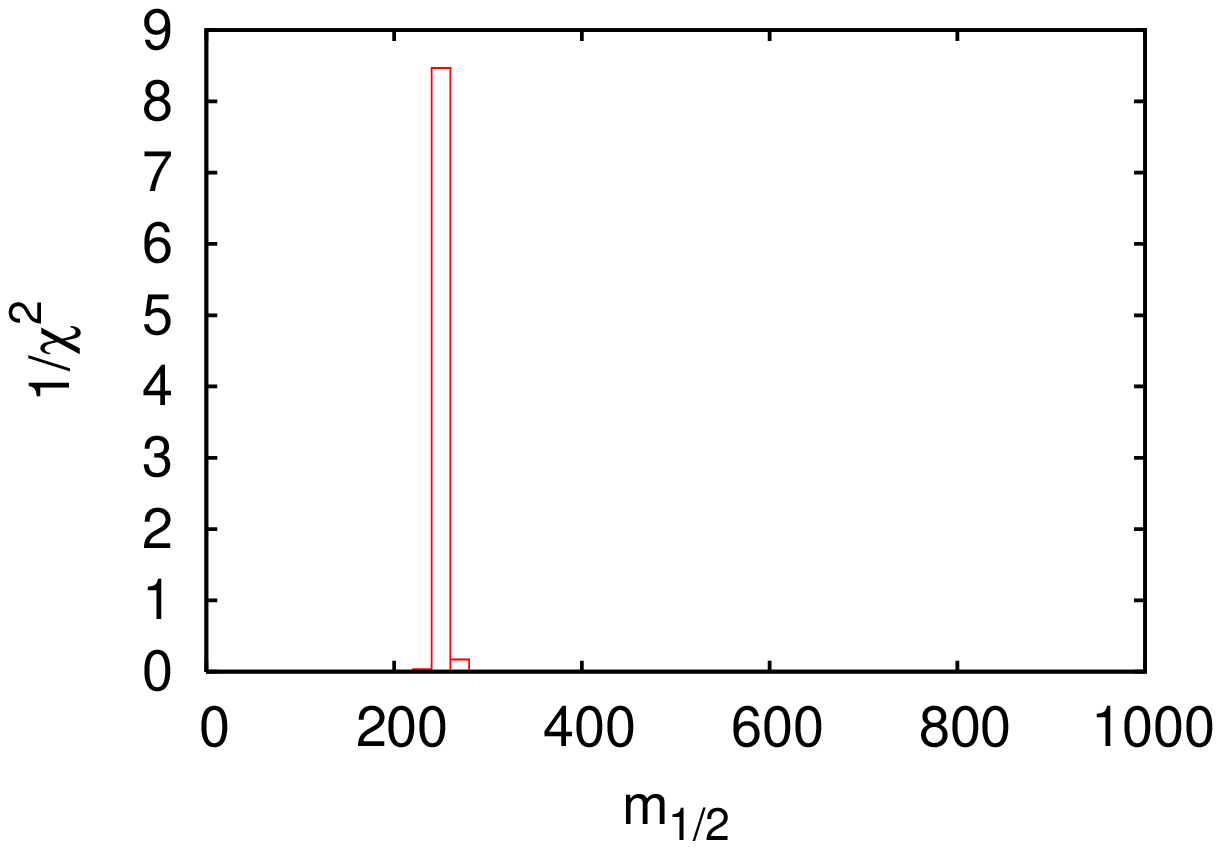}
\caption[]{SFitter output for MSUGRA in SPS1a.
  Upper left: list of the best log--likelihood values over the MSUGRA
  parameter space. Upper right:
  two--dimensional profile likelihood $\chi^2$ over the
  $m_0$--$m_{1/2}$ plane.  Lower: one--dimensional profile likelihoods
  $1/\chi^2$ for $m_0$ and $m_{1/2}$. All masses are given in GeV.}
\label{fig:sugra_map_f1}
\end{figure}

Looking for example at the parameter point SPS1a at the LHC, different
parameters are heavily correlated, some parameters are only poorly
constrained, and distinct different maxima in $\mathcal{L} \sim
\chi^2$ can differ by $\om(N)$, where $N$ is the number of
observables. Therefore, one would like to produce probability
distributions or likelihoods over subspaces of the model--parameter
space from the fully exclusive likelihood map. In other words,
unwanted dimensions of the parameter space are eliminated until only
one-- or two--dimensional `likelihoods' remain. The
likelihood cannot just be integrated unless a measure is defined in
the model space. This measure automatically introduces a bias and
leads to a Bayesian pdf.

Instead, in this section a profile likelihood is used: for each
(binned) parameter point in the $(n-1)$--dimensional space we explore
the $n$th direction which is to be removed
$\mathcal{L}(x_{1,...,n-1},x_n)$. The best value of $\mathcal{L}^{{\rm
max}(n)}$ is picked along this direction and its function value is
identified with the lower--dimensional parameter point
$\mathcal{L}(x_{1,...,n-1}) \equiv \mathcal{L}^{{\rm
max}(n)}(x_{1,...,n-1},x_n)$. Using this kind of projection most
notably guarantees that the best--fit points always survives to the
final representation, unless two of them belong to the same bin in the
reduced parameter space.\bigskip

For the MSUGRA case the likelihood map is computed over the
entire parameter space given a smeared LHC data set. This map covers
the model parameters $m_0, m_{1/2}, A_0, B, m_t$, where $B$ is later
traded for the weak--scale $\tan\beta$, as described in
Sec.~\ref{sec:sugra_high_scale}. Usually $\tan\beta$ will be shown,
because this parameter has a more obvious interpretation in the
weak--scale theory.\bigskip

The SFitter result is shown in Fig.~\ref{fig:sugra_map_f1}: a
completely exclusive map over the 5-dimensional parameter space is the
starting point. Combining 30 Markov chains 600000
model-parameter points are collected. For the renormalization group running 
SoftSUSY~\cite{softsusy} is used with an efficiency of $25 \cdots 30\%$, which
corresponds to a few hours of CPU time for each of the 30 chains. 
Because the resolution of the Markov chain is not sufficient to
resolve each local maximum in the log-likelihood map, an additional
maximization algorithm (Minuit's Migrad) starts at the best points of
the Markov chains to identify the local maxima. 

In Fig.~\ref{fig:sugra_map_f1} the best--fit points in
the MSUGRA parameter space are shown, as obtained from the 5-dimensional
likelihood map. For the SPS1a parameter point a general pattern
of four distinct maxima emerges in the likelihood: first, the trilinear
coupling can assume the correct value of around $-100$~GeV, but it can
also become large and positive $\sim 700$~GeV. This degeneracy is
correlated with a slight shift in the top mass, which means it will be
much less pronounced if the top quark mass is not part of the MSUGRA
parameters set. This correlation occurs through the light Higgs mass and
its strength largely depends on the theory error assumed for the Higgs
mass. Secondly, a similar feature is present for each sign of $\mu$,
correlated with a slight shift in $\tan\beta$, which compensate each
other in the neutralino--chargino sector. Such a degeneracy is expected,
because at the LHC only one of the two heavy neutralinos are observed.
Including the precise and more complete ILC measurements this degeneracy
should vanish.\bigskip

An example correlation between two model parameters is the
profile likelihood in the $m_0$--$m_{1/2}$ plane, after projecting
away the $A_0$, $B$, $\sign(\mu)$ and $m_t$ directions. The likelihood
maximum starts from the true values $m_0=100$~GeV and
$m_{1/2}=250$~GeV and continues into two branches. These branches
reflect the fact that extracting masses from kinematic endpoints
involves quadratic equations. Ignoring such correlations between
parameters the two-dimensional profile likelihood is projected onto each
of the two remaining directions. Both distributions show sharp maxima
of the profile likelihoods in the correct places, because the
resolution is not sufficient to resolve the four distinct solutions
for $A_0$ and ${\rm sign}(\mu)$. Note that all these profile
likelihood distributions are mathematically not probability
distributions, because projecting on a parameter subspace does not
protect the normalization of the original likelihood map (which can be
viewed as a probability distribution).\bigskip

\begin{figure}[t]
 \includegraphics[width=6cm]{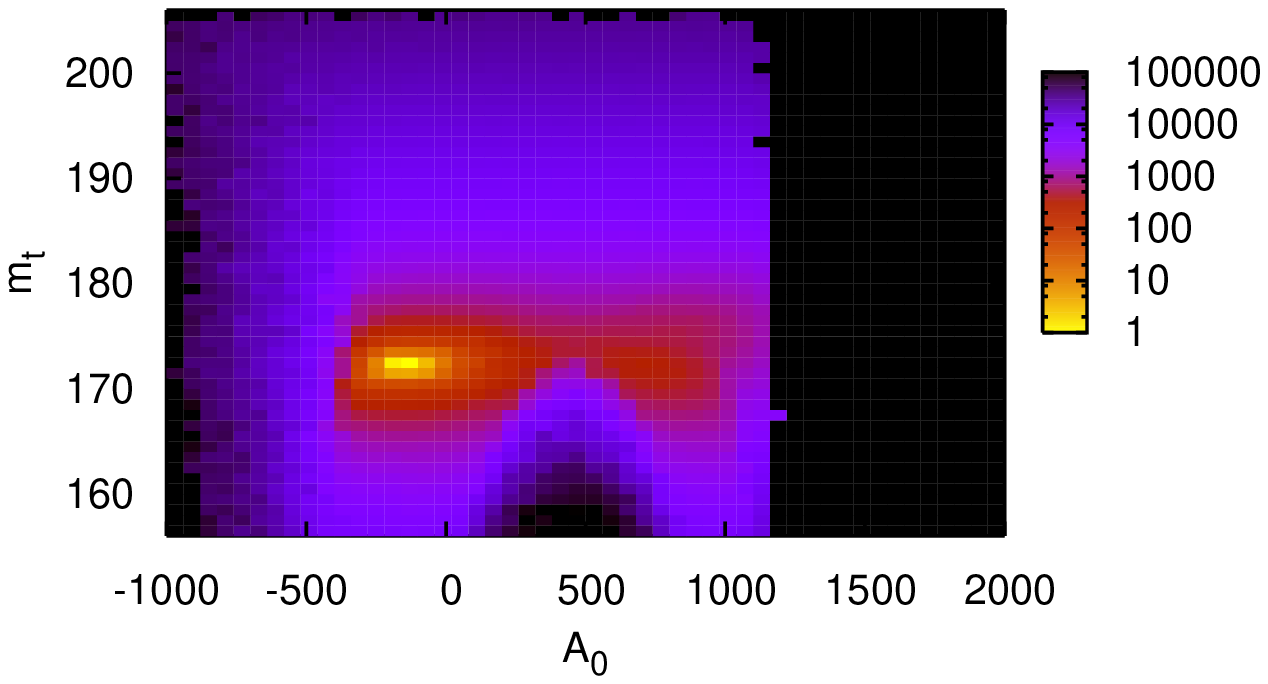} \hspace*{2cm}
 \includegraphics[width=6cm]{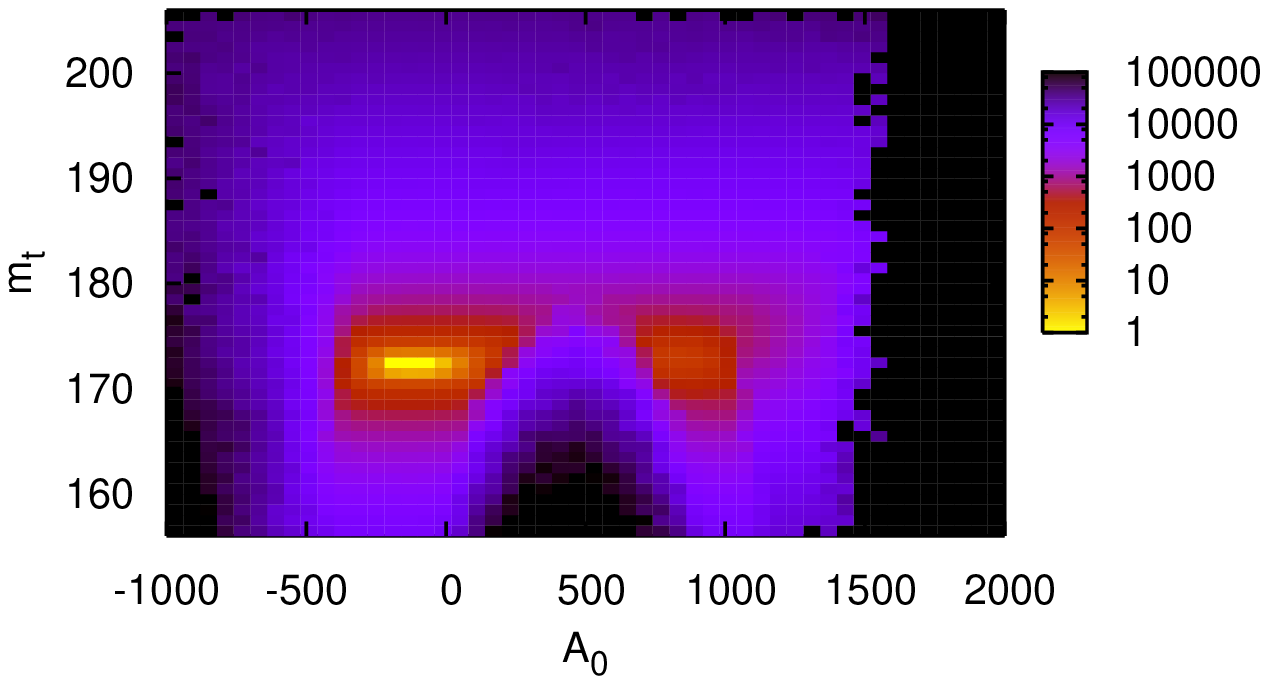} \\
 \includegraphics[width=5cm]{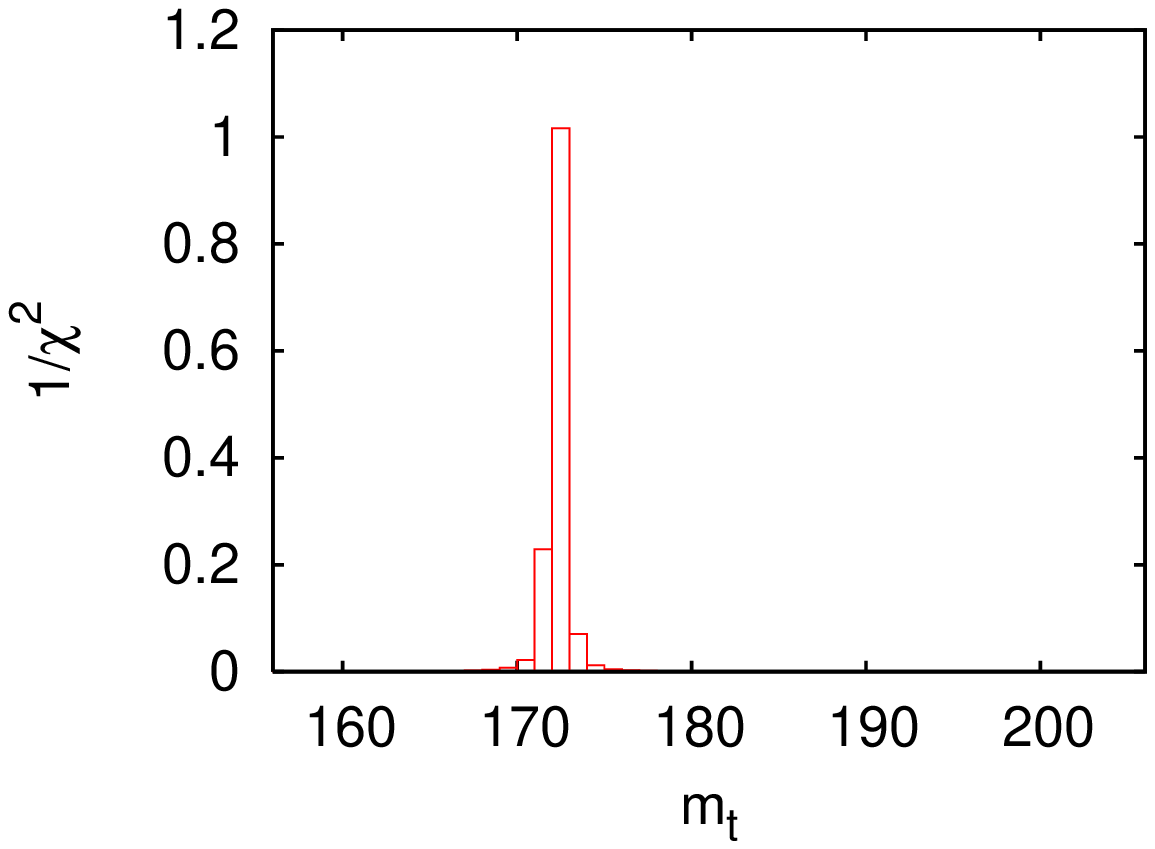} \hspace*{3cm}
 \includegraphics[width=5cm]{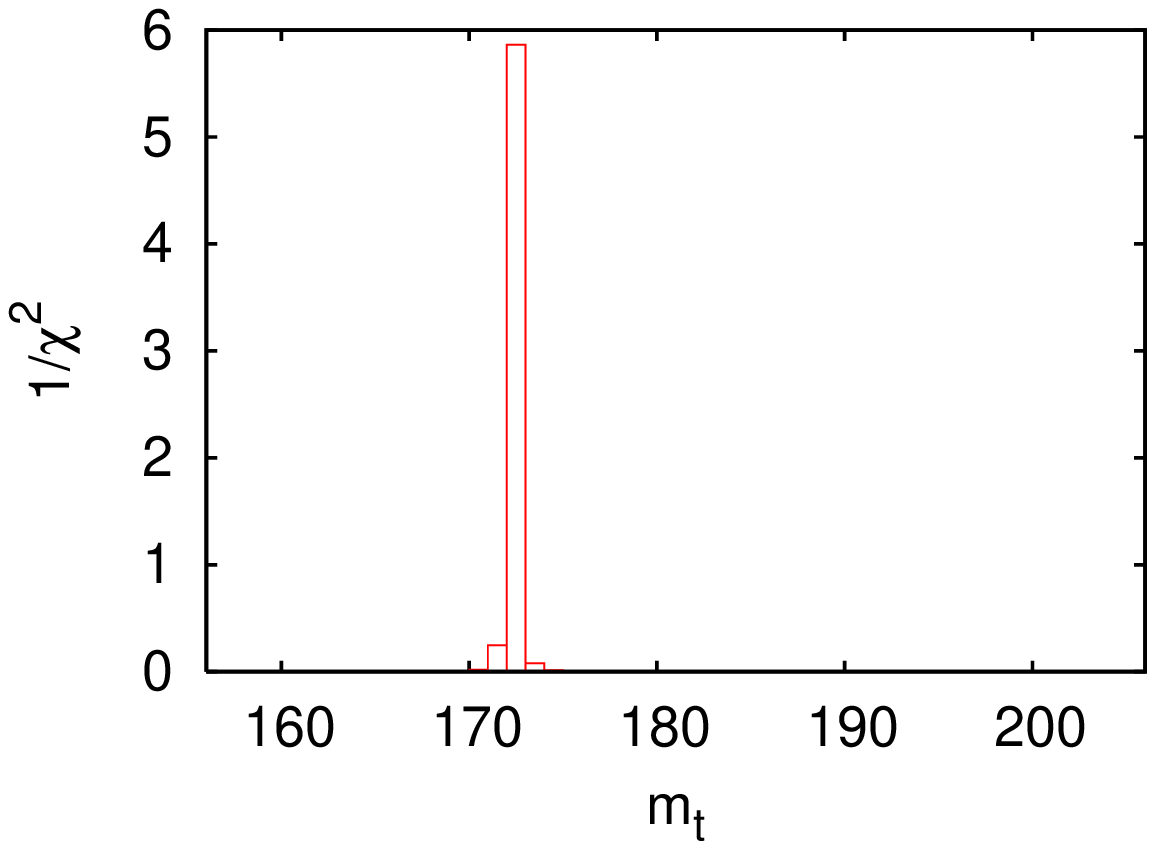} \\
 \includegraphics[width=5cm]{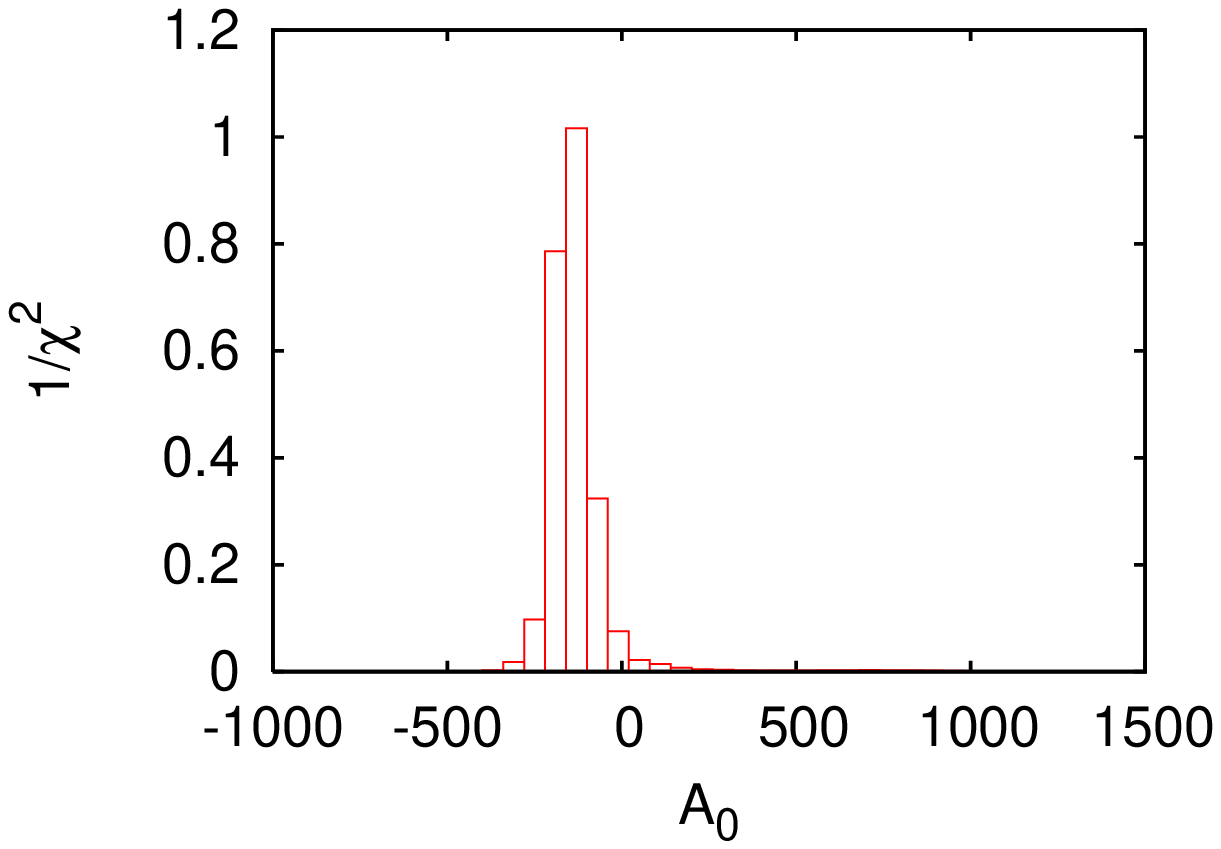} \hspace*{3cm}
 \includegraphics[width=5cm]{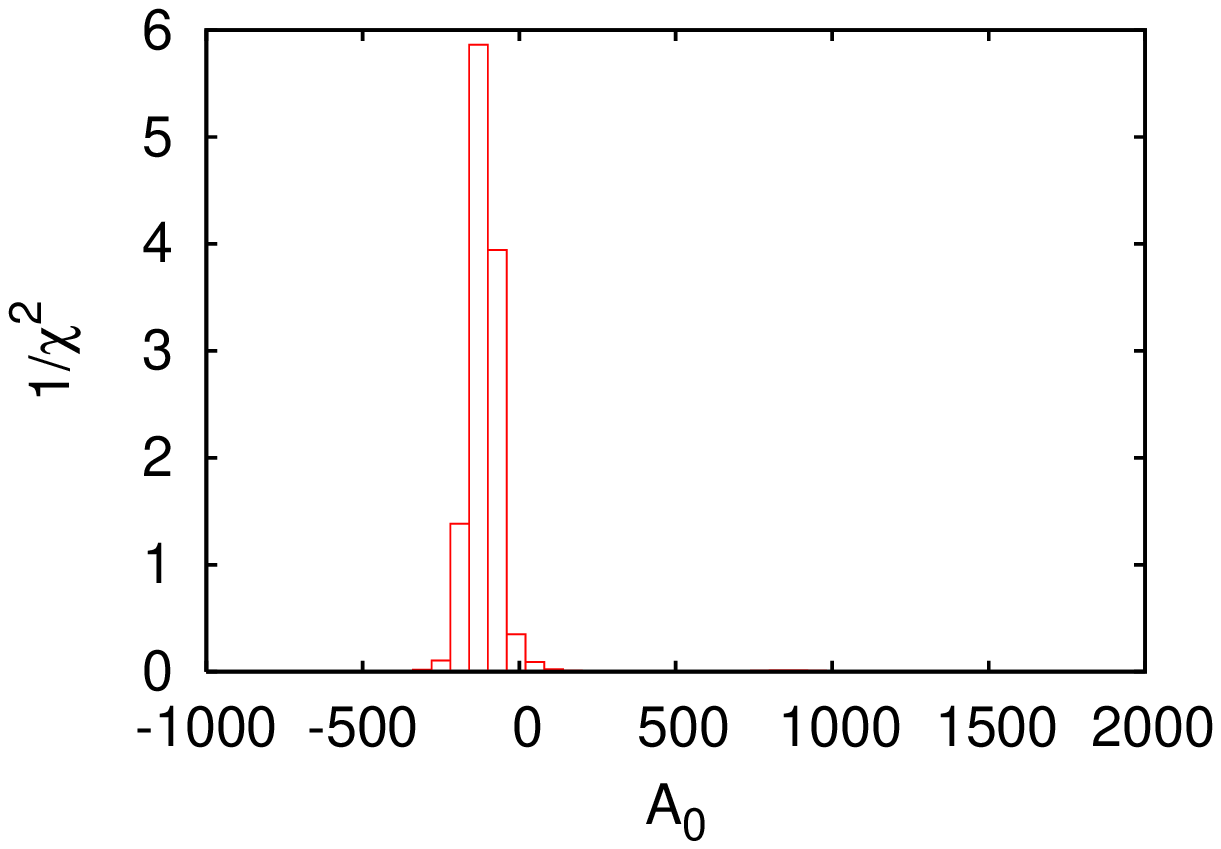} \\
\caption[]{SFitter output for $A_0$ and $y_t$. The two
 columns of one-- and two--dimensional profile likelihoods correspond
 to $\mu<0$ (left) and $\mu>0$ (right). A $\chi^2$ map is shown in the
 first row and $1/\chi^2$ distributions in the second and third.}
\label{fig:sugra_map_f2}
\end{figure}

Thus eliminating a dimension in the
parameter space means loss of information. Therefore, it is not
obvious that producing low--dimensional distributions from the
completely exclusive likelihood map is always sensible. An example is
the correlation of $m_t$ and $A_0$ --- as mentioned before
a strong correlation from the Higgs mass measurement is expected. 
Fig.~\ref{fig:sugra_map_f2} shows the two--dimensional and
one--dimensional profile likelihoods in the $m_t$--$A_0$ subspace.  In
the two columns the two signs of $\mu$ are separated; from the list of
maxima the best--fit points are expected to be roughly 1~GeV higher in
$m_t$ and $30\cdots 80$~GeV higher in $A_0$ for $\mu >0$.

Locally, the two--dimensional profile likelihoods around the maxima
show little correlation between $m_t$ and $A_0$. The correct value
around $A_0=-100$~GeV is preferred, but the alternative solution
around $A_0=900$~GeV is clearly visible. On top of this
double--maximum structure for both signs of $\mu$ there is a
parabola--shaped correlation between the $m_t$ and $A_0$. The apex of
the parabola is roughly 5~GeV above the best fits in $m_t$. This
correlation becomes invisible once one of the two parameter directions
are projected away and the one--dimensional profile
likelihoods are analyzed. The two alternative solutions do not appear in the $m_t$
histogram, because the alternative maximum is relatively unlikely and
because the two best--fit values for $m_t$ differ by a mere GeV. The
same is true for $A_0$ where only a tiny tail towards the
wrong solution can be seen.\bigskip


Since only one measurement smeared according to the gaussian
experimental errors is used for the parameter extraction shown in
Fig.~\ref{fig:sugra_map_f1}, the correct values do not have to
coincide with the best log-likelihood among the local maxima. As a
matter of fact, just changing the theory errors from the correct flat
to a possibly approximate gaussian shape can have an effect on the
ranking of maxima: for gaussian theory errors the $\chi^2$ values of
4.35, 26.1, 10.5, 22.6 appear in the order shown in
Fig.~\ref{fig:sugra_map_f1}. In other words, just smearing the
measurements can indeed shift the ordering of the best local maxima,
supporting our claim that a careful look at more than just
the best solution might make sense in a parameter space as complex as
MSUGRA.

Even if such inversions arise, the parameter determination can be
repeated with different (smeared) sets of observables. The frequency
with which the wrong parameter set corresponds to the lowest $\chi^2$
value is a measure how seriously degenerate the alternative maxima
are.

\subsection{Bayesian approach}
\label{sec:sugra_bayesian}

\begin{figure}[t]
 \begin{minipage}{6cm}
    \begin{tabular}{l|rrrrrr}
     $\chi^2$&$m_0$ &$m_{1/2}$ &$\tan\beta$&$A_0$&$\mu$&$m_t$ \\ \hline
     0.09  &102.0 & 254.0 & 11.5 & -95.2  & $+$ & 172.4 \\
     1.50  &104.8 & 242.1 & 12.9 &-174.4  & $-$ & 172.3 \\
     73.2  &108.1 & 266.4 & 14.6 & 742.4  & $+$ & 173.7 \\
    139.5  &112.1 & 261.0 & 18.0 & 632.6  & $-$ & 173.0 \\
          \dots
     \end{tabular}
 \end{minipage} 
 \hspace*{2cm}
 \begin{minipage}{6cm}
 \includegraphics[width=6cm]{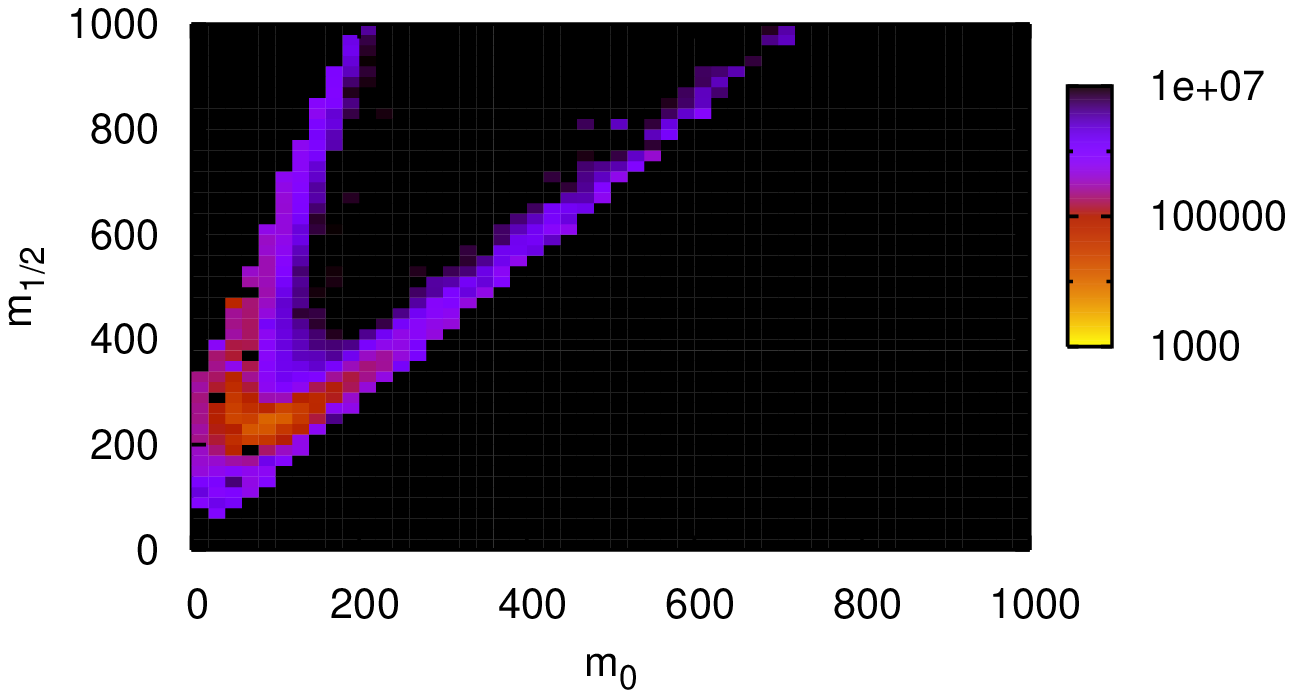}
 \end{minipage} \\
  \includegraphics[width=5cm]{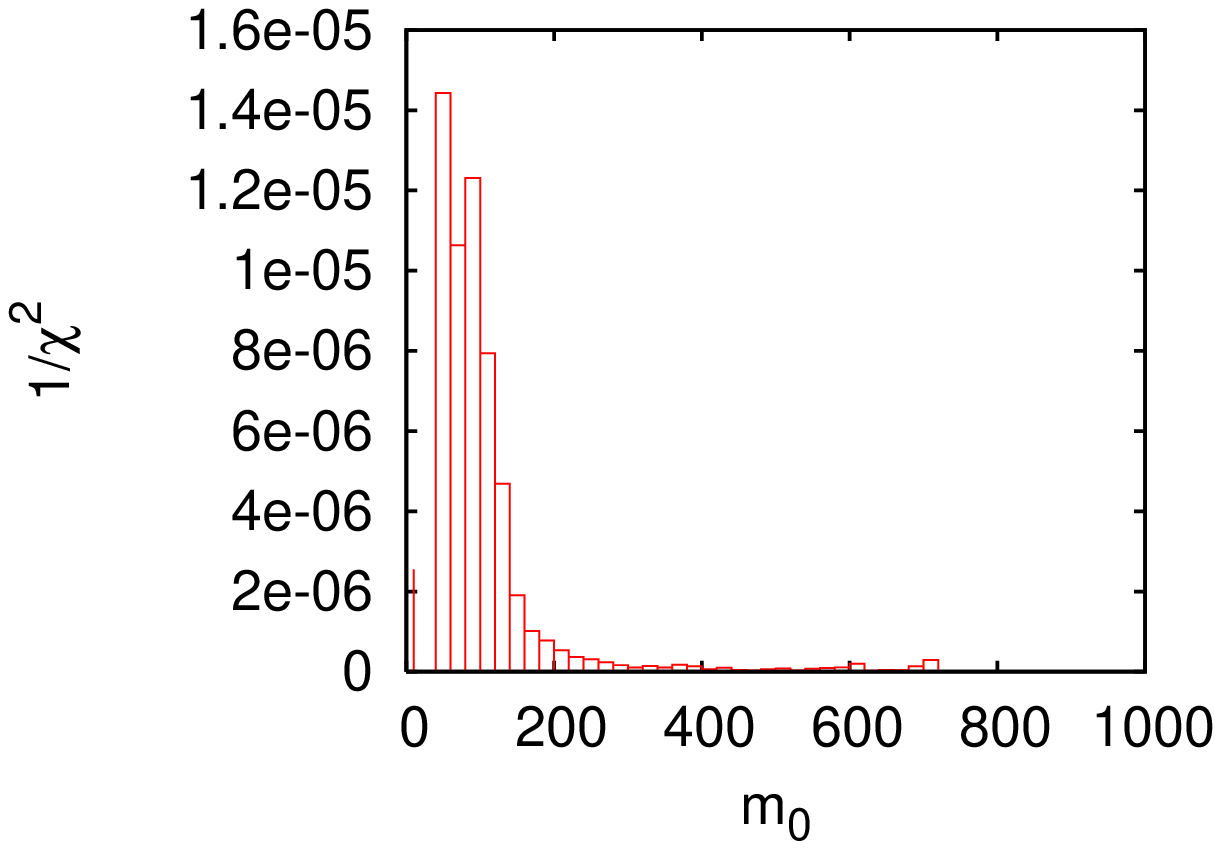} \hspace*{2cm}
  \includegraphics[width=5cm]{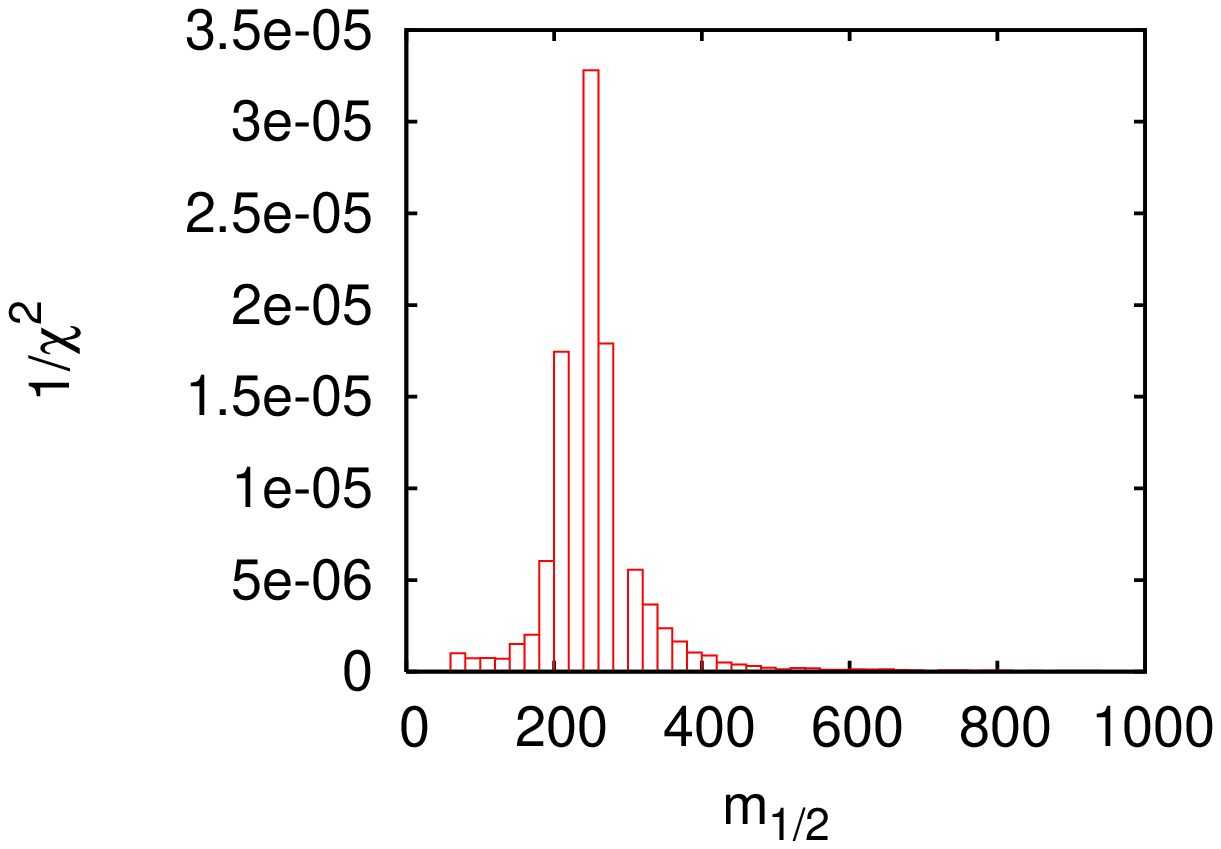}
\caption[]{SFitter output for MSUGRA in SPS1a.
  Upper left: list of the largest log--likelihood values over the
  MSUGRA parameter space. Upper right:
  two--dimensional Bayesian pdf $\chi^2$ over the $m_0$--$m_{1/2}$
  plane marginalized over all other parameters.  Lower:
  one--dimensional Bayesian pdfs $1/\chi^2$ for $m_0$ and $m_{1/2}$.
  All masses are given in GeV.}
\label{fig:sugra_map_b1}
\end{figure}

A likelihood analysis as presented in the last section is
unfortunately not designed to produce probability distributions for
model parameters. This means it will not answer questions of the kind:
in the light of electroweak precision constraints and dark matter
constraints, what sign of $\mu$ is preferred in
MSUGRA~\cite{dark_side}? Note that this is not the same question as:
what is the relative difference in the likelihood for the two best
points on each side of $\mu$. To answer the first question 
the likelihood over each of the two halves of the parameter
space needs to be integrated over. 
All parameter dimensions except for $\mu$ must be integrated
over to compute the pdf for $\mu$ given the data. For such an
integration leading to lower--dimensional probability distributions a
measure has to be introduced, the (Bayesian) prior. This prior has its
advantages, but it can also lead to unexpected effects, as shown in
the following~\cite{ben}.\bigskip

One might argue that such questions are irrelevant because the goal is
to find the correct, \ie the most likely parameter
point. On the other hand, asking for a reduced--dimensionality
probability density could well be a very typical situation in the LHC
era. Questions like: what kind of linear collider
should be built given LHC data? What is the most likely mechanism for
dark--matter annihilation? How to detect dark matter? 
deserve well--defined answers.\bigskip

As discussed before, shifting from a frequentist to a Bayesian
approach does not affect the main part of the SFitter program. Or in
other words, SFitter produces Bayesian probability distributions or
profile likelihoods without any preference.  While not strictly
necessary in a Bayesian analysis, the top--likelihood points from
Fig.~\ref{fig:sugra_map_f1} also appear in the Bayesian results shown
in Fig.~\ref{fig:sugra_map_b1}. The second panel in
Fig.~\ref{fig:sugra_map_b1} now shows a two--dimensional
representation of the Bayesian pdf over the MSUGRA parameter
space. All parameter dimensions except for $m_0$ and $m_{1/2}$ are
marginalized using flat priors.  The only slight complication arises
from the treatment of $B$ or $\tan\beta$, as described in
Sec.~\ref{sec:sugra_high_scale}. Unless explicitly stated otherwise
the prior is flat in the high--scale mass parameter $B$. The
results are typically shown in terms of $\tan\beta$, because this
parameter is easier to interpret at the weak scale.

In the two--dimensional pdf shown in Fig.~\ref{fig:sugra_map_b1} the
same two--branch structure appears as for the profile
likelihood. However, there are two differences: first, the area around
the true parameter point is less pronounced in the Bayesian pdf,
compared to the profile likelihood. In the integration over a
direction in parameter space noise gets collected from regions with a
finite but insignificant likelihood. This noise washes out the peaked
structures, while the profile likelihood by construction keeps mainly
these best--fit structures. This effect also considerably smears the
one--dimensional Bayesian pdf distributions in $m_0$ and $m_{1/2}$.

Secondly, the branch structure is more pronounced in
Fig.~\ref{fig:sugra_map_b1}.  While in the profile likelihood the area
between the two branches is filled by single good parameter points in
the parameters projected away, the Bayesian marginalization provides
`typical' likelihood values in this region which in general does not
fit the data as well.\bigskip

\begin{figure}[t]
 \includegraphics[width=6cm]{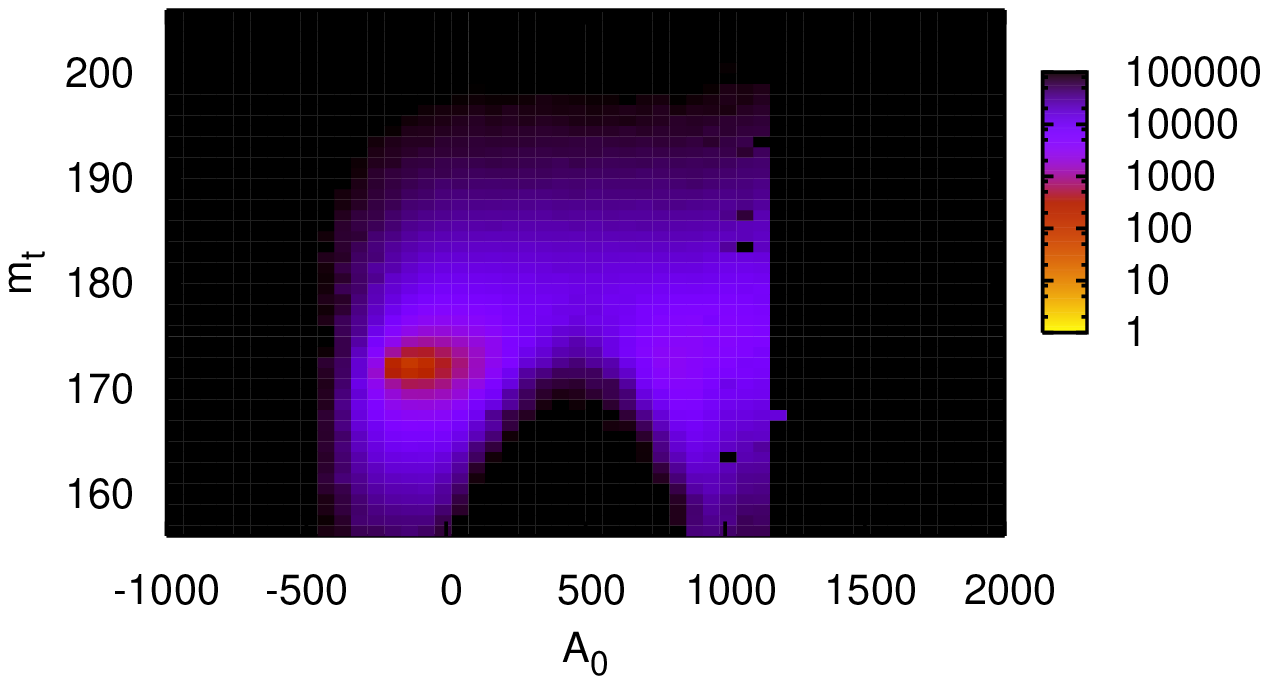} \hspace*{2cm}
 \includegraphics[width=6cm]{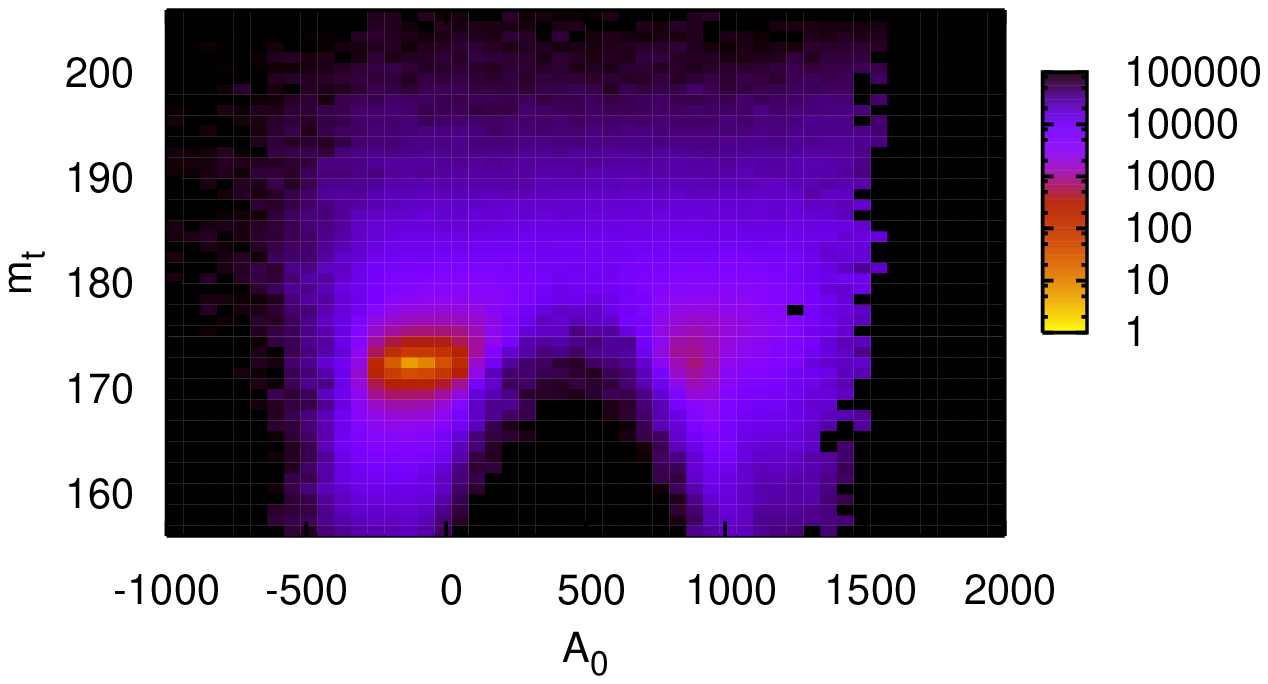} \\
 \includegraphics[width=5cm]{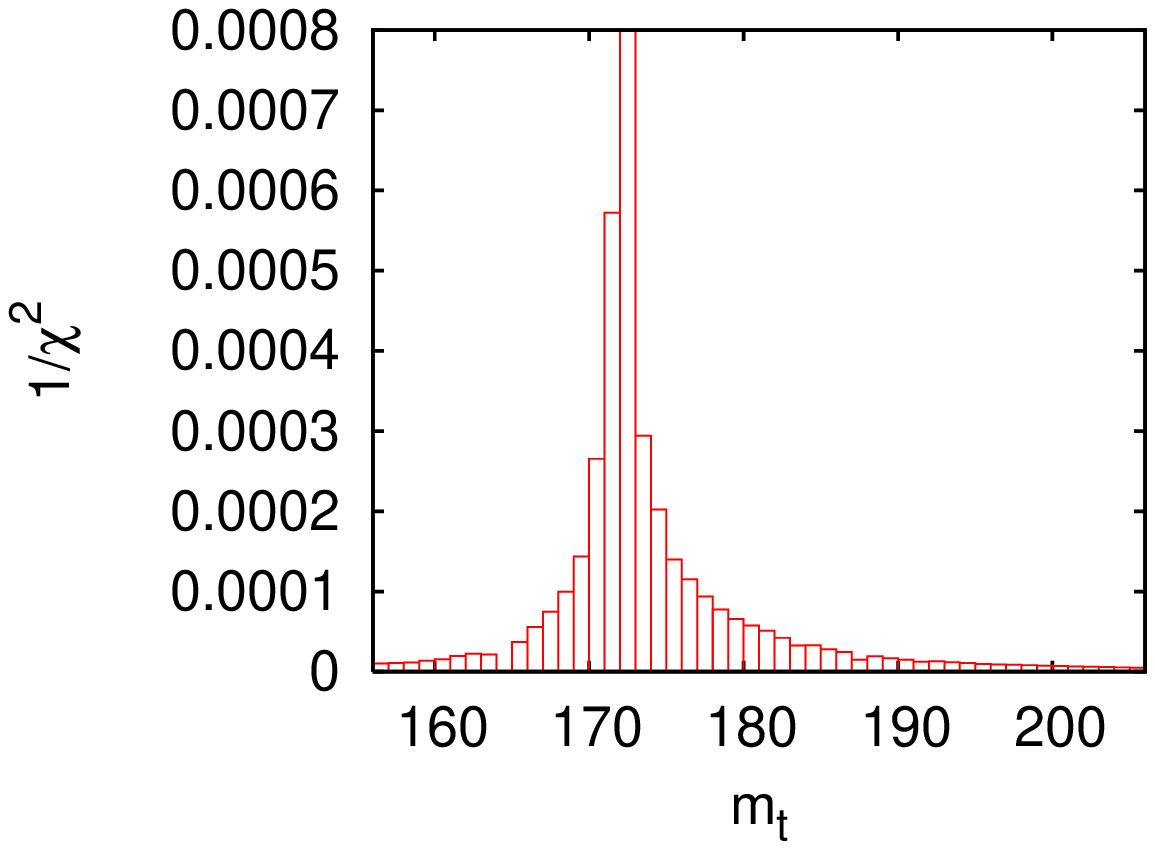} \hspace*{3cm}
 \includegraphics[width=5cm]{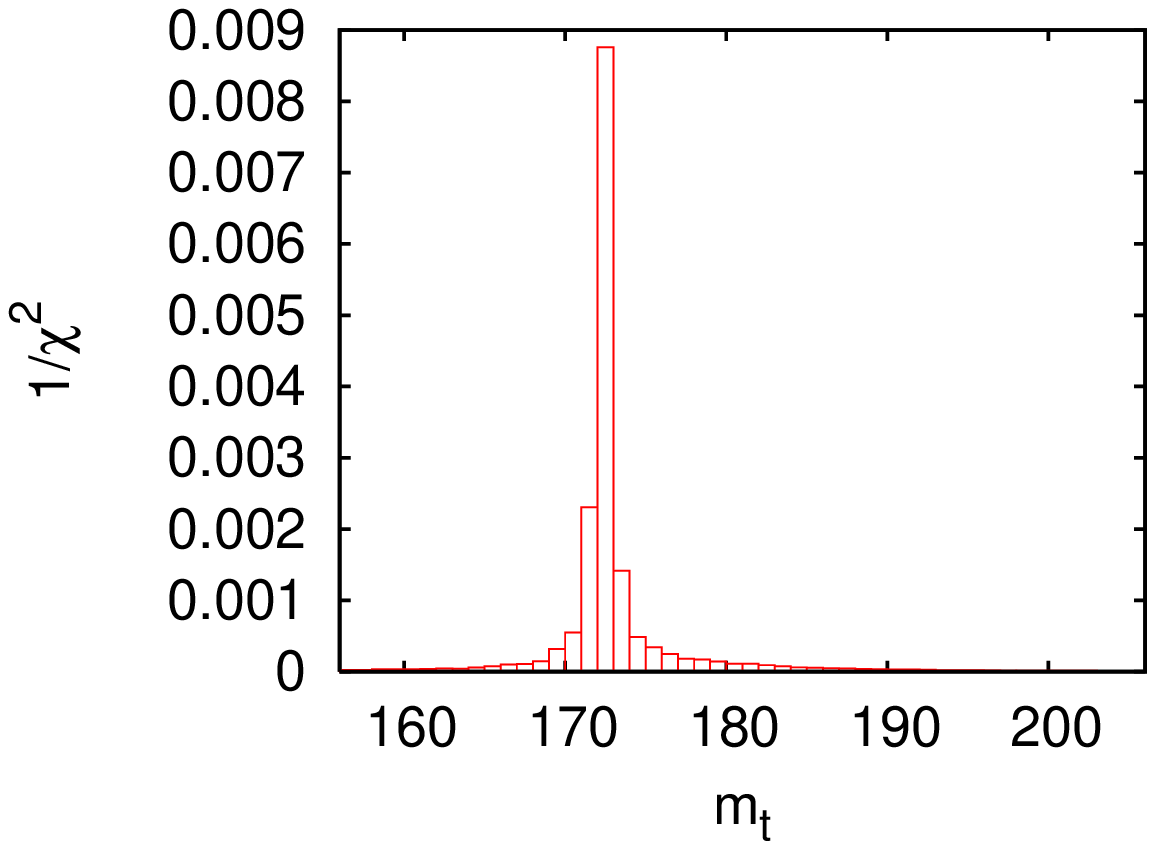} \\
 \includegraphics[width=5cm]{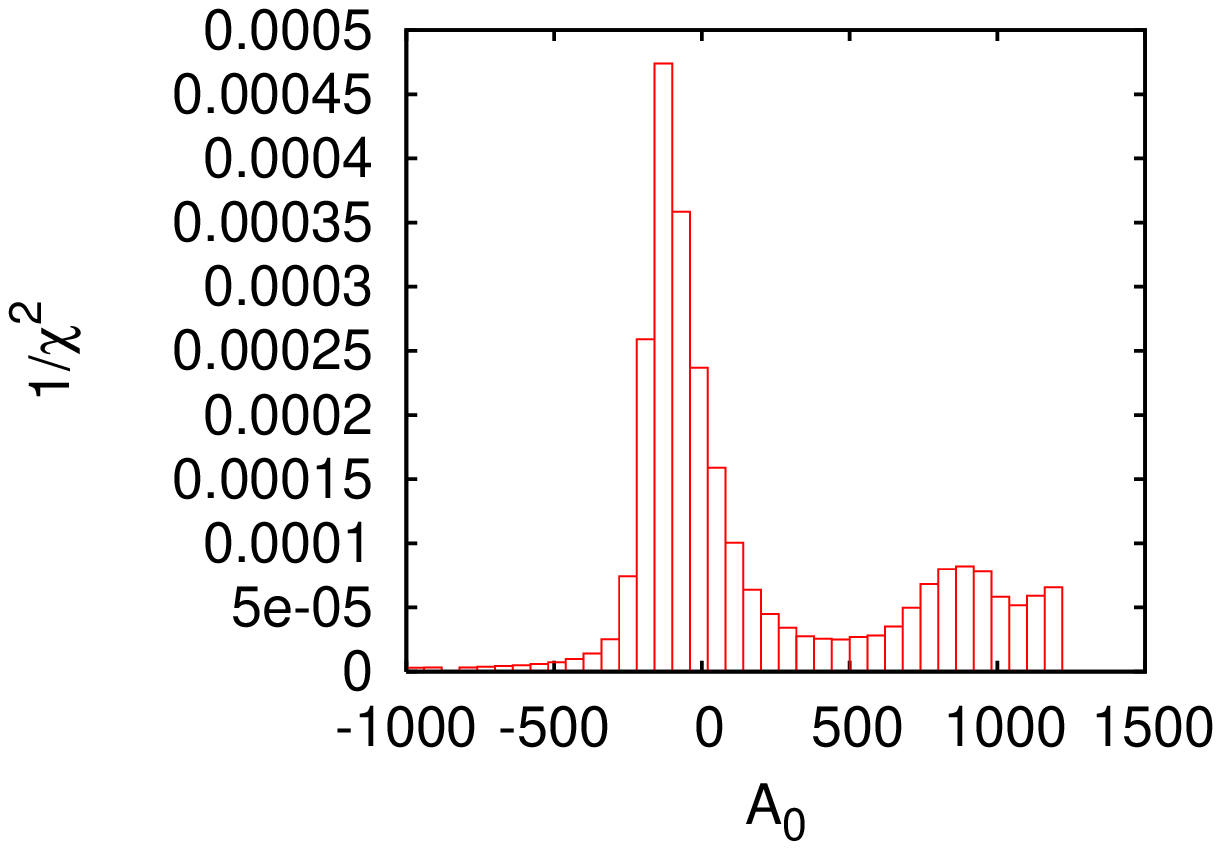} \hspace*{3cm}
 \includegraphics[width=5cm]{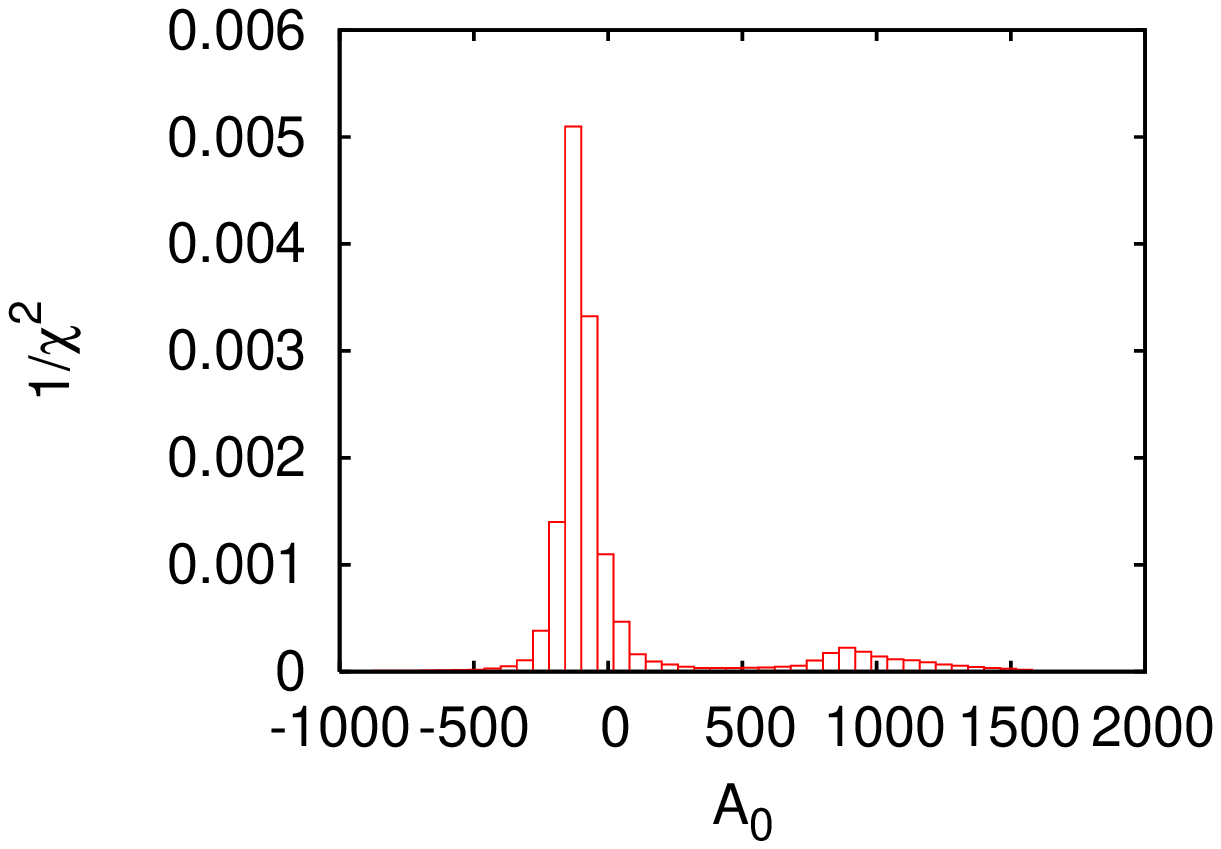} \\
\caption[]{SFitter output for $A_0$ and $y_t$. The two
 columns of marginalized Bayesian pdfs correspond to $\mu<0$ (left)
 and $\mu>0$ (right). For illustration purposes the parameters $m_0$
 and $m_{1/2}$ are only marginalized around their best--fit values.
 We show a $\chi^2$ map in the first row and $1/\chi^2$ distributions
 in the second and third.}
\label{fig:sugra_map_b2}
\end{figure}

Again in complete analogy to the likelihood analysis the study of the
correlation of $m_t$ and $A_0$ serves as an example of how
marginalizing parameters can weaken the understanding of the parameter
space, independent of the frequentist or Bayesian approach. 
Fig.~\ref{fig:sugra_map_b2} shows the Bayesian pdfs for $m_t$ and
$A_0$. Because of the strongly peaked likelihood map in the $m_0$ and
$m_{1/2}$ directions a full marginalization is not applied in these
directions. Instead, the mass parameters $m_0$ and $m_{1/2}$ are
marginalized only in a frame $\pm 2$~GeV and $\tan\beta$ is varied by
$\pm 1.5$, always around the best--fit point for each sign of
$\mu$. This additional constraint or bias can be useful when producing
a marginalized Bayesian pdf for comparably poorly measured
parameters. In order not to be mislead it is necessary to explicitly
check that the partly marginalized parameters $m_0, m_{1/2},
\tan\beta$ are not significantly correlated with the remaining $m_t$
and $A_0$.

In $m_t$ the Bayesian pdf is not symmetric with respect to the central
values for each sign of $\mu$. This asymmetry of the tails arises
from the parabola shape of the $m_t$--$A_0$ correlation. The
large--likelihood region around the apex becomes more important than
the far--away arms of the parabola after marginalizing $A_0$. This is
a typical volume effect in Bayesian statistics. At first sight these
asymmetric tails of the Bayesian pdf for $m_t$ seem to disagree with
its profile likelihood, but it is a physics effect, \ie a correlation
marginalized away.  This result is useful when it comes to trying to
resolve such a correlation, but by no means problematic.

Comparing the profile likelihood and the Bayesian pdf for $A_0$ the
volume effects significantly enhance the relative weight of the
secondary maximum at $A_0 \sim 800$~GeV. Moreover, comparing the
likelihood scales for $\mu<0$ and (the correct) $\mu>0$, the relative
enhancement of the Bayesian pdf is almost an order of magnitude, while
the binned best--fit points differ by only a factor 5 for the profile
likelihood.

\subsection{Purely high--scale model}
\label{sec:sugra_high_scale}

\begin{figure}[t]
 \includegraphics[width=6cm]{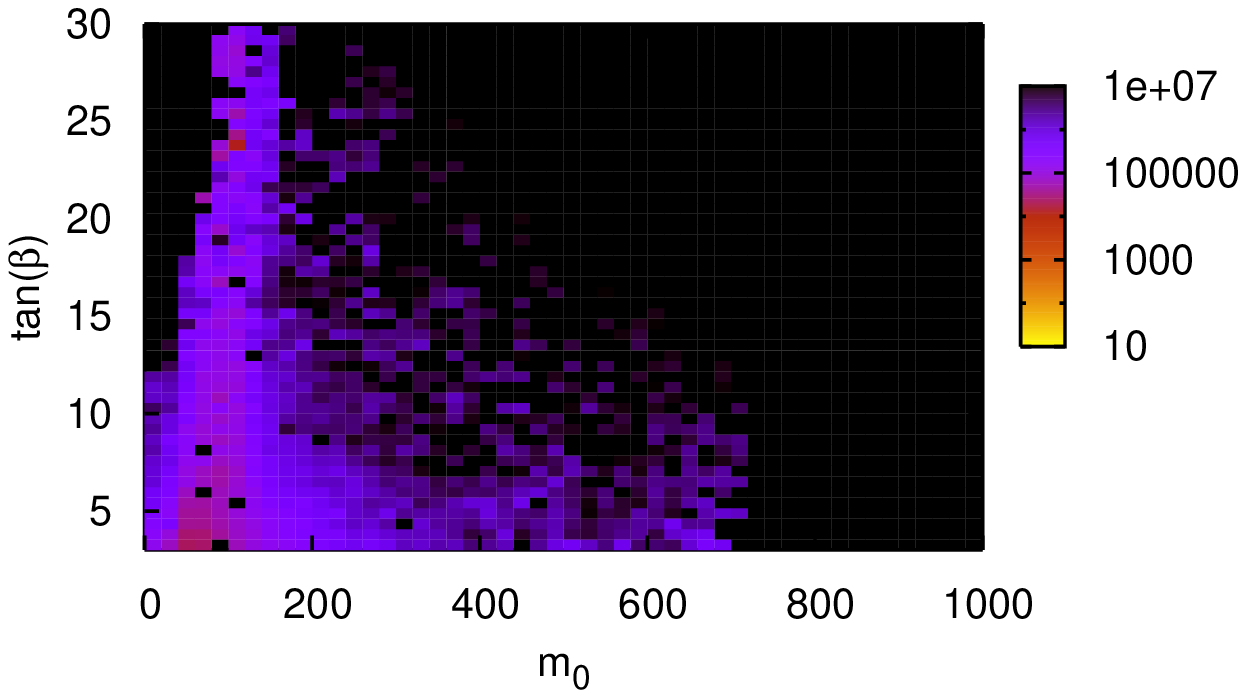} \hspace*{2cm}
 \includegraphics[width=5cm]{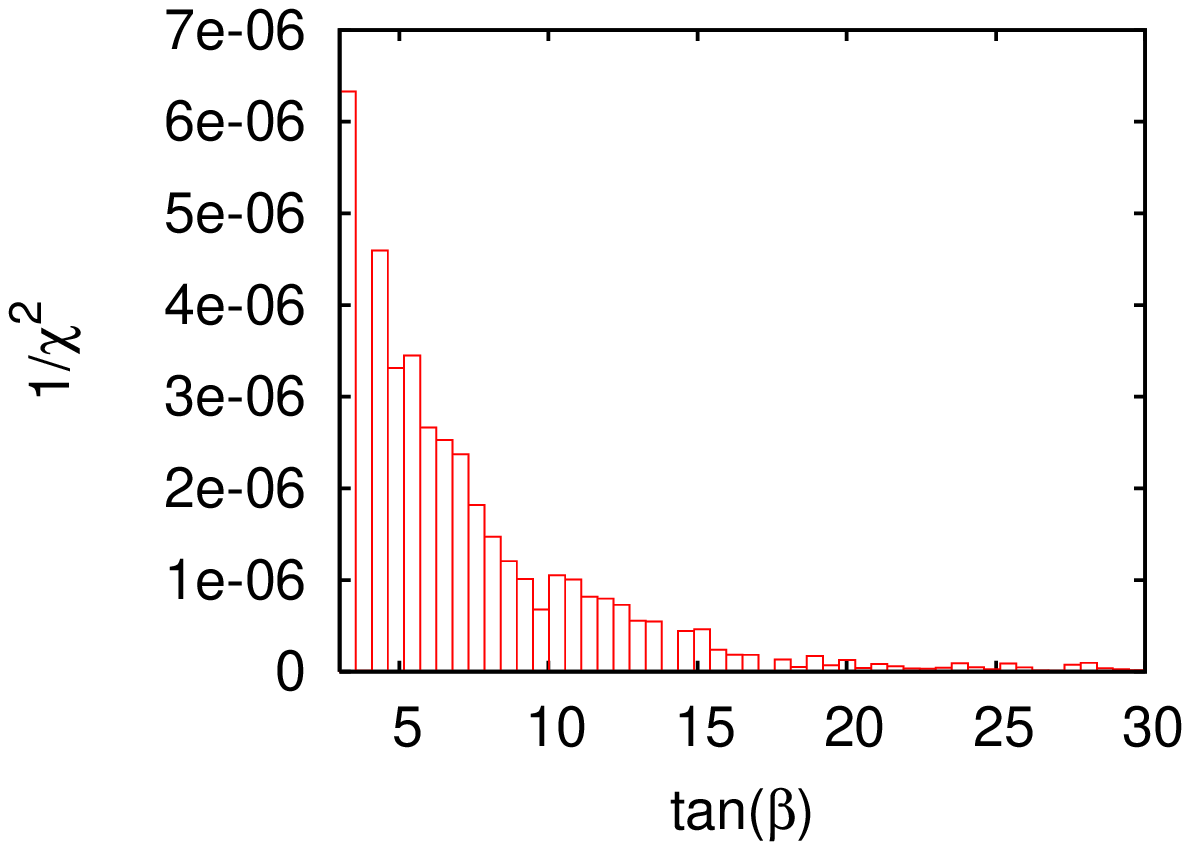} \\
 \includegraphics[width=6cm]{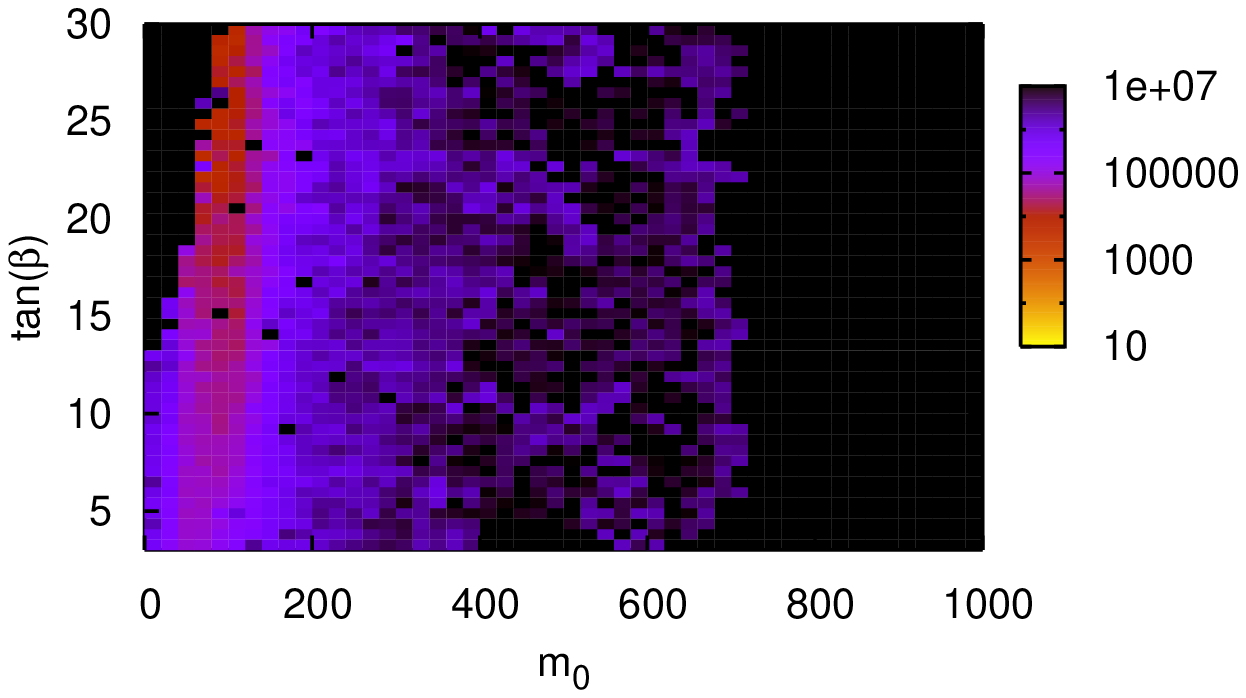} \hspace*{2cm}
 \includegraphics[width=5cm]{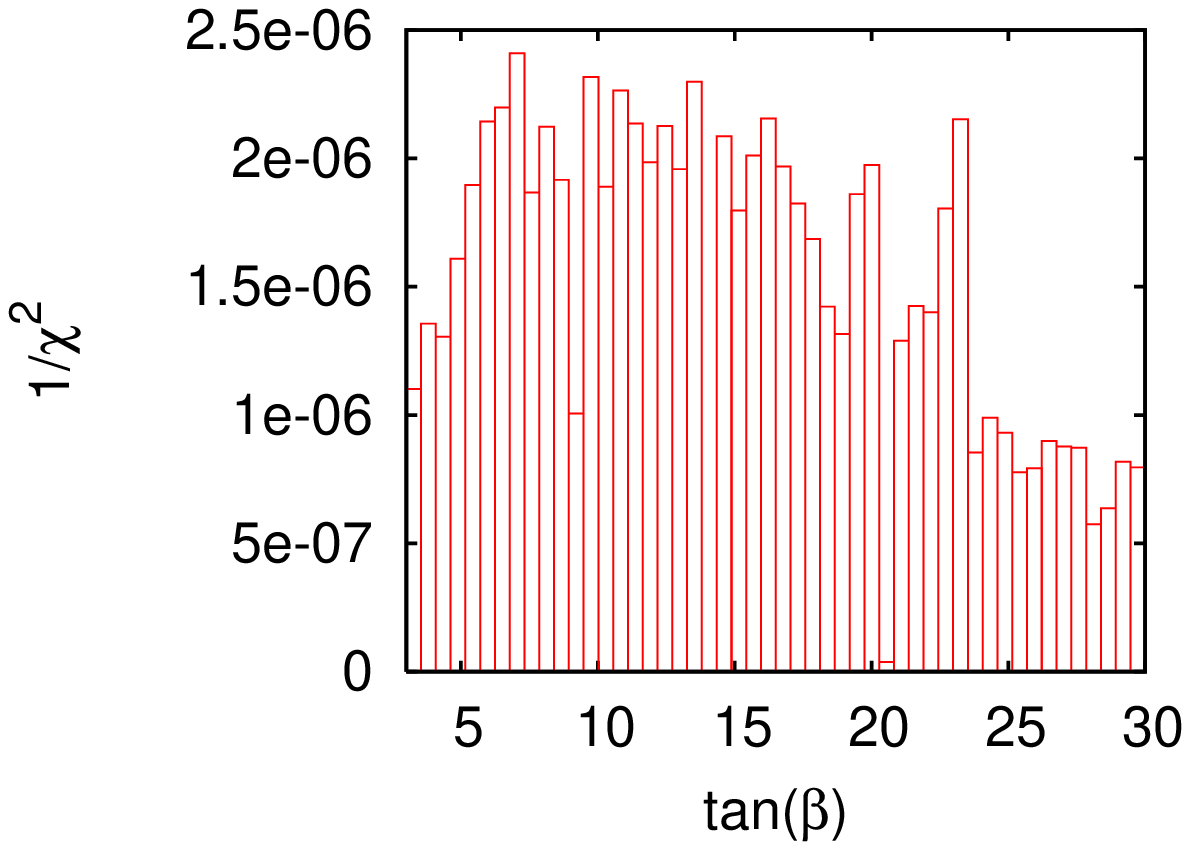} \\
 \includegraphics[width=6cm]{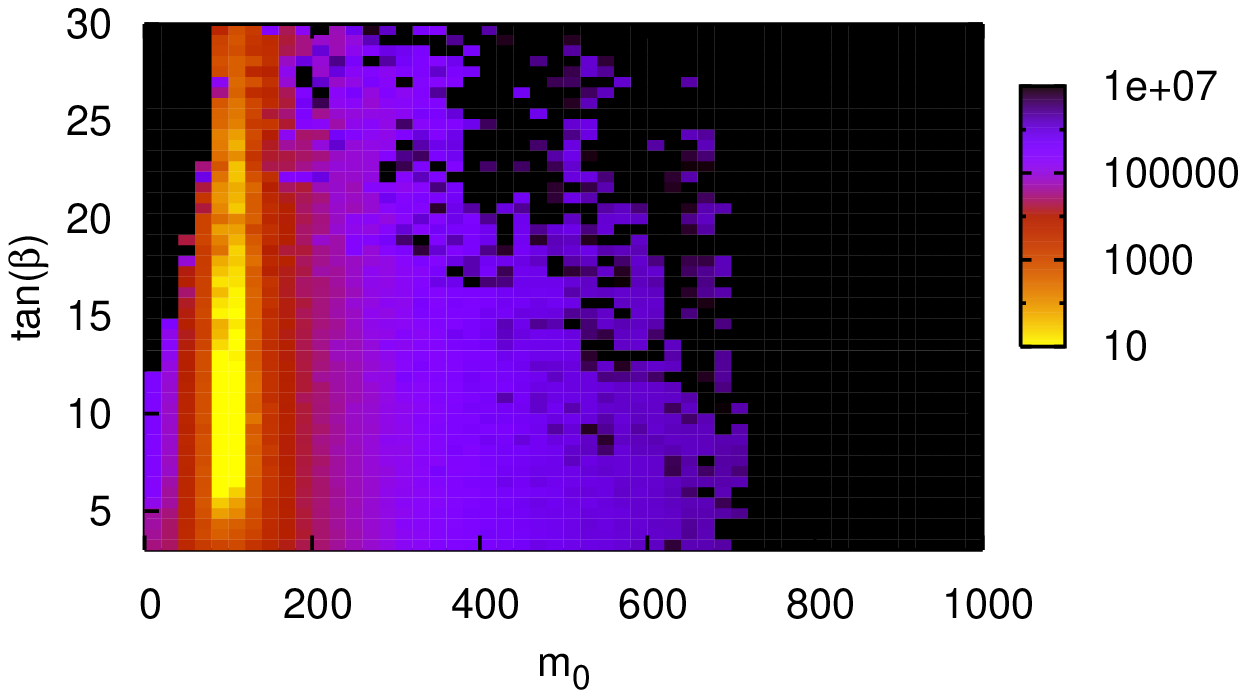} \hspace*{2cm}
 \includegraphics[width=5cm]{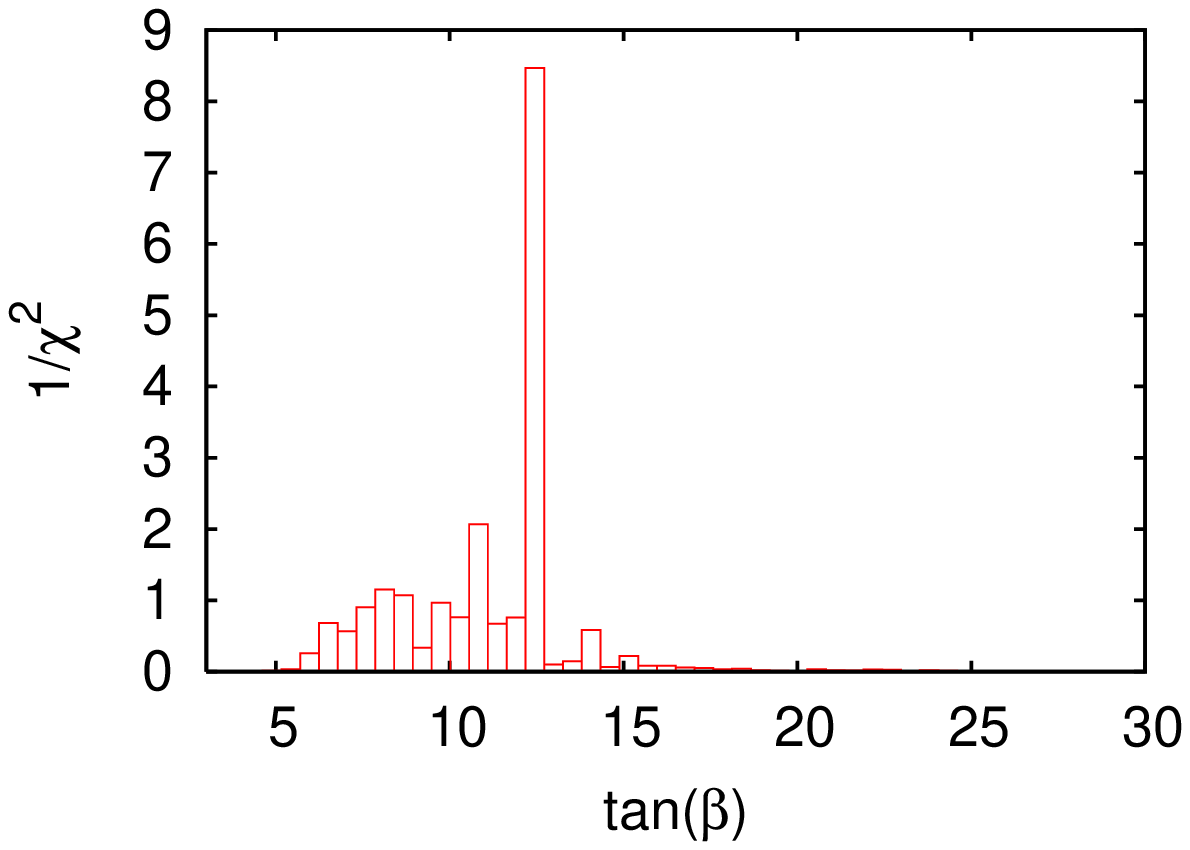} 
\caption[]{SFitter output for $\tan\beta$. The first row shows 
  Bayesian pdfs with a flat prior in $B$, the second row Bayesian pdfs
  with a flat prior in $\tan\beta$, and the last row profile
  likelihoods.}
\label{fig:sugra_high_scale}
\end{figure}

Strictly speaking, the usual set of MSUGRA model parameters contain 
the high--scale mass
parameters $m_0, m_{1/2}, A_0$, and on the other hand contain the
weak--scale ratio of vacuum expectation values $\tan\beta=v_2/v_1$,
which explicitly assumes radiative electroweak symmetry breaking.
Minimizing the potential in the directions of both vevs gives the two
conditions~\cite{drees_martin}:
\begin{alignat}{5}
\mu^2 &=  \frac{m^2_{H,2} \sin^2 \beta - m^2_{H,1} \cos^2 \beta}
               {\cos 2 \beta}
        - \frac{1}{2} m_Z^2 \notag \\
2 B \mu &= \tan 2 \beta \; \left( m^2_{H,1} - m^2_{H,2} 
                           \right)
           + m_Z^2 \sin 2 \beta
\end{alignat}
The masses $m_{H,j}$ correspond to the two Higgs doublets in the
type-II two--Higgs doublet model of the MSSM. $H_1$ has a tree--level
coupling only to down--type fermions, while $H_2$ couples to up--type
fermions only. The mass--squared parameter $B \mu$ appears in front of
mixed terms of the kind $H_1^0 H_2^0$. Assuming electroweak symmetry
breaking usually $m_{H,j}$ and $\tan\beta$ are used to compute the
mass parameters $B$ and $\mu$, assuming the well measured Standard
Model parameter $m_Z$. In MSUGRA the two scalar Higgs masses at the
high scale are given by $m_0$, so in fact only $m_0$ and $\tan\beta$
are used.

A well-motivated alternative is to replace $\tan\beta$ with $B$ as a
high--scale input parameter together with $m_0$ and compute
$\tan\beta$ and $\mu$ (modulo its sign) assuming electroweak symmetry
breaking and the $Z$ mass. This approach has the advantage that all
input parameters are high--scale mass parameters. This does not make a
difference for frequentist profile-likelihood map, but in a Bayesian
approach taking into account volume effects it does. \bigskip

To illustrate the effects of flat priors either in $B$ or in
$\tan\beta$ the Bayesian pdfs and the profile likelihoods are shown in
the $m_0$--$\tan\beta$ plane and the one--dimensional $\tan\beta$
distributions in Fig.~\ref{fig:sugra_high_scale}. From the best--fit
points in Fig.~\ref{fig:sugra_map_f1} even after
including theory errors the correct value for $\tan\beta$ can be
determined from the set of LHC measurements. However, the first row of
plots in Fig.~\ref{fig:sugra_high_scale} clearly shows that with a
flat prior in $B$ the one--dimensional Bayesian pdf is largely
dominated by noise and by a bias towards as small as possible
$\tan\beta$. This bias is simply an effect of the flat prior in
$B$. Switching to a flat prior in $\tan\beta$, noise effects are still
dominant, but the maximum of the one--dimensional Bayesian pdf is in
the correct place. As expected, the profile likelihood picks the
correct central value of $\tan\beta \sim 12$ for the smeared parameter
point. 

\subsection{Errors on parameters}
\label{sec:sugra_smear}

Once a best-fit point has been determined from any set of
measurements, the question arises what the precision of the
determination of the parameters is. First the case for LHC
measurements is studied and then the impact of the ILC is evaluated.

\subsubsection{LHC: masses vs kinematic endpoints}
\label{sec:sugra_lhc}

\begin{table}[t]
\begin{small}
\begin{tabular}{|l|r|c|cccc|}
\hline
            & SPS1a  & $\Delta^{\rm theo-exp}_{\rm zero}$
                     & $\Delta^{\rm expNoCorr}_{\rm zero}$ 
                     & $\Delta^{\rm theo-exp}_{\rm zero}$ 
                     & $\Delta^{\rm theo-exp}_{\rm gauss}$ 
                     & $\Delta^{\rm theo-exp}_{\rm flat}$ \\
\hline
            &        & masses 
                     & \multicolumn{4}{c|}{endpoints} \\
\hline
$m_0$       & 100    & 4.11 & 1.08 & 0.50 & 2.97 & 2.17 \\
$m_{1/2}$   & 250    & 1.81 & 0.98 & 0.73 & 2.99 & 2.64 \\
$\tan\beta$ & 10     & 1.69 & 0.87 & 0.65 & 3.36 & 2.45 \\
$A_0$       & -100   & 36.2 & 23.3 & 21.2 & 51.5 & 49.6 \\
$m_t$       & 171.4  & 0.94 & 0.79 & 0.26 & 0.89 & 0.97 \\
\hline
\end{tabular}
\end{small}
\caption[]{Best--fit results for MSUGRA at the LHC derived from 
  masses and endpoint measurements with absolute errors in GeV.  The
  big columns correspond to mass and endpoint measurements.  The
  subscript represents neglected, (probably approximate) gaussian or
  proper flat theory errors. The experimental error includes
  correlations unless indicated otherwise in the superscript. The top
  mass is quoted in the on-shell scheme.}
\label{tab:sugra_mass_edge}
\end{table}

To determine the central values and the errors on the fundamental
parameters two different approaches are available for the LHC
measurements. Either the kinematical endpoints or the particle masses
(from a fit to the endpoints without any model
assumptions~\cite{edges,per}) can serve as data. The first
question is how an extraction of the MSUGRA model parameters from
kinematic endpoints listed in Tab.~\ref{tab:edges} compares to an
extraction from the mass measurements listed in
Tab.~\ref{tab:mass_errors}.\bigskip

Because the extraction of masses from endpoints is highly correlated,
both approaches are only equivalent if the complete correlation matrix
of masses is taken into account. For the experimental errors the mass
determination from edges introduces non-trivial correlations in the
masses, whereas the theory is essentially uncorrelated in masses, but
non trivially correlated in the endpoints.

Numerically, theory errors cannot be neglected. In particular, the
determination of $\tan\beta$ and $A_0$ largely relies on the light
Higgs mass, which can be computed in perturbation
theory~\cite{m_h}. This calculation has a parametric error, \eg from
the top Yukawa, and a systematic error due to unknown higher
orders. The parametric errors are correlated with the direct mass
measurements, which means they do not enter as theory errors from the
Higgs mass calculation. The remaining theory error on the light Higgs
mass due to unknown higher--order terms can be estimated to lie around
2~GeV~\cite{m_h}. For the top pole--mass measurement an
experimental error of 1~GeV is expected at the LHC and therefore used in
the analysis. As long as the experimental
error stays above roughly a GeV, the theory error on the top mass from
the unknown renormalization scheme of $m_t$ at a hadron
collider~\cite{theo_mt} should be small, $\lambda_{\rm QCD}
\ll$~GeV.

For supersymmetric partner masses in MSUGRA theory errors arise mostly
from the limited perturbative order of the renormalization group
running~\cite{rge_errors}. Moreover, at the weak scale higher--order
corrections have to be taken into account when converting Lagrangian
mass parameters into physical masses. The combined theory errors are
estimated to an uncorrelated $1\%$~$(3\%)$ for weakly (strongly)
interacting particles~\cite{bagger, weak_corrections}. If a parameter
point does not predict one of the endpoints included in the set of
observables, the likelihood of this parameter point is set to
zero.\bigskip

The errors on the MSUGRA parameters for different assumptions are
shown in Tab.~\ref{tab:sugra_mass_edge}.  Changing from mass
measurements to endpoints measurements (for gaussian experimental
errors and no correlations) improves the errors by a factor of 
more than three
for $m_0$ and a factor two for the gaugino mass parameter $m_{1/2}$.
This improvement arises from the absence of the correlation matrix
between the mass observables. If this matrix were known, the results
would be similar. As a next step, again using only experimental
errors, but taking into account the correlation of the systematic
energy--scale errors (JES and LES) a further improvement of a
factor two for the common scalar mass parameter and a slight
improvement for the gaugino mass parameter is observed. This comparison shows that
to obtain the best precision from the LHC data, it is important to
correctly estimate the correlation between the observables.

The impact of theory errors on the parameter determination is shown in
the next columns where first the gaussian (approximate) and then the flat
(proper) theory error is studied. For the well--measured scalar and gaugino
masses $m_{1/2}$ the theory error increases the small purely
experimental error considerably. For the ratio of the vacuum
expectation values $\tan\beta$ the theory error on the Higgs mass
becomes the dominant source of error, because the experimental
precision on the Higgs mass measurement is almost a factor 10 better
than its theory error. In the SPS1a parameter point the two different
techniques of treating the theory error give the same results
within 20\%. Note that the precision of the top mass parameter as part
of the SUSY ensemble is slightly better than the direct
top mass measurement alone.\bigskip

\begin{table}[t]
\begin{small}
\begin{tabular}{|c|rrrrr|}
\hline
             & $m_0$ & $m_{1/2}$ &  $\tan\beta$ & $A_0$ & $m_t$ \\
\hline
$m_0$        &     1 & 0.485 & 0.523 & 0.042 & 0.063 \\
$m_{1/2}$    &       &     1 &-0.100 & 0.648 & 0.449 \\
$\tan\beta$  &       &       &     1 &-0.467 &-0.192 \\
$A_0$        &       &       &       &     1 & 0.495 \\
$m_t$        &       &       &       &       &     1 \\
\hline
\end{tabular} \hspace*{5mm}
\begin{tabular}{|c|rrrrr|}
\hline
             & $m_0$ & $m_{1/2}$ &  $\tan\beta$ & $A_0$ & $m_t$ \\
\hline
$m_0$        &     1 & 0.501 & 0.432 & 0.094 & 0.214 \\
$m_{1/2}$    &       &     1 &-0.206 & 0.740 & 0.720 \\
$\tan\beta$  &       &       &     1 &-0.401 &-0.256 \\
$A_0$        &       &       &       &     1 & 0.648 \\
$m_t$        &       &       &       &       &     1 \\
\hline
\end{tabular}
\end{small}
\caption[]{The (symmetric) correlation matrix of all SUSY
parameters in the MSUGRA fit using endpoint measurements at the LHC
and including approximate gaussian (left panel) and proper flat (right
panel) theory errors.}
\label{tab:sugra_correlations}
\end{table}

As expected, the correlation matrix between the different MSUGRA
parameters is by no means diagonal. In
Tab.~\ref{tab:sugra_correlations} $m_{1/2}$ and $\tan\beta$ are
largely uncorrelated, as are $A_0$ and $\tan\beta$. The latter is
somewhat unexpected in the light of the Higgs--mass measurement, but
it can be understood by the pseudo-fixpoint behavior of $A_t$ as a
function of $A_0$ and by the fact that the important parameter in the
Higgs mass calculation is the light stop mass, which depends
critically on $m_0$ and slightly on $m_{1/2}$~\cite{drees_martin}. The
two mass parameters $m_0$ and $m_{1/2}$ are strongly correlated
through the renormalization group running of the squark and slepton
masses. Similarly, $A_0$ and $m_{1/2}$ are strongly
correlated.

Through most of this analysis SoftSUSY~\cite{softsusy} is the workhorse 
for the renormalization--group evolution to link the high--scale MSUGRA model
parameters with the weak--scale masses and other observables,
including some higher--order corrections. As a consistency check on
the theory errors, the observables were calculated with
SoftSUSY, but the model parameters were determined with
SuSpect~\cite{suspect}. While the central values are shifted as
expected, they are compatible within 3$\sigma$, thus giving confidence
that the estimated theory errors cover at least the different
theoretical calculations.\bigskip

\begin{figure}[t]
 \includegraphics[width=5cm]{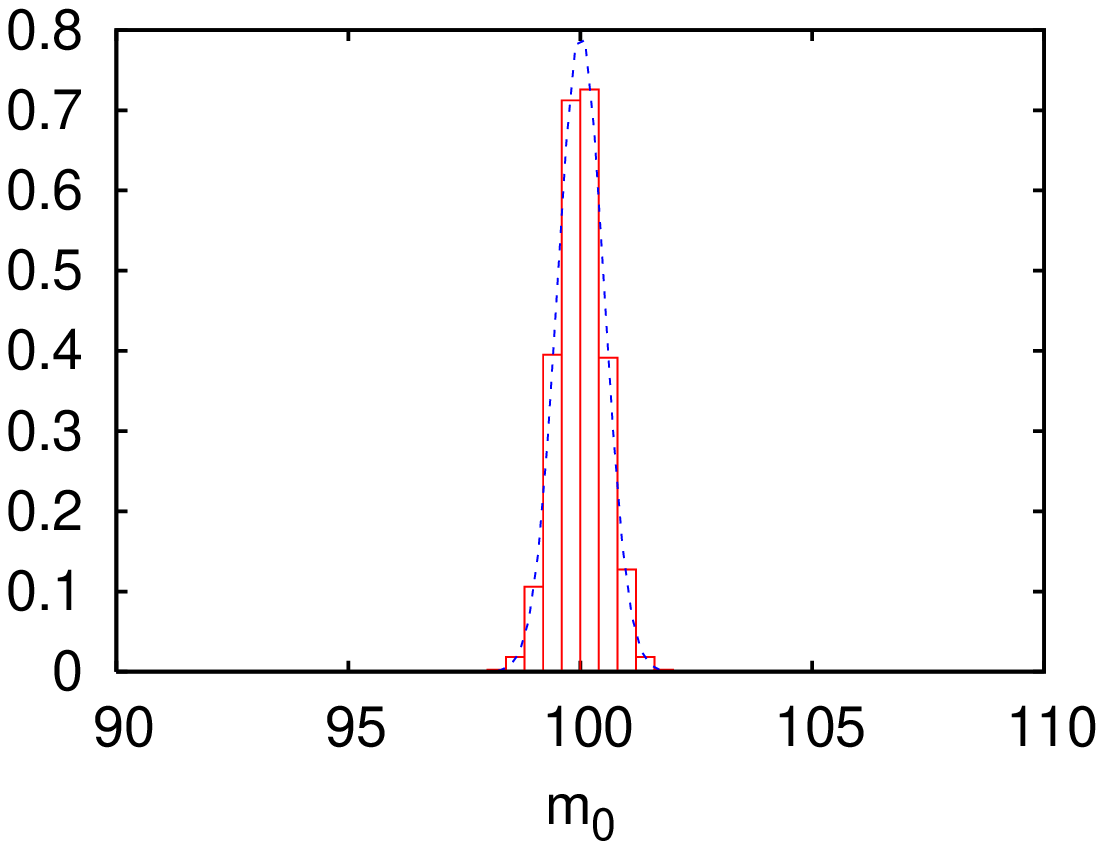} \hspace*{2cm}
 \includegraphics[width=5cm]{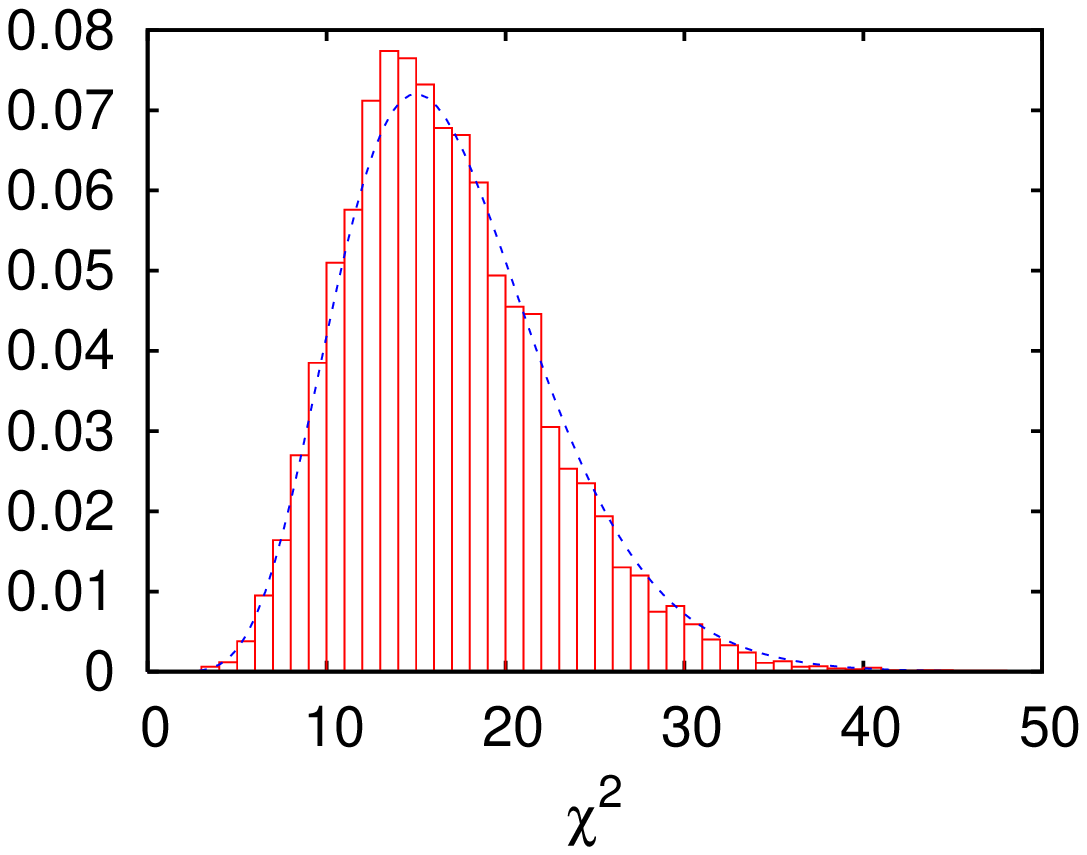} \\
 \includegraphics[width=5cm]{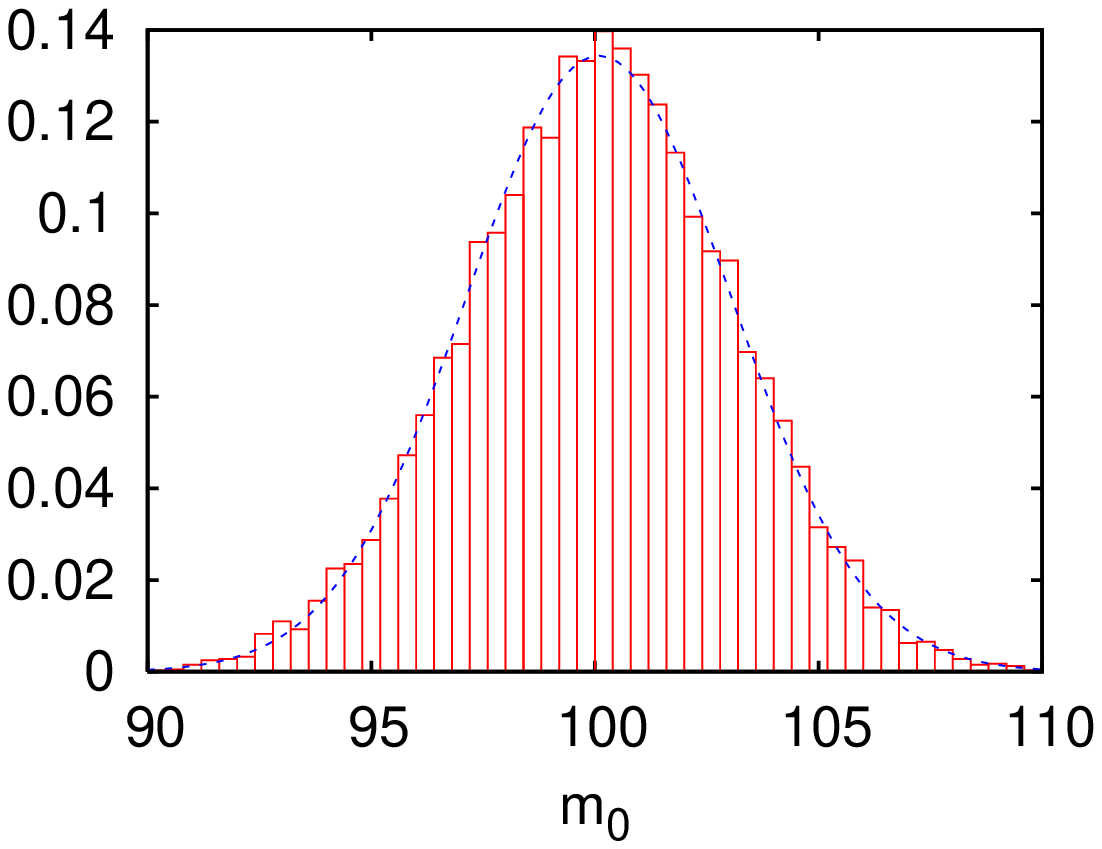} \hspace*{2cm}
 \includegraphics[width=5cm]{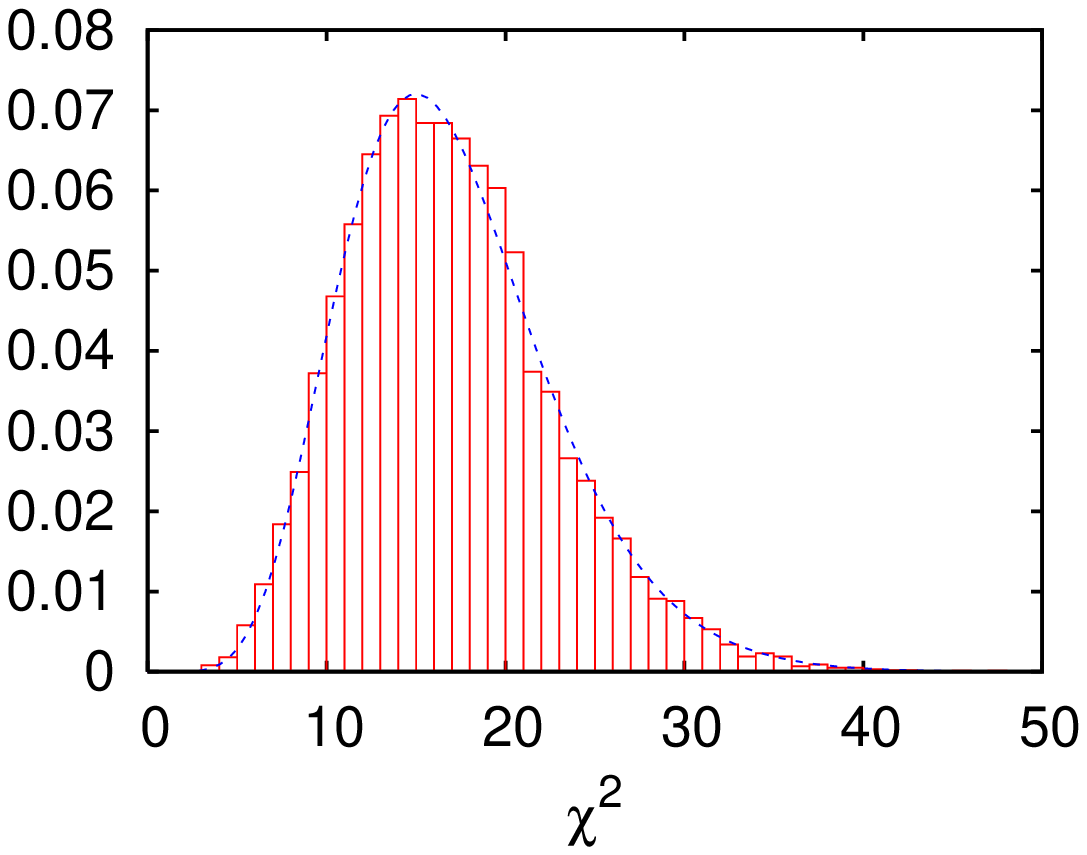} \\
 \includegraphics[width=5cm]{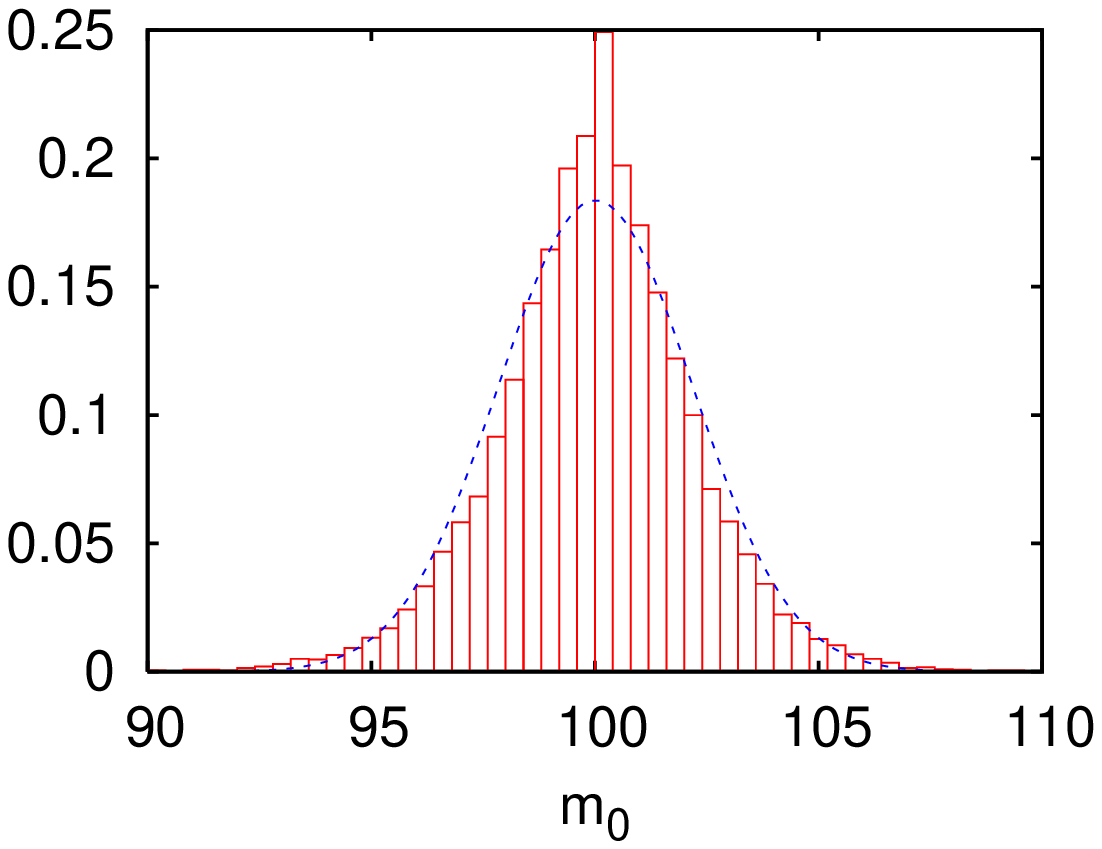} \hspace*{2cm}
 \includegraphics[width=5cm]{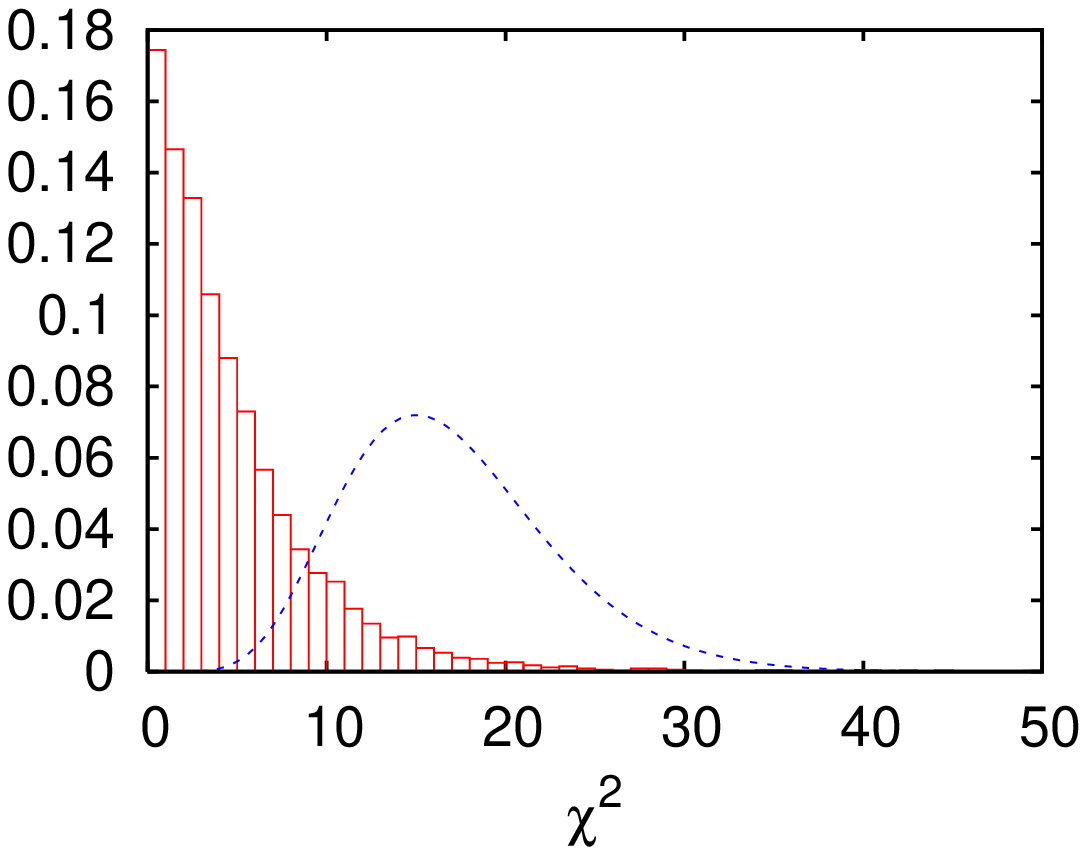} \\
\caption[]{SFitter output for $m_0$ and $\chi^2$. For different 
 assumptions for the theoretical error (neglected,
 gaussian and flat theoretical error from top to bottom) histograms 
 for 10000 pseudo--measurements are shown. The dotted blue line shows 
 a fitted gaussian for the $m_0$-plots and a $\chi^2$-distribution with
 16 degrees of freedom for the $\chi^2$-plots, respectively.}
\label{fig:sugra_smear}
\end{figure}

The distribution of 10000 individually run best--fit results to
smeared data samples (pseudo--measurements) is shown in
Fig.~\ref{fig:sugra_smear}. Such a histogram is simply the numerical
simulation of error propagation~\cite{cp_phases} and should in the
gaussian case reproduce the same result as a convolution of the
different gaussian errors. For the first two rows only
gaussian experimental errors are assumed and (hopefully approximate) gaussian
theory errors. Both of the resulting distributions for $m_0$ are
gaussian, as are all the other distributions not shown here. For the
third row the correct flat theory errors are shown. The $m_0$
distribution is now slightly too narrow to be gaussian. On the other
hand, all one--dimensional distributions are surprisingly
similar to gaussian. However, this just reflects the central limit
theorem, namely that if a distribution is probed often enough a
gaussian distribution will be observed, independent of the shape of
the errors.

Depending on the relative impact of the different errors and on the
detailed correlations, a non--gaussian behavior can be more or less
pronounced for a finite number of attempts. For example,
$m_{1/2}$ is dominantly gaussian, even including flat theory errors,
while the $A_0$ distribution is wide and not gaussian at all. As a
check the distribution of the log--likelihood $\chi^2$ was computed 
and compared to the gaussian assumption. For neglected or gaussian
theory errors the log-likelihood distribution matches a $\chi^2$
distribution with the correct number of degrees of freedom. For flat
theory errors the prescription effectively removes measurements which
are within the theory--error bands from the counting of the degrees of
freedom, thereby lowering the effective value of $\chi^2$.\bigskip

In the list of measurements listed in Tab.~\ref{tab:mass_errors} the
LHC will only identify three out of four neutralinos --- the
third--heaviest neutralino will be missed due to its higgsino
nature. Higgsino--neutralino couplings to light--flavor fermions and
sfermions are largely suppressed and can only be observed in cascade
decays through gauge bosons or possibly a Higgs~\cite{edges}.  The
question is what happens if the fourth--heaviest neutralino is wrongly
labeled as third--heaviest. SFitter indeed finds a best--fitting
parameter point to fit this data set. This point is slightly shifted
in $m_0$ and $m_{1/2}$ by up to 1~GeV.  The largest difference between
the correctly and wrongly assigned parameter points occurs in
$\tan\beta$, which is shifted by about~2. The $\chi^2$ value remains
reasonable for both points.

While at first sight the set looks like a {\sl bona fide} alternative
minimum, it can easily be discarded using LHC data. Having determined
the `wrong' model parameters, the full spectrum and couplings can be
predicted. In particular, the fourth neutralino now has a mass of
about 400~GeV. For example, more squark decays to $\chi_4$ than to
$\chi_3$ are predicted for this `wrong' parameter point, in
contradiction to the data sample. Unfortunately, distinguishing such
discrete alternative descriptions rely on signatures which should have
to be seen. At the LHC, what can and what cannot be seen is determined
by Standard Model backgrounds and detector effects, which makes an
automated answering algorithm unrealistic.

\subsubsection{Impact of the ILC}
\label{sec:sugra_ilc}

\begin{table}[t]
\begin{tabular}{|l|r|ccc|ccc|}
\hline
            & SPS1a  
                     & $\Delta_{\rm endpoints}$ 
                     & $\Delta_{\rm ILC}$ 
                     & $\Delta_{\rm LHC+ILC}$ 
                     & $\Delta_{\rm endpoints}$ 
                     & $\Delta_{\rm ILC}$ 
                     & $\Delta_{\rm LHC+ILC}$ \\
\hline
            &        & \multicolumn{3}{c|}{exp. errors}
                     & \multicolumn{3}{c|}{exp. and theo. errors} \\
\hline
$m_0$       & 100    & 0.50 & 0.18  & 0.13  & 2.17 & 0.71 & 0.58 \\
$m_{1/2}$   & 250    & 0.73 & 0.14  & 0.11  & 2.64 & 0.66 & 0.59 \\
$\tan\beta$ & 10     & 0.65 & 0.14  & 0.14  & 2.45 & 0.35 & 0.34 \\
$A_0$       & -100   & 21.2 & 5.8   & 5.2   & 49.6 & 12.0 & 11.3 \\
$m_t$       & 171.4  & 0.26 & 0.12  & 0.12  & 0.97 & 0.12 & 0.12 \\
\hline
\end{tabular}
\caption[]{Best--fit results for MSUGRA at the LHC (endpoints) and including
  ILC measurements. Only absolute errors are given. The LHC results
  correspond to Tab.~\ref{tab:sugra_mass_edge}, including flat theory
  errors.}
\label{tab:sugra_ilc}
\end{table}

Combining LHC data with data from a future linear collider shifts the
focus even further into the determination of the errors on the MSUGRA
parameters. As shown in Tab.~\ref{tab:sugra_ilc} the errors on the
parameters from ILC measurements alone are already considerably
smaller than the LHC errors. This is true for all MSUGRA parameters,
because for example the missing gluino--mass measurement at the ILC is
not necessary because the weak gaugino masses are known. The general
improvement of the errors is expected, since mass measurements at the
ILC are about an order of magnitude more precise. The resulting
improvement in precision on the model parameters is about a factor~5.
Combining ILC and LHC measurements in MSUGRA only leads to a marginal
additional improvement of the errors, even though squarks and gluinos
largely escape the ILC analyses. The reason is that 
the precision of $m_0$ (simple error calculation) is 
dominated by the slepton masses alone.  Comparing
the LHC+ILC errors with and without theory errors show the margin for
the improvement of theory predictions, justifying the SPA
project~\cite{slha}. \bigskip


The correlation between the parameter measurements is different once
the ILC measurements are included. For example, $A_0$ and $\tan\beta$
are now largely correlated. Such a correlation appears in the
measurement of the off--diagonal entries of the scalar mixing matrices
as well as in $m_h$. In contrast to the LHC measurement, the top
Yukawa is now largely uncorrelated with all MSUGRA parameters, because
it can be independently determined using the $0.12$~GeV measurement of
the physical top mass.

\section{Weak--scale MSSM Lagrangian}
\label{sec:mssm}

If supersymmetry or other new physics is observed at the TeV scale the
weak--scale Lagrangian should be determined from data.  High--scale
models for example of SUSY breaking then have to be inferred from this
TeV--scale data~\cite{sfitter,fittino,inverse}.  This problem is what
SFitter is really designed to solve, after being tested extensively in
the lower--dimensional MSUGRA parameter space.\bigskip

The complete parameter space of the MSSM can have more than 100
parameters. However, at experiments like the LHC some
new--physics parameters can be fixed because no information on them is
expected. This for example includes $CP$ phases~\cite{cp_phases} or
non-minimal flavor violation~\cite{susy_flavor} for weak--scale
high-$p_T$ measurements at the LHC. It also includes the first and
second generation trilinear couplings $A_{l,u,d}$, which in minimal
flavor violation are multiplied by the corresponding Yukawa coupling
and which beyond minimal flavor violation are very strongly
constrained.

Because at the LHC flavor information is difficult to obtain on light
quarks, we use an average squark mass for left and right handed
scalars. The different handedness can be distinguished through their
appearance in different cascades. The right--handed squark typically
decays directly to the bino and a quark, while the left--handed squark
has a sizeable coupling to the wino, leading to the usual long decay
chain.  Unfortunately, in the currently experimentally simulated LHC
data set there is little information on the stop--chargino
sector~\cite{mihoko}.  Without this information, any combination of
$B$ physics data with high-$p_T$ LHC data will fall short --- we
postpone a detailed discussion of this problem to a later
paper~\cite{with_flavor}. In the lepton sector electrons can easily be
separated from muons. A possible unification of the first two
generations can then be determined from data~\cite{sleptons}.

The third--generation trilinear couplings $A_{\tau, b}$ can in
principle play a role as off--diagonal entries in the down--type mass
matrices. However, they are multiplied by the corresponding Yukawas
and compete with the term $\mu \tan\beta \sim (60~\gev)^2$.  Seeing
effects of the trilinear coupling would require $A_b \gtrsim
1400$~GeV, so for a low-$\tan\beta$ parameter point $A_{\tau, b}$ have
no impact on the likelihood around the correct or alternative
best--fitting points. The same $A_{\tau, b}$ appear as parameters in
the computation of the light MSSM Higgs mass, but again their effect
is negligible compared to for example $A_t$~\cite{m_h}. There is a
slim possibility that the stau mixing angle and with it $A_\tau$ might
be determined in cascade decays similarly to the usual UED--SUSY spin
analysis~\cite{pmz_spin}, but this analysis has not yet been
experimentally confirmed.\bigskip

Properly including $m_t$ this leads to the effective 19-dimensional
weak--scale MSSM parameter space listed for example in
Tab.~\ref{tab:mssm_secondary}. Obviously, the assumption of parameters
being irrelevant for the MSSM likelihood map can and has to be tested.
Moreover, the SFitter analysis will show that more than just
the trilinear $A$ parameters turn out to be invisible at the
LHC.\bigskip

In contrast to the MSUGRA model $\tan\beta$ is used as a parameter
in the Higgs sector and not $B$, because all MSSM parameters are
defined at the weak scale assuming electroweak symmetry breaking.  In
other words, $\tan\beta$ and $m_A$ are the two Higgs--sector
parameters in the MSSM analysis. Looking at the currently confirmed
LHC measurements none of the heavy
Higgs bosons with masses of the order $\om(m_A)$ would be seen in
SPS1a, which is not good news for the parameter determination in the
Higgs sector.

Because computing the mass spectrum in the weak--scale MSSM does not
require any shift in scales, \ie it does not involve renormalization
group running or large logarithms, a smaller theory error for the
on--shell particle masses should be assumed. As a rough estimate 
a relative error of $1\%$ for the masses of strongly
interacting particles and $0.5\%$ for weakly interacting
particles~\cite{bagger, weak_corrections} are used, plus a $2\%$ non-parametric
error on the light MSSM Higgs boson~\cite{m_h}. Just as in
Sec.~\ref{sec:sugra_errors} the correct flat theory
errors, eq.(\ref{eq:flat_errors}), are used for the determination of the errors on
model parameters.\bigskip

\subsection{MSSM likelihood map}
\label{sec:mssm_lhc}

SFitter approaches the problem of the higher--dimensional MSSM
parameter space in analogy to the MSUGRA case, but now organized in
four steps:

\begin{enumerate}
\item First, SFitter produces a set of Markov chains over the
  entire parameter space. The proposal function is constant, allowing
  the algorithm to cover the entire MSSM space without focusing on
  the resolution of local likelihood maxima.  Starting from the best
  five points in this Markov chains Minuit resolves the local maxima
  in the likelihood map. This procedure ensures that there is no bias
  from starting points in the subsequent analysis. This step~1 can be
  repeated with different proposal functions, depending on the purpose
  of the Markov chain SFitter computes.
  
\item In a second step the Markov chains and the additional
  high--resolution Minuit algorithm are limited to the
  gaugino--higgsino subspace $M_1, M_2, M_3, \mu, \tan\beta$ and
  $m_t$.  Again, the proposal function is flat, focusing on the scan
  for local maxima in the likelihood map. For the 15 best local maxima
  in this subspace their resolution is improved by Minuit.
  
\item For the best point(s) in the gaugino--higgsino subspace these
  coordinates are then fixed.  The step-3 Markov chain probes the
  additional scalar parameter space around the local maxima in the
  gaugino--higgsino space, assuming a Breit--Wigner proposal function
  with a width of $1\%$ of the entire range in each direction. The
  resolution of the five best points is improved by Minuit.
  
\item Finally, Minuit traces the correlations between the
  gaugino--higgsino parameter space and the remaining scalar mass
  parameters. Once the global best--fitting parameter point is
  identified the errors on all parameters are determined using the
  usual smeared set of pseudo measurements and flat theory errors.

\end{enumerate}

All steps in the SFitter strategy are either Markov chains to globally
probe the parameter space (with a flat or a Breit--Wigner proposal
function), or a Minuit hill climber to identify the likelihood maxima
with high resolution. This approach can be applied to any problem
involving a high--dimensional parameter space, but the details of
course have to be adjusted.\bigskip

The large number of maxima mapped out in the second step corresponds
to the expectations from the MSUGRA model: starting from the true
parameter point an alternative solution with a switched sign in $\mu$
should exist. In the MSSM the hierarchy of $M_1$, $M_2$ and $|\mu|$
can be interchanged, which altogether can give $\om(10)$
distinct maxima in the likelihood map. To allow for additional
structures or several best points in the Markov chain to correspond to
the same local maximum, we increase the number of likelihood
maxima returned after step~2 to 15.

Last but not least, just as in the MSUGRA case alternative
likelihood maxima triggered by correlations between the rather poorly
measured parameters $A_t$, $\tan\beta$ and the right--handed stop mass
are expected.
One could imagine that secondary maxima appear in the $A_t$ - $m_t$
plane, like it happened in the MSUGRA case. However, this correlation
is not clearly visible in the MSSM because of a lack of direct
measurements in the stop sector.  \bigskip

\begin{table}
\begin{tabular}{|l|rrrr|rrrr|}
\hline
                     &\multicolumn{4}{c|}{$\mu<0$}&\multicolumn{4}{c|}{$\mu>0$}\\
\hline
                     &      &      &      &      &SPS1a&      &      &       \\
\hline \hline
$M_1$                &  96.6& 175.1& 103.5& 365.8&  98.3& 176.4& 105.9& 365.3\\
$M_2$                & 181.2&  98.4& 350.0& 130.9& 187.5& 103.9& 348.4& 137.8\\
$\mu$                &-354.1&-357.6&-177.7&-159.9& 347.8& 352.6& 178.0& 161.5\\
$\tan\beta$          &  14.6&  14.5&  29.1&  32.1&  15.0&  14.8&  29.2&  32.1\\
\hline
$M_3$                & 583.2& 583.3& 583.3& 583.5& 583.1& 583.1& 583.3& 583.4\\
$M_{\tilde{\tau}_L}$ & 114.9&2704.3& 128.3&4794.2& 128.0& 229.9&3269.3& 118.6\\
$M_{\tilde{\tau}_R}$ & 348.8& 129.9&1292.7& 130.1&2266.5& 138.5& 129.9& 255.1\\
$M_{\tilde{\mu}_L}$  & 192.7& 192.7& 192.7& 192.9& 192.6& 192.6& 192.7& 192.8\\
$M_{\tilde{\mu}_R}$  & 131.1& 131.1& 131.1& 131.3& 131.0& 131.0& 131.1& 131.2\\
$M_{\tilde{e}_L}$    & 186.3& 186.4& 186.4& 186.5& 186.2& 186.2& 186.4& 186.4\\
$M_{\tilde{e}_R}$    & 131.5& 131.5& 131.6& 131.7& 131.4& 131.4& 131.5& 131.6\\
$M_{\tilde{q}3_L}$   & 497.1& 497.2& 494.1& 494.0& 495.6& 495.6& 495.8& 495.0\\
$M_{\tilde{t}_R}$    &1073.9& 920.3& 547.9& 950.8& 547.9& 460.5& 978.2& 520.0\\
$M_{\tilde{b}_R}$    & 497.3& 497.3& 500.4& 500.9& 498.5& 498.5& 498.7& 499.6\\
$M_{\tilde{q}_L}$    & 525.1& 525.2& 525.3& 525.5& 525.0& 525.0& 525.2& 525.3\\
$M_{\tilde{q}_R}$    & 511.3& 511.3& 511.4& 511.5& 511.2& 511.2& 511.4& 511.5\\
$A_t$ $(-)$          &-252.3&-348.4&-477.1&-259.0&-470.0&-484.3&-243.4&-465.7\\
$A_t$ $(+)$          & 384.9& 481.8& 641.5& 432.5& 739.2& 774.7& 440.5& 656.9\\
$m_A$                & 350.3& 725.8& 263.1&1020.0& 171.6& 156.5& 897.6& 256.1\\
$m_t$                & 171.4& 171.4& 171.4& 171.4& 171.4& 171.4& 171.4& 171.4\\
\hline
\end{tabular}
\caption{List of the eight best--fitting points in the MSSM likelihood
  map with two alternative solutions for $A_t$.
  All masses are given in GeV. The $\chi^2$ value for all points 
  is approximately the same, so the ordering of the table is arbitrary. 
  The parameter point closest to the correct point is labeled as SPS1a.}
\label{tab:mssm_secondary}
\end{table}

In analogy to the MSUGRA analysis general features of the log-likelihood
map of the MSSM parameter space are studied before proceeding with
profile likelihood or Bayesian probability distributions.  Finally the
proper error analysis is performed. The first question is the presence
of alternative likelihood maxima in the MSSM parameter space.

Tab.~\ref{tab:mssm_secondary} lists the secondary local maxima in
the likelihood map, focusing on the neutralino--chargino
sector. These entries appear as a distinct secondary maximum in step~2
of SFitter. Each of them goes through steps~3 and 4, where 
it is explicitly checked that for a given value of $m_t$ no 
secondary likelihood maxima in the scalar sector alone turn up. 
In step~4 the resolution on the local maxima is improved and the
residual correlation between the neutralino--chargino and the scalar
sectors are evaluated.\bigskip

The most interesting feature in the different best-fitting points
listed in Tab.~\ref{tab:mssm_secondary} is the structure of the
neutralino sector. For a fixed sign of $\mu$ four equally good
solutions are found, which can be classified by the ordering of the mass
parameters: $M_1 < M_2 < |\mu|$ is the correct MSUGRA--type solution.
The reverse ordering of the two gaugino masses $M_2 < M_1 < |\mu|$ is
equally likely. In both cases the missing neutralino will be a
higgsino. Apart from these two light--gaugino scenarios the
second--lightest neutralino can be mostly a higgsino, which
corresponds to $M_1 < |\mu| < M_2$ and $M_2 < |\mu| < M_1$. Note that
given the set of LHC measurements the two gaugino masses can always be
switched as long as there are no chargino constraints. The one
neutralino which cannot be a higgsino is the LSP, because in that case
the $\mu$ parameter would also affect the second neutralino mass and
would have to be heavily tuned with the gaugino masses. Such a
solution does not have a comparable log-likelihood to the other
$2\times 4$ scenarios.

In spite of the different gaugino and higgsino contents, the physical
masses of the three visible neutralinos are the same in all points
listed in Tab.~\ref{tab:mssm_secondary}, as is the precisely measured
light Higgs mass. The shift in $\tan\beta$ for the correct SPS1a
parameter point is an effect of the smeared data set combined with the
rather poor constraints on this parameter and is within
the error bar (see later in this section).\bigskip

Looking at Tab.~\ref{tab:edges} there is an important feature of the set
of measurements: there are 22 measurements, counting the measurements
involving $m_{\tilde{l}}$ separately for electrons and muons.
Using these naively it should be possible to
completely constrain a 19-dimensional parameter space. However, the
situation is more complicated. These 22 measurements are constructed
from only 15 underlying masses. The additional measurements will
resolve ambiguities and improve errors, but they will not constrain
any additional parameters.  Looking at the set of measurements and at
Tab.~\ref{tab:mssm_ilc} with the errors, five model parameters turn out
to be not well constrained.  One problem which has already been
discussed is the heavy Higgs mass $m_A$. The next poorly determined
parameters are $M_{\tilde{t}_R}$ and $A_t$.  These parameters occur in
the stop sector, but none of them appear in any of the edge
measurements.

Moreover, there is no good direct measurement of $\tan\beta$.  Looking at
the neutralino and sfermion mixing matrices any effect in
changing $\tan\beta$ can always be accommodated by a corresponding
change in another parameter. This is particularly obvious in the
poorly measured stau sector. There only the lighter of
$M_{\tilde{\tau}_L}$ or $M_{\tilde{\tau}_R}$ is determined from the kinematic
endpoint of the $\tau\tau$ invariant--mass distribution. The heavier
mass parameter and $\tan\beta$ can compensate each other's effects
freely.  In contrast, the light--flavor slepton masses for all maxima
are identical.  This is an effect of the cascade measurements, which
very strongly constrain the mass difference between the neutralinos
and the light--flavor sleptons.\bigskip

There is exactly one measurement which strongly links these otherwise
unconstrained parameters, the mass of the lightest Higgs boson $m_h$.
This leaves a four--dimensional surface with a constant
log-likelihood. As the dependence between the different parameters is
highly non-linear, this limits the range in these parameters.  Outside
this surface the Higgs mass does not reach the measured value (or
other elementary constraints like non-tachyonic stops are violated) no
matter what the other parameters are. Therefore a
meaningful error can still be assigned 
to at least some of the parameters, while others turn
out to be basically undetermined.\bigskip

The parameter points in Tab.~\ref{tab:mssm_secondary} should therefore
be seen as a `typical' set of different solutions for these
parameters. The common link, the lightest Higgs mass, illustrates the
dependence on the individual parameters.

To illustrate the effect of the minimum surface two values
for $A_t$ are quoted in the table of minima. One of them appears as a solution of
the minimization procedure, while the other one is generated by an
additional step where every parameter except $A_t$ is kept fixed. 
The minimization is started from the original value for $|A_t|$ but
with a flipped sign. This procedure gives only
one additional solution.  The significant shift in $|A_t|$ shows the
sizeable correlations with the other parameters. Its origin is the
stop contribution to the lightest Higgs mass which contains sub--leading
terms linear in $A_t$. As a matter of fact, in other supersymmetric
parameter points where $\mu / \tan\beta$ is of the same order as $A_t$
much larger terms linear in $A_t$ would appear, while in SPS1a the
linear contributions of $A_t$ to $m_h$ are strongly suppressed compared
to the quadratic terms.

The two alternative solutions with flipped signs of $A_t$ are
particularly interesting, since two alternative MSUGRA solutions have 
already been observed in section~\ref{sec:sugra_likelihood}.  There, the
lack of measurements is compensated by the requirement of parameter
unification at the GUT scale. In the general MSSM an
alternative solution exists even if all parameters except for $A_t$
are kept fixed. If the four--dimensional minimum surface can be
constrained by further measurements, this degeneracy will vanish and
correlations will require the other parameters to shift, in order to
accommodate two distinct point--like minima.  The prime candidate for
such a shift is the top mass, as known from the SUGRA study.\bigskip

Technically, searching for alternative local maxima in
the log-likelihood map it is much easier to use gaussian theory
errors. Of course, this assumption is an approximation and cannot be
used to quote errors on the parameter points. Moreover, it can be
misleading when it comes to ranking the alternative solutions
according to their log-likelihood. On the other hand, switching from
gaussian to flat theory errors will only lead to a higher degeneracy
of the log-likelihood because of the flat behavior of $\chi^2$ and 
already for gaussian theory errors all alternative solutions
are equally likely. Flat theory errors do
not lead to additional alternative likelihood maxima or structures in
the likelihood map. In particular, they do not change the statement,
that the lightest neutralino has to be a gaugino to explain the
cascade--decay measurements.

As discussed in the MSUGRA case, these different interpretations of
the LHC data set could at least in part be disentangled by additional
channels which should open for different `wrong' mass parameters.

\subsection{Alternative mass assignment}
\label{sec:mssm_alternative}

\begin{table}[t]
\begin{tabular}{|l|rrr||l|rrr|}
\hline
                     &   SPS1a  & correct    & inverted       &    
                     &   SPS1a  & correct    & inverted       \\ \hline \hline
$M_1$                &    103.1 &     102.1 & 101.6          &
$M_2$                &    192.9 &     193.6 & 191.0          \\
$M_3$                &    577.9 &     582.0 & 582.1          &
$\tan\beta$          &     10.0 &       7.2 &   7.8          \\
$m_A$                &    394.9 &     394.0 & 299.3          &
$\mu$                &    353.7 &     347.7 & 369.3          \\ \hline
$M_{\tilde{e}_L}$    &    194.4 &     192.3 & 192.3          &
$M_{\tilde{e}_R}$    &    135.8 &     134.8 & 134.8          \\
$M_{\tilde{\mu}_L}$  &    194.4 &     191.0 & 191.0          &
$M_{\tilde{\mu}_R}$  &    135.8 &     134.7 & 134.7          \\
$M_{\tilde{\tau}_L}$ &    193.6 &     192.9 & 185.7          &
$M_{\tilde{\tau}_R}$ &    133.4 &     128.1 & 129.9          \\
$M_{\tilde{q}_L}$    &    526.6 &     527.0 & 527.1          &
$M_{\tilde{q}_R}$    &    508.1 &     514.8 & 514.9          \\
$M_{\tilde{q}3_L}$   &    480.8 &     477.9 & 478.5          &
$M_{\tilde{t}_R}$    &    408.3 &     423.6 & 187.6          \\
                     &          &\multicolumn{2}{c||}{}      & 
$M_{\tilde{b}_R}$    &    502.9 &     513.7 & 513.2          \\ \hline
$A_{l1,2}$           &   -251.1 &\multicolumn{2}{c||}{fixed 0}&
$A_\tau$             &   -249.4 &\multicolumn{2}{c|}{fixed 0}\\
$A_{d1,2}$           &   -821.8 &\multicolumn{2}{c||}{fixed 0}&
$A_b$                &   -763.4 &\multicolumn{2}{c|}{fixed 0}\\
$A_{u1,2}$           &   -657.2 &\multicolumn{2}{c||}{fixed 0}&
$A_t$                &   -490.9 &    -487.7 & -484.9         \\ \hline
$m_t$                &    171.4 &     172.2 &  172.2         &
                     &          &\multicolumn{2}{c|}{}        \\ \hline
\end{tabular}
\caption[]{Result for the MSSM parameter determination using the 
 LHC endpoint measurements assuming either the third or fourth
 neutralino to be missing. The log-likelihood for both points is
 almost identical. All masses are given in GeV.}
\label{tab:mssm_swapped}
\end{table}

Another test of general features of the MSSM likelihood just based on
best--fitting points is to exchange the two heavy neutralinos in the
LHC measurements as discussed in section~\ref{sec:sugra_lhc} for MSUGRA.
For this comparison the time--consuming error
estimate at the end of step~4 is neglected and the log-likelihood values
for the two best--fitting points are compared.  The results for the two fits with
the correct and swapped neutralino mass assignments are shown in
Tab.~\ref{tab:mssm_swapped}. After the discussion in the last section
it is not surprising that the likelihood for the two hypotheses in
their best--fitting points is not significantly different. There are
small shifts in all parameters entering the neutralino mass matrix,
but none of them appear significant. The central values for the four
neutralino masses move from $\{98.5,175.7,353.5,374.9\}$~GeV to
$\{98.5,175.8,375.0,393.3\}$~GeV.  The correctly identified fourth
neutralino in the first set has the same mass as the third neutralino
in the swapped case.

The consistent shift in the extracted value of $\tan\beta$ is an
effect of the smeared parameter point. The relatively large shift in
the heavy Higgs mass between the two scenarios looks more dramatic
than it is. When taking into account the error on this parameter
shown in Tab.~\ref{tab:mssm_ilc} this shift will
turn out to be well within the error bands and largely reflect different
starting values combined with a flat log-likelihood distribution in
$m_A$. Even though the heavy Higgs mass is
vastly different between the two cases, the light Higgs mass in both
best--fitting points is identical.  This means that for the typical
LHC precision the parameter point SPS1a is in the decoupling limit of
the heavy MSSM Higgs states.\bigskip

It might be possible to search for higgsinos in cascade decays
involving gauge bosons. Such a measurement could remove this
degeneracy, namely the mis-identification for example of three out of
four neutralinos. The same would be true if chargino masses could be
included in the analysis, which are not part of the standard SPS1a
sample~\cite{mihoko}.

\subsection{Profile likelihood and Bayesian probability}
\label{sec:mssm_lowdim}

The organization of SFitter in the MSSM case implies that it is not
possible to produce a high--resolution Markov chain for the entire
19-dimensional MSSM parameter space. The only Markov chain covering
the entire space is obtained at the end of step~1, and will be 
fairly coarse. On the other hand,
a dense--coverage log-likelihood map of the MSSM
parameter space as for the MSUGRA space cannot be produced because of the
large number of dimensions. This means that the analysis has to follow
two paths in parallel, namely the analysis of global features using a
Markov chain and the analysis of local features using additional
Minuit-type algorithms described in the Appendix.\bigskip

The Markov chain produced in step~1 covers the entire MSSM parameter
space. It should be used to compute lower--dimensional profile
likelihoods or Bayesian probability distributions, following the
discussion in Sec.~\ref{sec:sugra}. The problem is that to guarantee
coverage of the entire MSSM parameter space a flat proposal
function is used, which reduces the acceptance probability below the per-mille
level.  This acceptance rate is fine for the intended purpose, namely
to define an unbiased starting point for the maximum searches while
making sure that no regions of parameter space are missed.  In a
repeat of step~1 a more appropriate proposal function can be used,
for example a Breit--Wigner shape, with a width of one percent of the
total parameter range in each direction.\bigskip

A slight technical complication is that weighted Markov chains require
an accurate estimate of the size of excluded regions, \ie regions with
$\chi^2=0$. For example, the measurements of a mass difference in
Tab.~\ref{tab:edges} includes the sign of this mass difference.
Parameter points with an inverted mass hierarchy are assigned a zero
log-likelihood, which means one measurement can remove half of the
entire parameter space. This feature of the kinematic endpoints
reduces the relative volume of valid points in the exclusive
log-likelihood map to a very small fraction and introduces large
absolute errors on the determined size of this fraction. At this
stage, these statistical fluctuations dominate the behavior of the
marginalized Bayesian probabilities. To illustrate the log-likelihood
map the number of points per bin,
\ie the traditional Markov chain algorithm, is used. For a small fraction of
allowed parameter points this distribution is statistically
more stable. As a drawback, only the relative size of entries in the
log-likelihood map is significant.\bigskip

\begin{figure}[t]
 \psfrag{e}{\tiny$\mathsf{M}_{\tilde{e}_\mathsf{L}}$}
 \psfrag{L}{}
 \includegraphics[width=6cm]{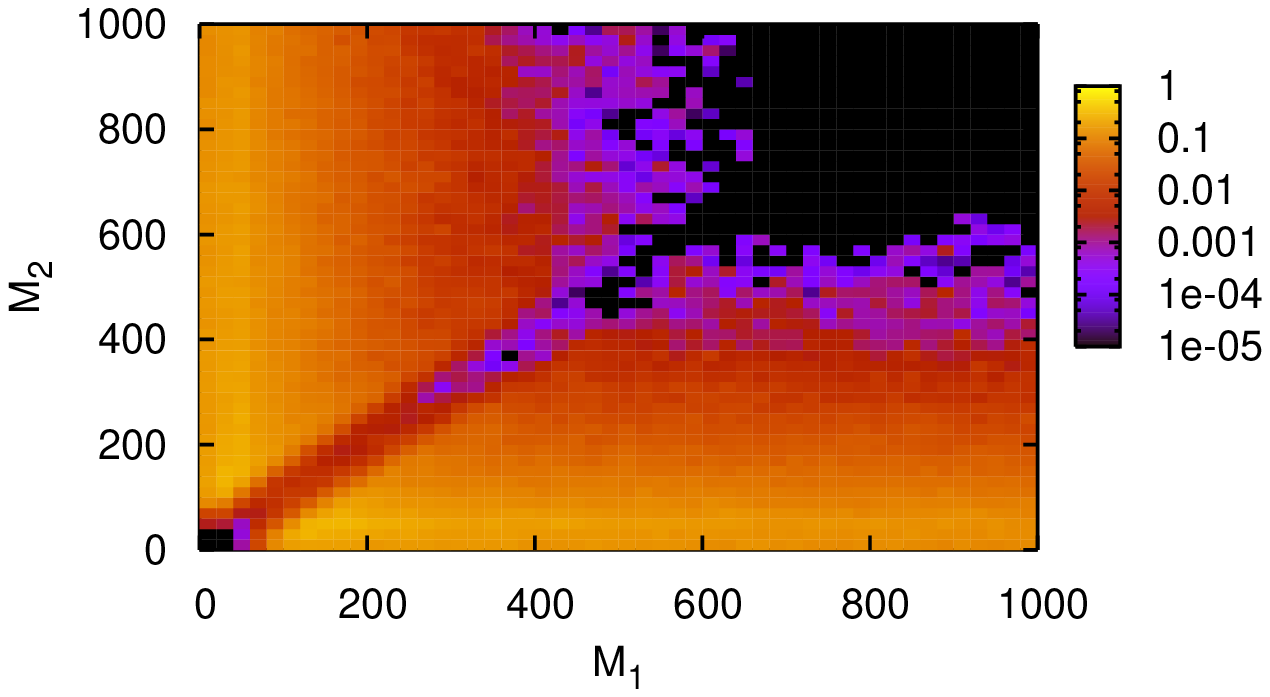} \hspace*{2cm}
 \includegraphics[width=6cm]{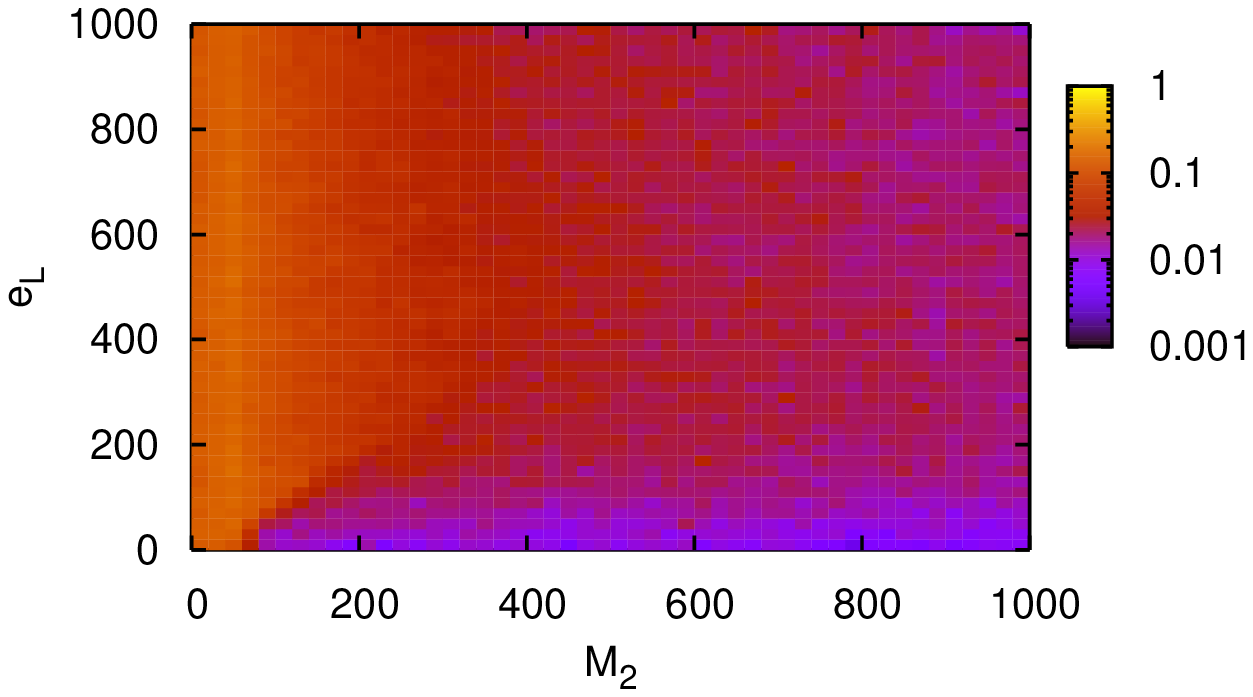} \\
 \includegraphics[width=5cm]{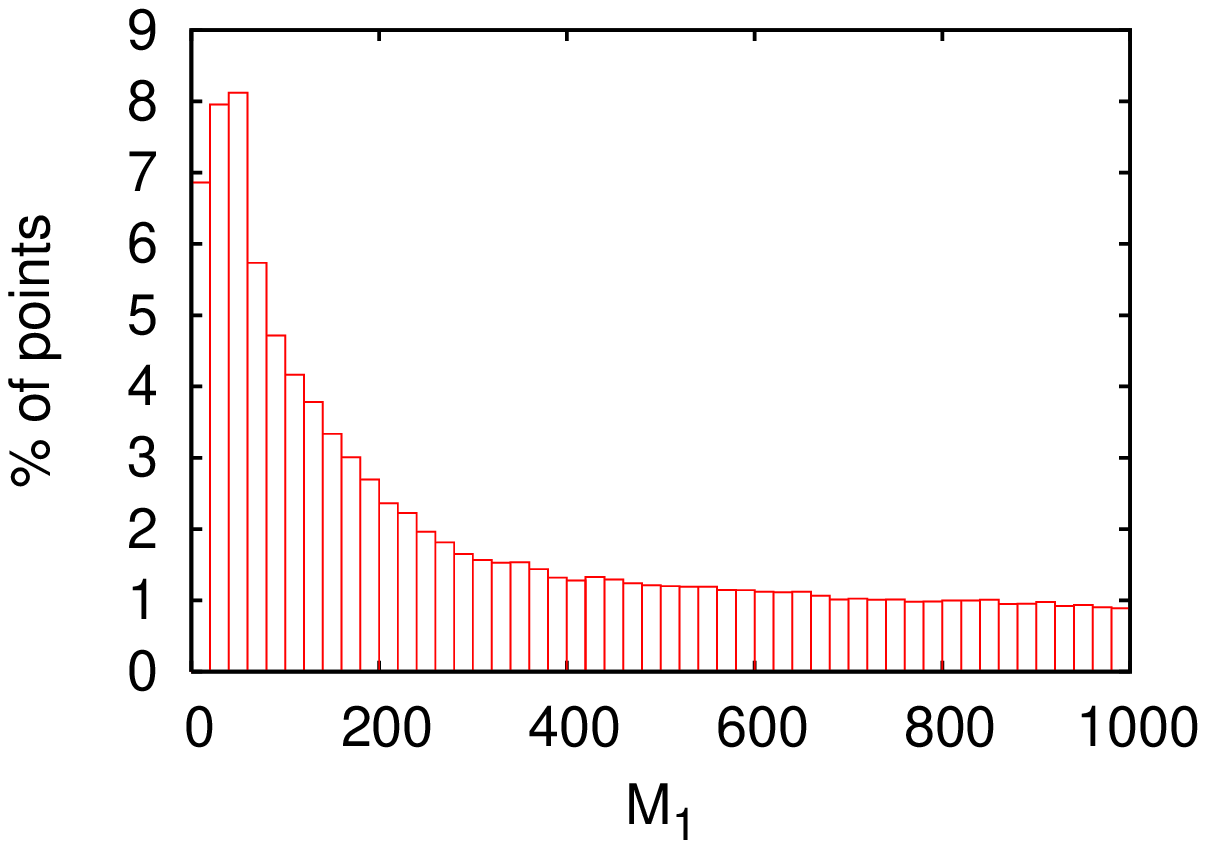} \hspace*{2cm}
 \includegraphics[width=5cm]{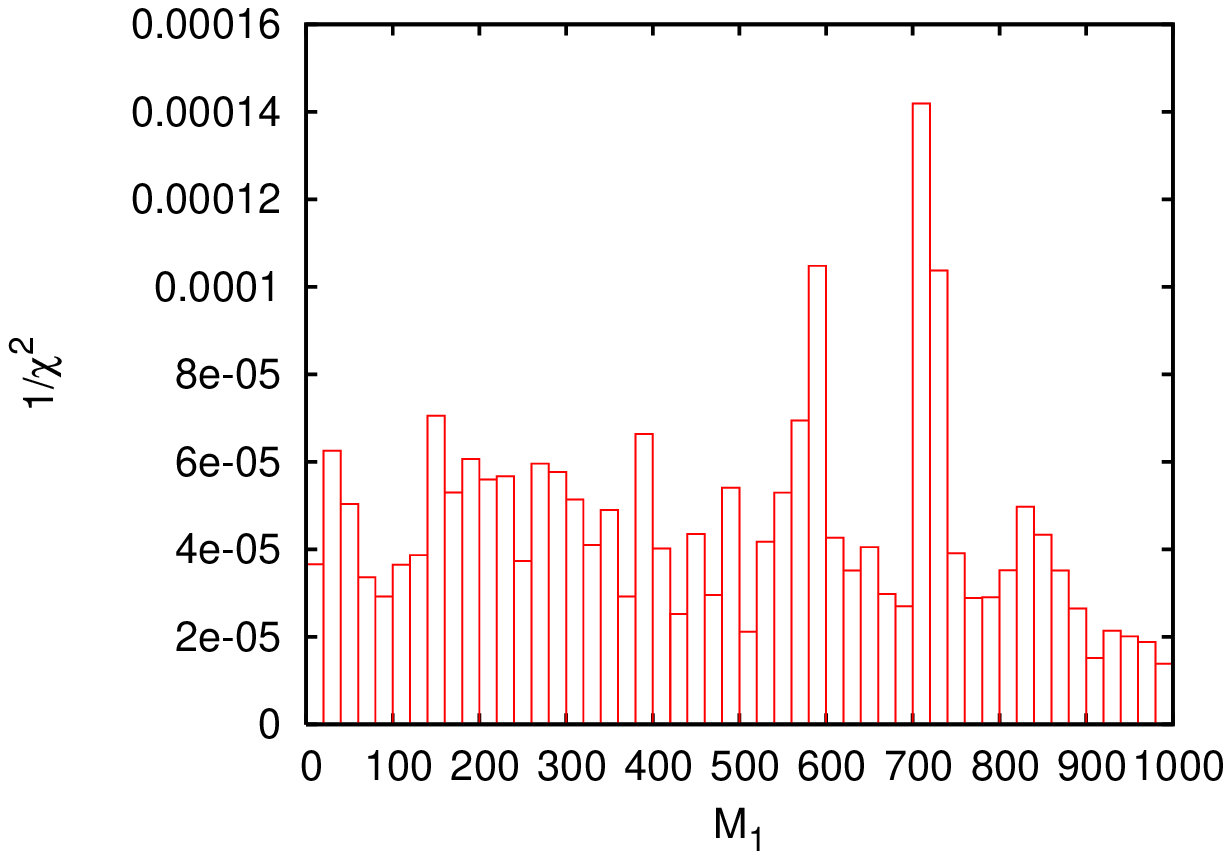} 
\caption[]{Marginalized Bayesian likelihoods (first three panels)
  and profile likelihood (bottom--right panel) for the complete MSSM
  parameter space (step~1) from SFitter. A Breit--Wigner
  proposal function is used to produce a Markov chain with $10^7$ points.}
\label{fig:mssm_map_full}
\end{figure}

In Fig.~\ref{fig:mssm_map_full} the marginalized Bayesian pdf is shown
for selected MSSM parameters using an exclusive likelihood map with a
Breit--Wigner proposal function. The two-dimensional $M_1-M_2$ plane
shows two branches, where one of the two gauginos has to form the
lightest neutralino. The second-lightest neutralino can be either a
gaugino or a higgsino. In the latter case the gaugino mass which does
not fix the LSP mass can either determine the last remaining
neutralino mass or it can essentially decouple.  In the
two-dimensional distribution a decoupled $M_1$ corresponds for example
to small $M_2$ giving the correct LSP mass and a higgsino--like
second-lightest neutralino.  In the one--dimensional distribution for
$M_1$ there is a broad peak at the correct value, and a washed--out
extended tail to large values. This tail is not a noise effect, but
corresponds to the described decoupling. The same $M_1$ distribution
computed as a profile likelihood illustrates the problem with the
Markov chain from step~1: in comparison to the Bayesian pdf from the
non--weighted Markov chain the profile likelihood is dominated by
noise.

The selectron and the wino masses in the second panel of
Fig.~\ref{fig:mssm_map_full} are uncorrelated, which in retrospect
justifies the 4-step organization of SFitter. Because of the explicit
appearance of the gluino--sbottom mass difference in the list of
measurements, Tab.\ref{tab:edges}, the gaugino--higgsino sector and
the scalar sector are if at all correlated through the gluino ---
which means that $M_3$ could as well be held fixed in step~2. 
This has little effect on the final result, but the
gluino--sbottom correlation will be the dominant effect in step~4 of
the SFitter strategy.\bigskip

\begin{figure}[t] 
 \psfrag{e}{\tiny$\mathsf{M}_{\tilde{e}_\mathsf{L}}$}
 \psfrag{q}{\tiny$\mathsf{M}_{\tilde{q}_\mathsf{L}}$}
 \psfrag{L}{}
 \includegraphics[width=6cm]{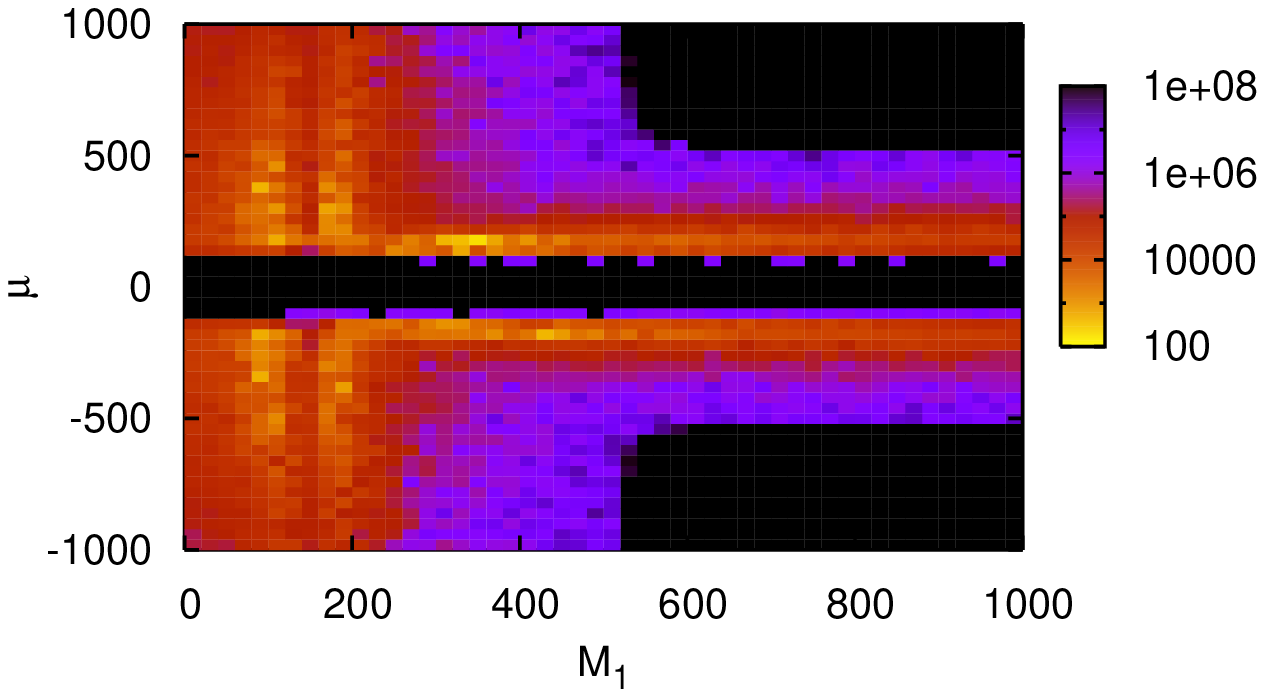} \hspace*{2cm}
 \includegraphics[width=6cm]{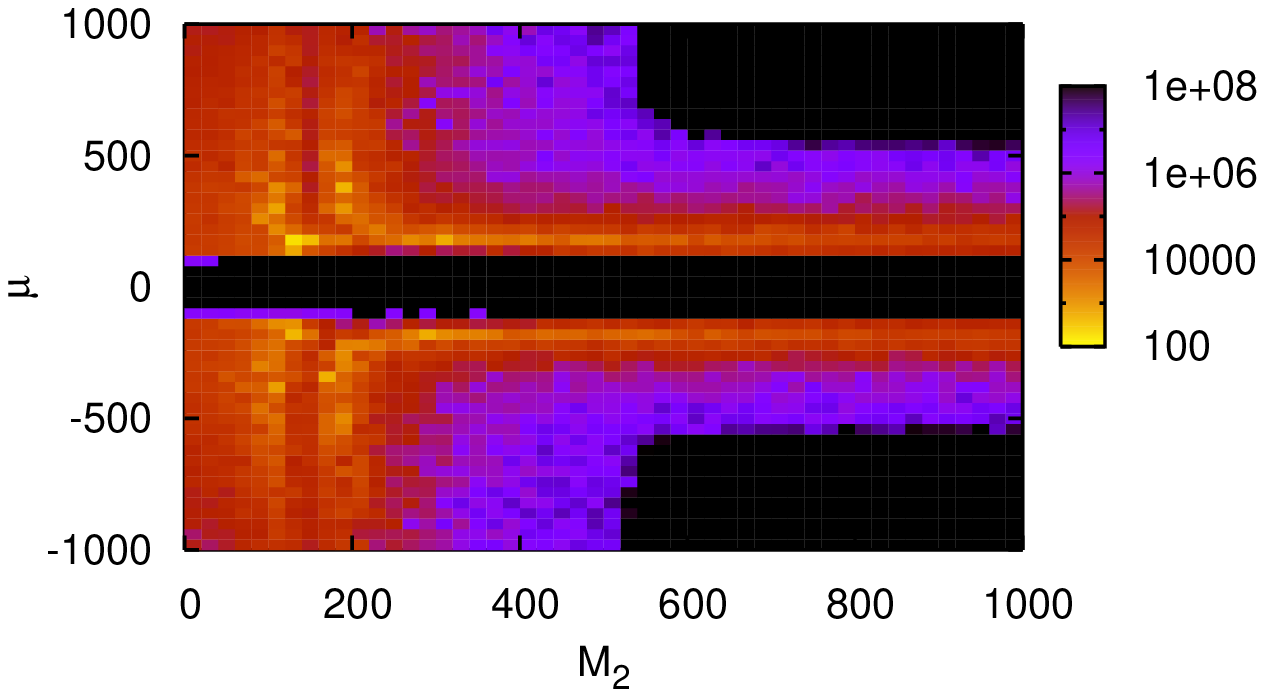} \\ 
 \includegraphics[width=6cm]{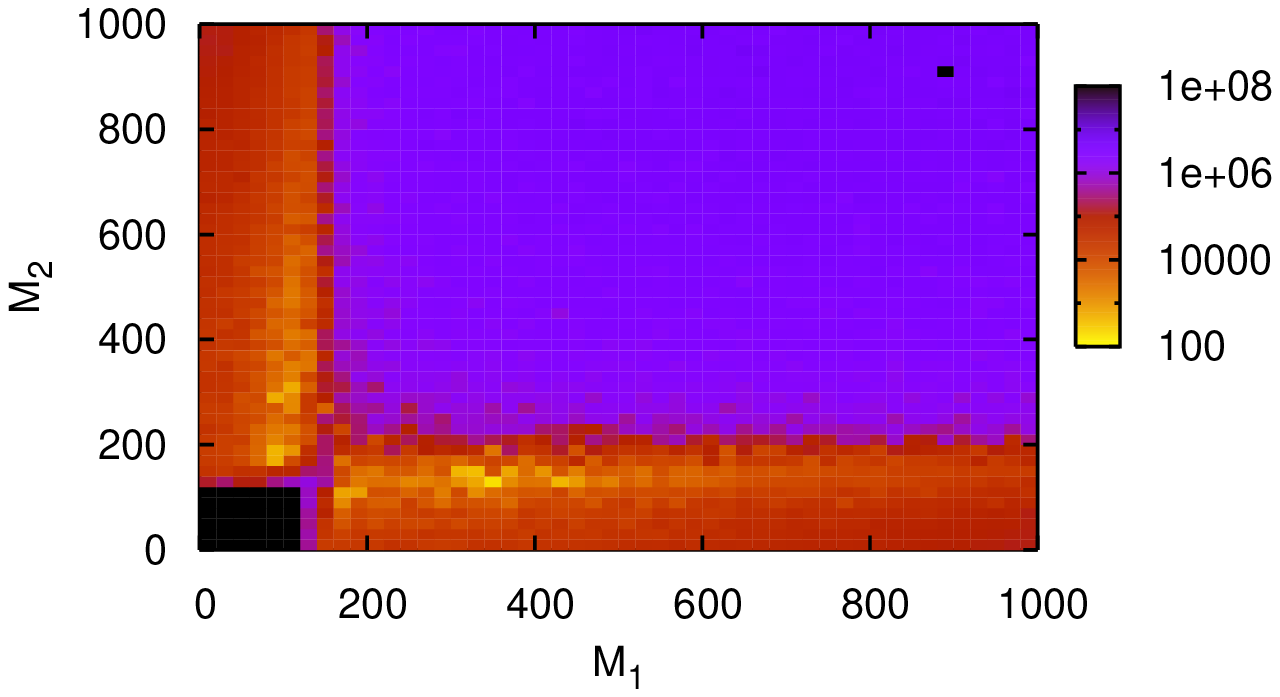} \hspace*{2cm}
 \includegraphics[width=6cm]{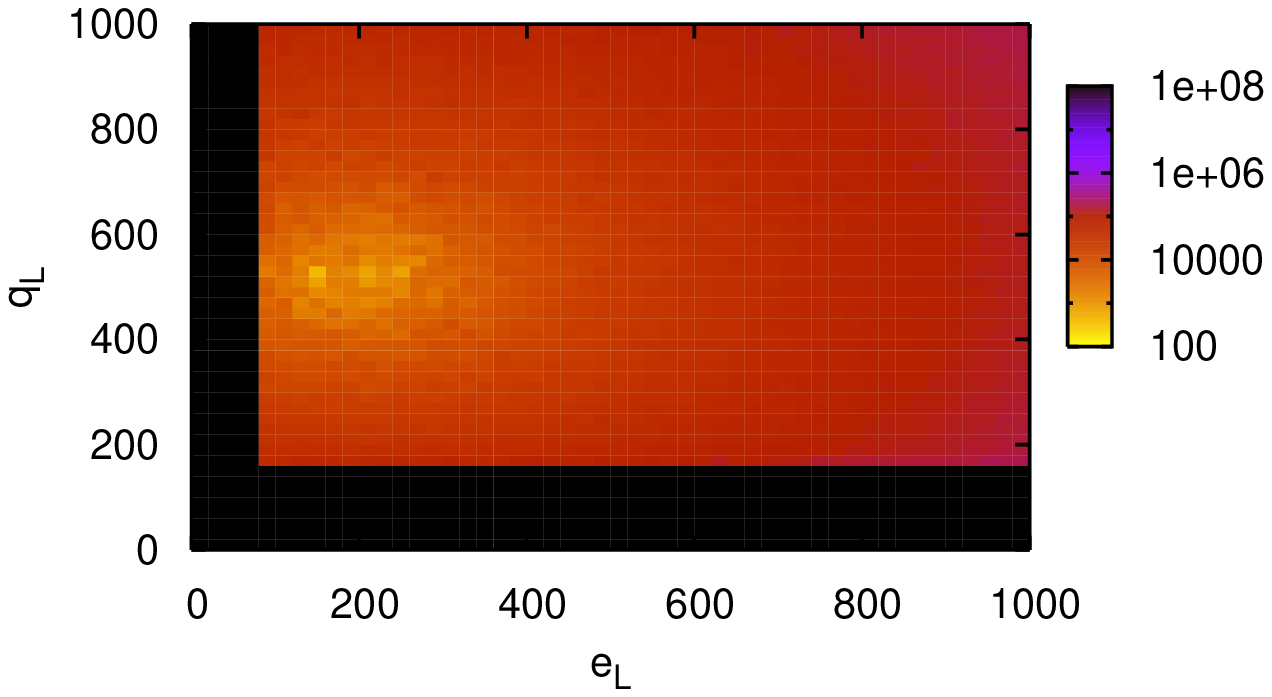} \\
 \includegraphics[width=5cm]{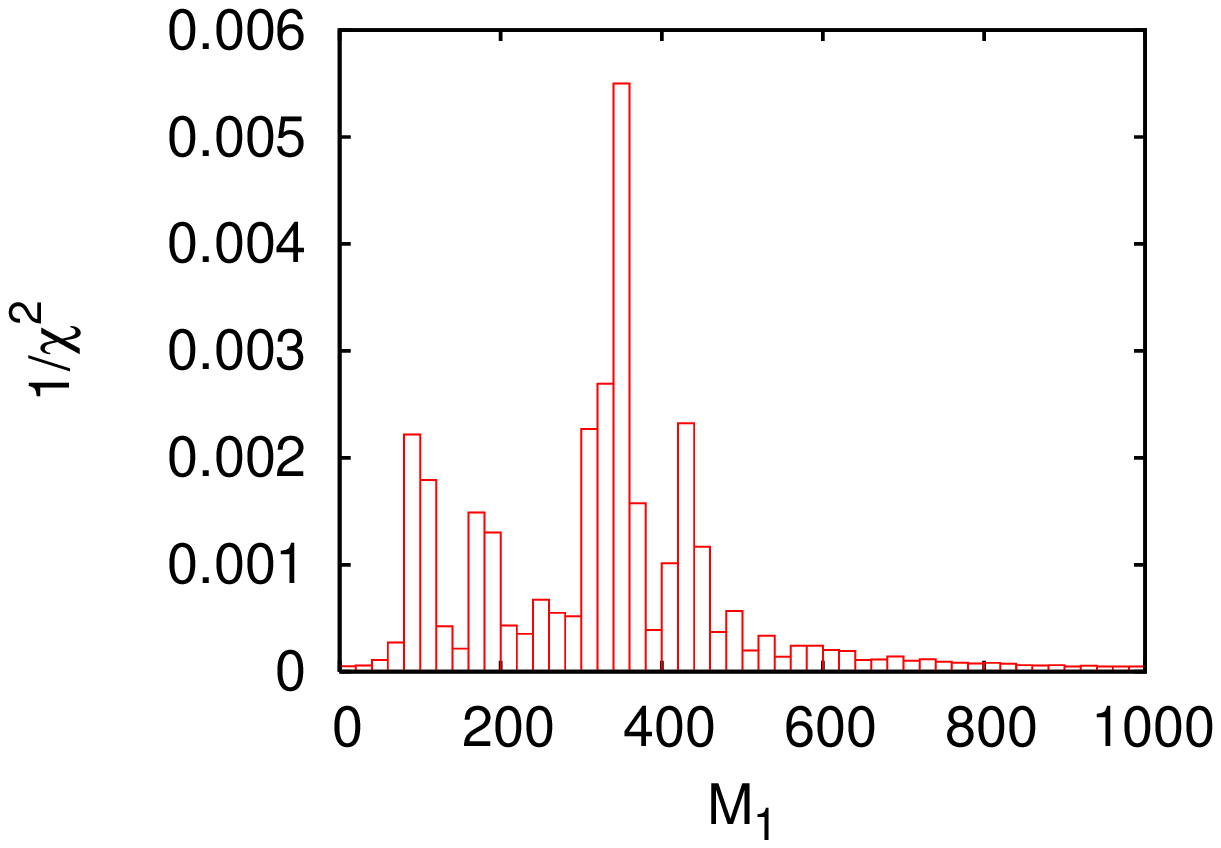} \hspace*{2cm}
 \includegraphics[width=5cm]{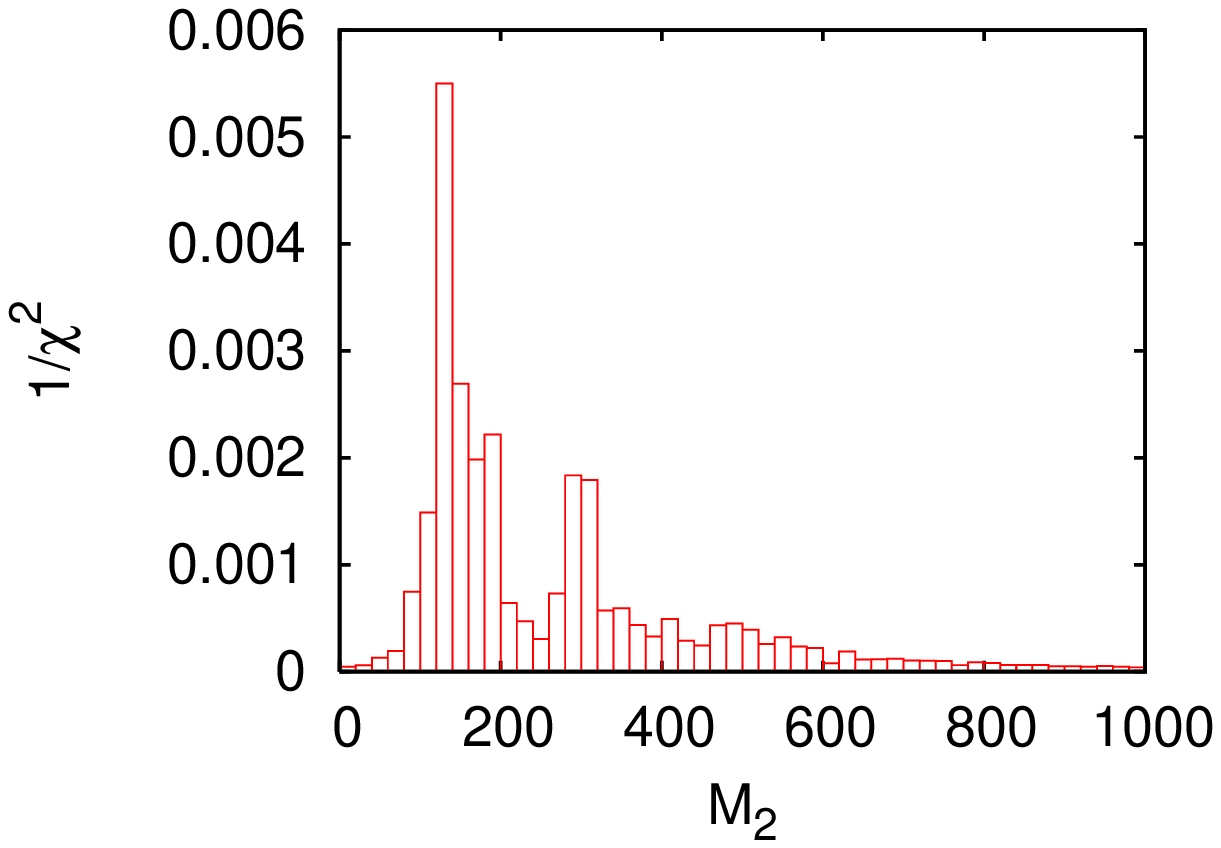} \\
 \includegraphics[width=5cm]{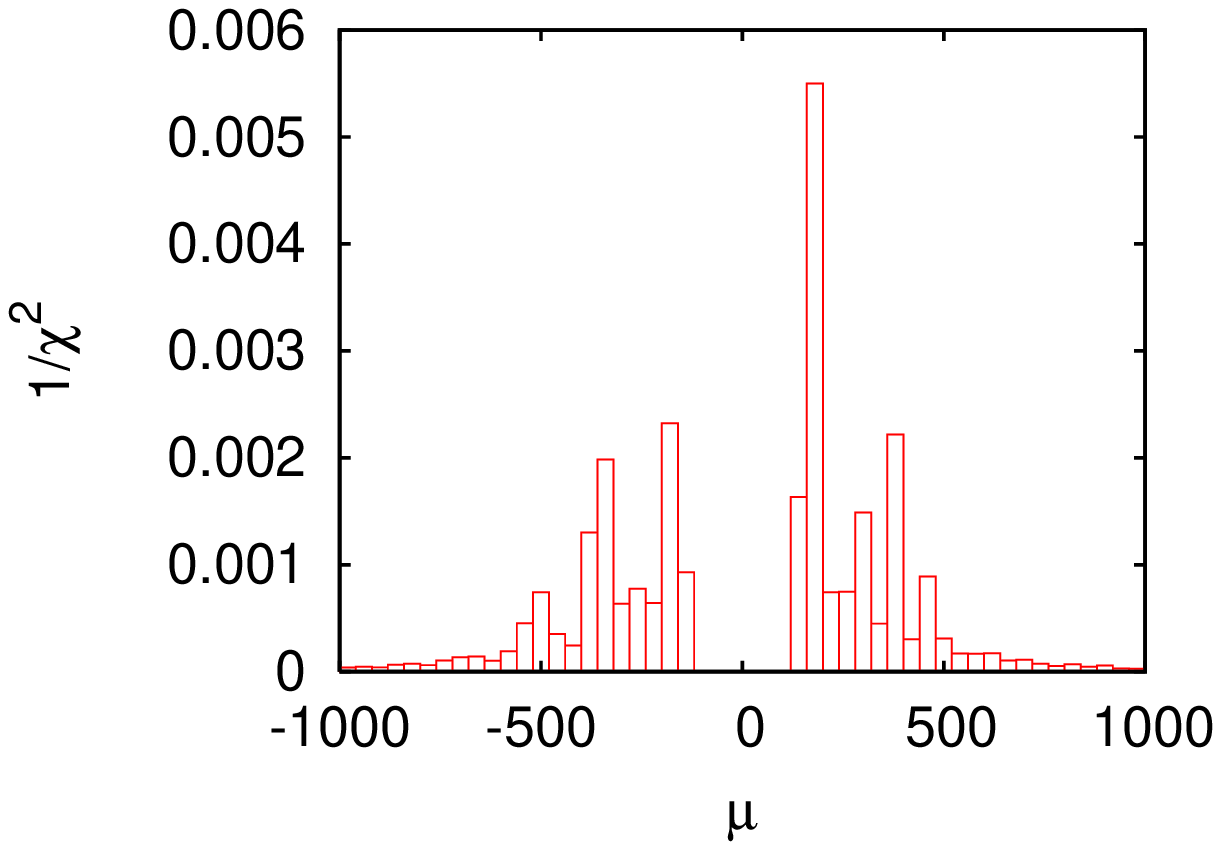} \hspace*{2cm}
 \includegraphics[width=5cm]{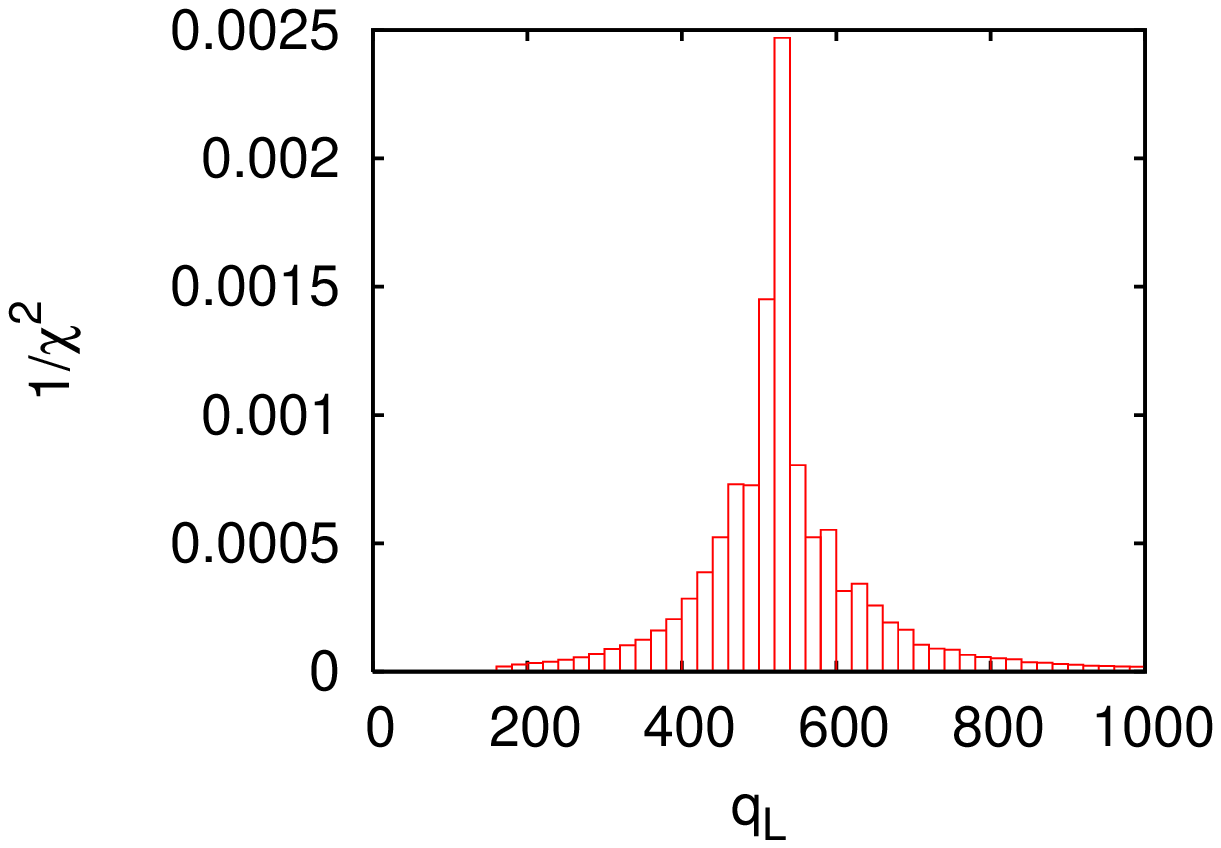} 
\caption[]{Profile likelihoods for the MSSM from SFitter.
 The distributions of the neutralino sector are derived from the
 log-likelihood map of the neutralino sector alone, using the Markov
 chain after step~2 in the SFitter strategy.}
\label{fig:mssm_map_f}
\end{figure}

Given the lack of correlations between the neutralino--chargino sector
and the scalar sector illustrated by Fig.~\ref{fig:mssm_map_full}, 
information from the Markov chain can be extracted in the
neutralino--chargino sector which SFitter computes in step~2. Fixing
all scalar parameters is equivalent to scanning them over their
orthogonal parameter space, provided the correlation between the
sectors are negligible, \ie the dimensions of the parameter space are
indeed orthogonal.  In Fig.\ref{fig:mssm_map_f} profile
likelihoods (as defined in Sec.~\ref{sec:sugra}) are shown for $M_{1,2}$ and
$\mu$. In the $M_1-M_2$ plane the same structure as in
Fig.~\ref{fig:mssm_map_full} is observed: one of the two gaugino
masses corresponds to the measured LSP mass while the other gaugino
mass can in principle decouple. In the $M_{1,2}-\mu$ plane
the three neutralino masses can be identified in the $M_{1,2}$ directions. For
light $M_{1,2}$ the higgsino mass parameter $|\mu|$ can be large,
while for one heavy gaugino $|\mu|$ is constrained to be small.

The one--dimensional profile likelihood for example for $M_1$ again
shows these three options with peaks around 100, 200 and 350~GeV,
corresponding to the three measured neutralino masses. The peak above
400~GeV is an alternative log-likelihood maximum which does not
correspond to a measured neutralino mass. For $M_2$ there is again the
100~GeV peak, where the LSP is a wino. The correct solution around
200~GeV is merged with the first maximum, while the third peak around
300~GeV corresponds to at least one light higgsino. In the profile
likelihood for $\mu$ the two signs of $\mu$ both
produce reasonable results. The 100~GeV range does not show a
distinctive peak because it would require the two lightest neutralinos
to be higgsinos, which means a high degree of tuning in all other
parameters. However, peaks around 200~GeV are clearly observed and in the
heavy--neutralino range for both signs of $\mu$.\bigskip

Because the Markov chains for the neutralino--chargino sector are
distinct, no information on the correlations between
the two sectors after step~1 of our SFitter strategy is available. Using only the
scalar--sector Markov chain from step~3 a small correlation is present in
the two scalar masses occurring in the squark cascades. They are in
principle slightly correlated through the kinematic endpoints from the
left--handed squark decay, but noise effects numerically dominate the
profile likelihood. The one--dimensional profile likelihood for the
squark mass parameter, however, is clearly peaked around the correct
value.\bigskip

The combination of these two Markov chains is of course not suited to
extract properly normalized probability distributions, because the
scalar sector is simply fixed to some best--fit values out of
step~1. On the other hand, these incomplete Markov chains show that
our likelihood map for the MSSM parameter space works and contains the
relevant structures, but that after step~1 it is somewhat
noisy.\bigskip

\begin{figure}[t] 
 \psfrag{e}{\tiny$\mathsf{M}_{\tilde{e}_\mathsf{L}}$}
 \psfrag{q}{\tiny$\mathsf{M}_{\tilde{q}_\mathsf{L}}$}
 \psfrag{L}{}
 \includegraphics[width=6cm]{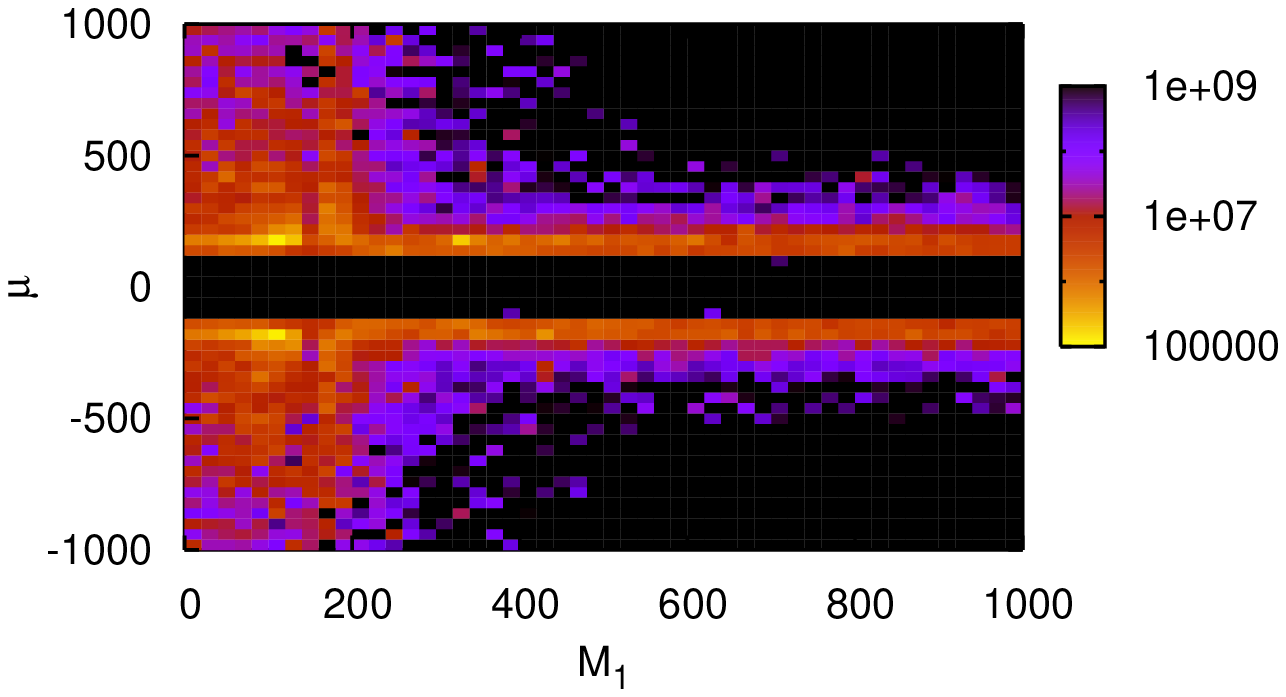} \hspace*{2cm}
 \includegraphics[width=6cm]{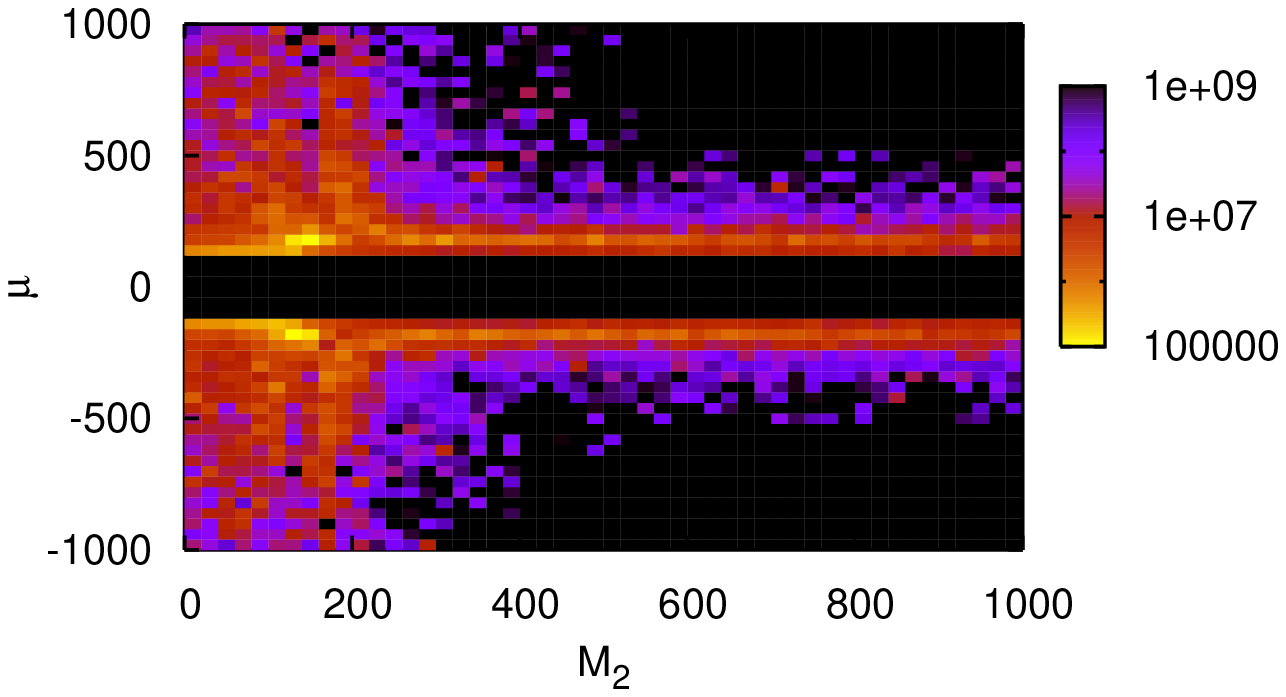} \\
 \includegraphics[width=6cm]{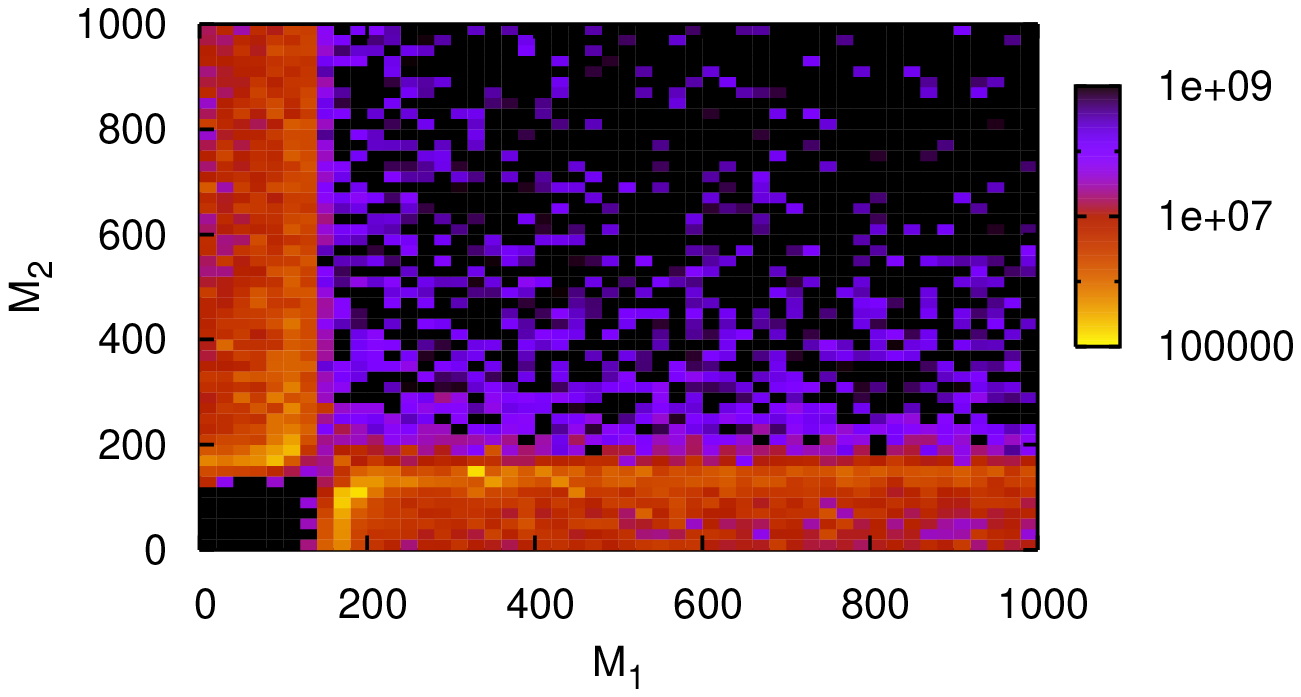} \hspace*{2cm}
 \includegraphics[width=6cm]{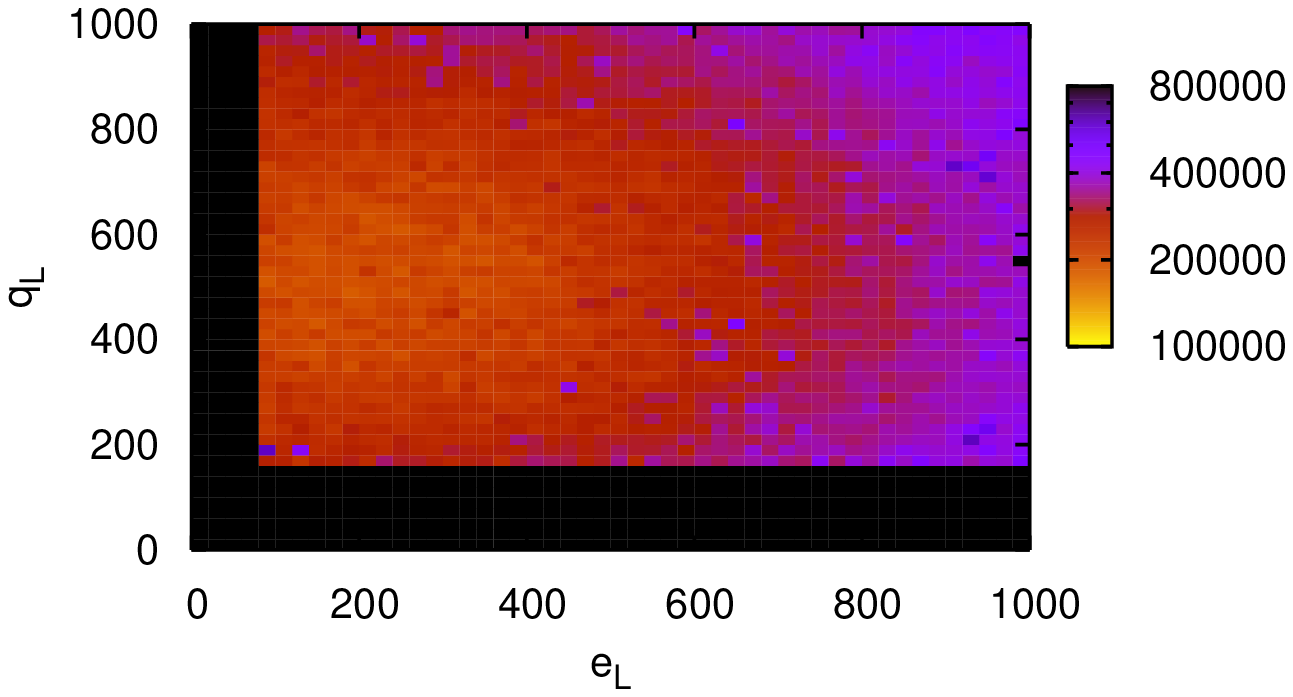} \\
 \includegraphics[width=5cm]{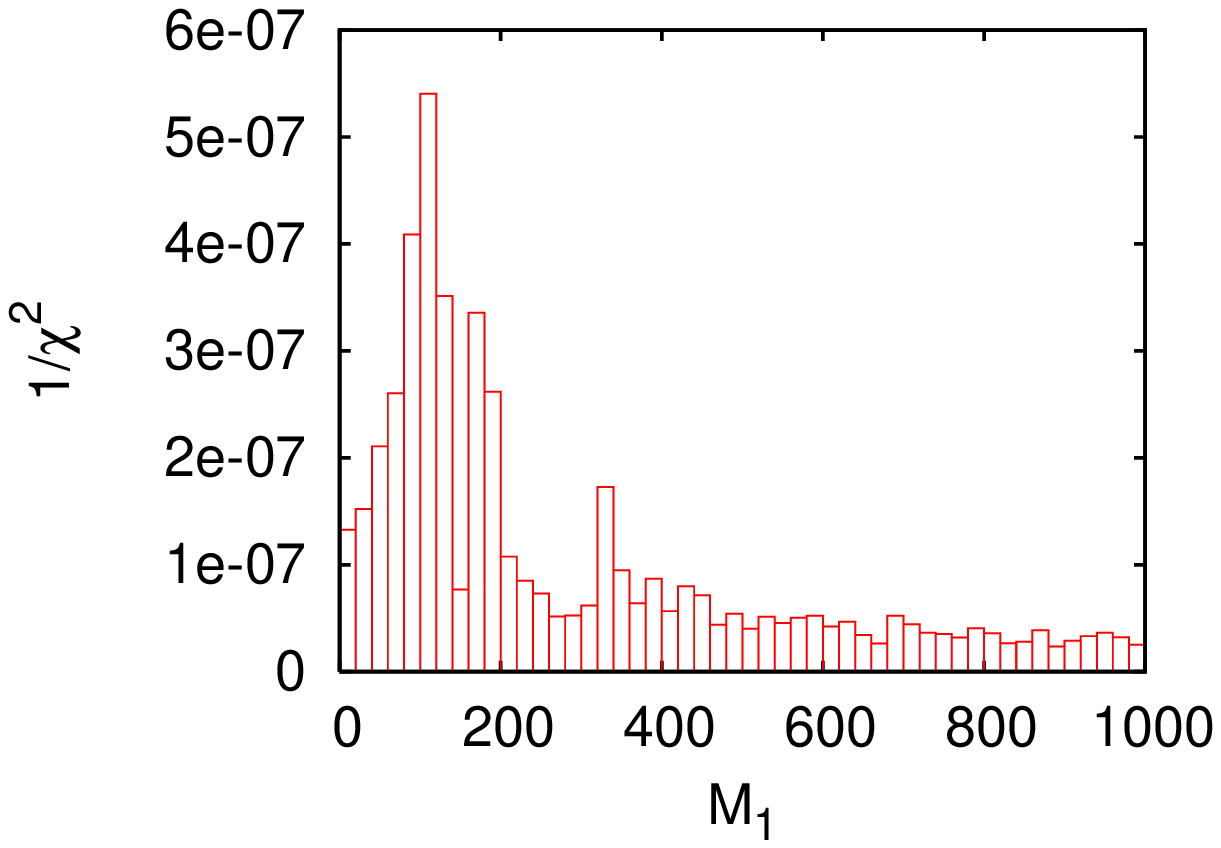} \hspace*{2cm}
 \includegraphics[width=5cm]{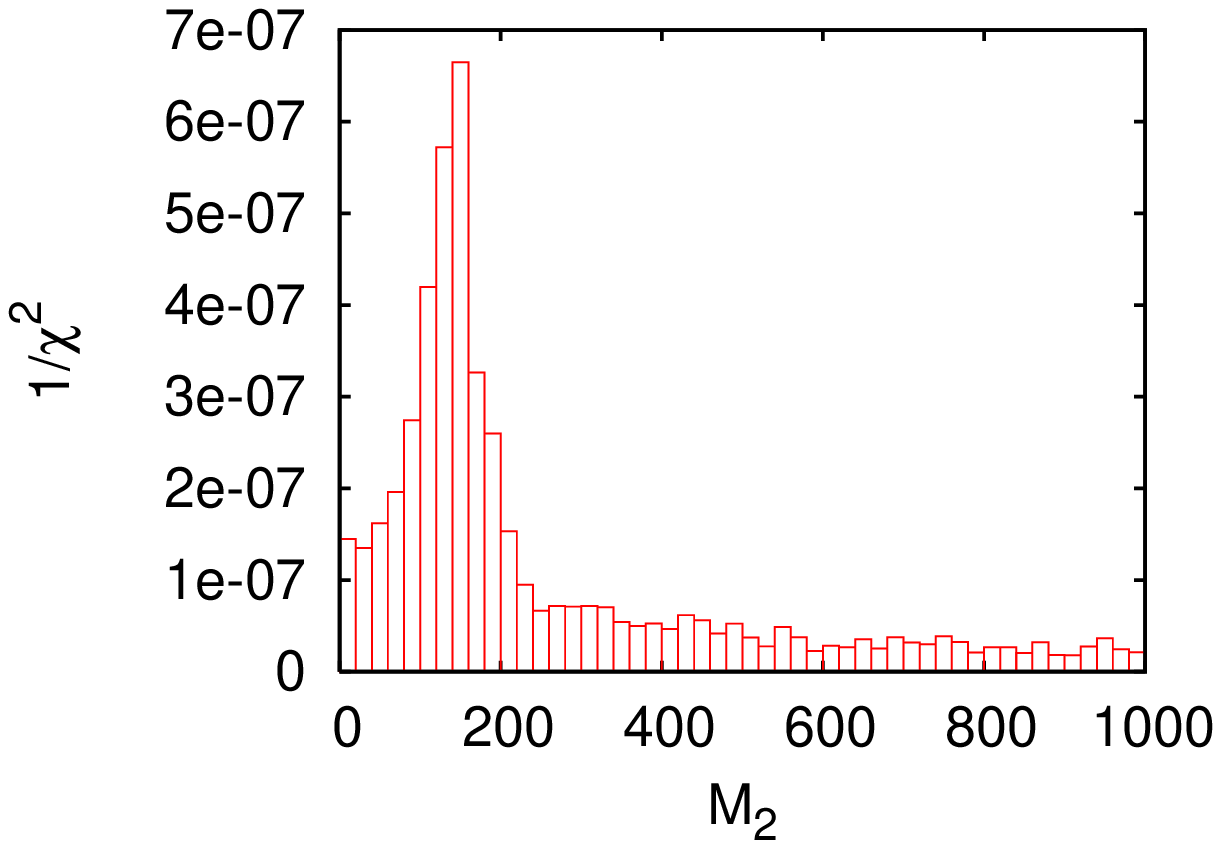} \\
 \includegraphics[width=5cm]{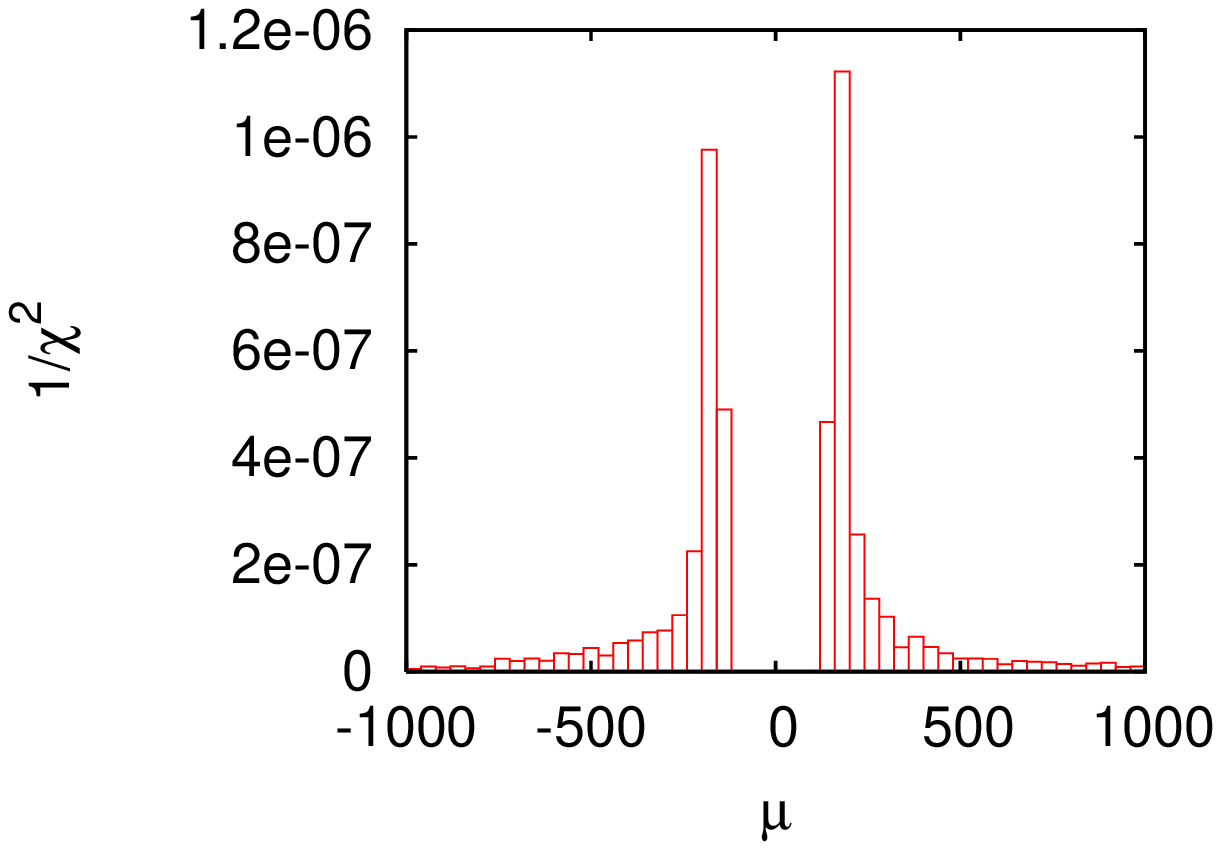} \hspace*{2cm}
 \includegraphics[width=5cm]{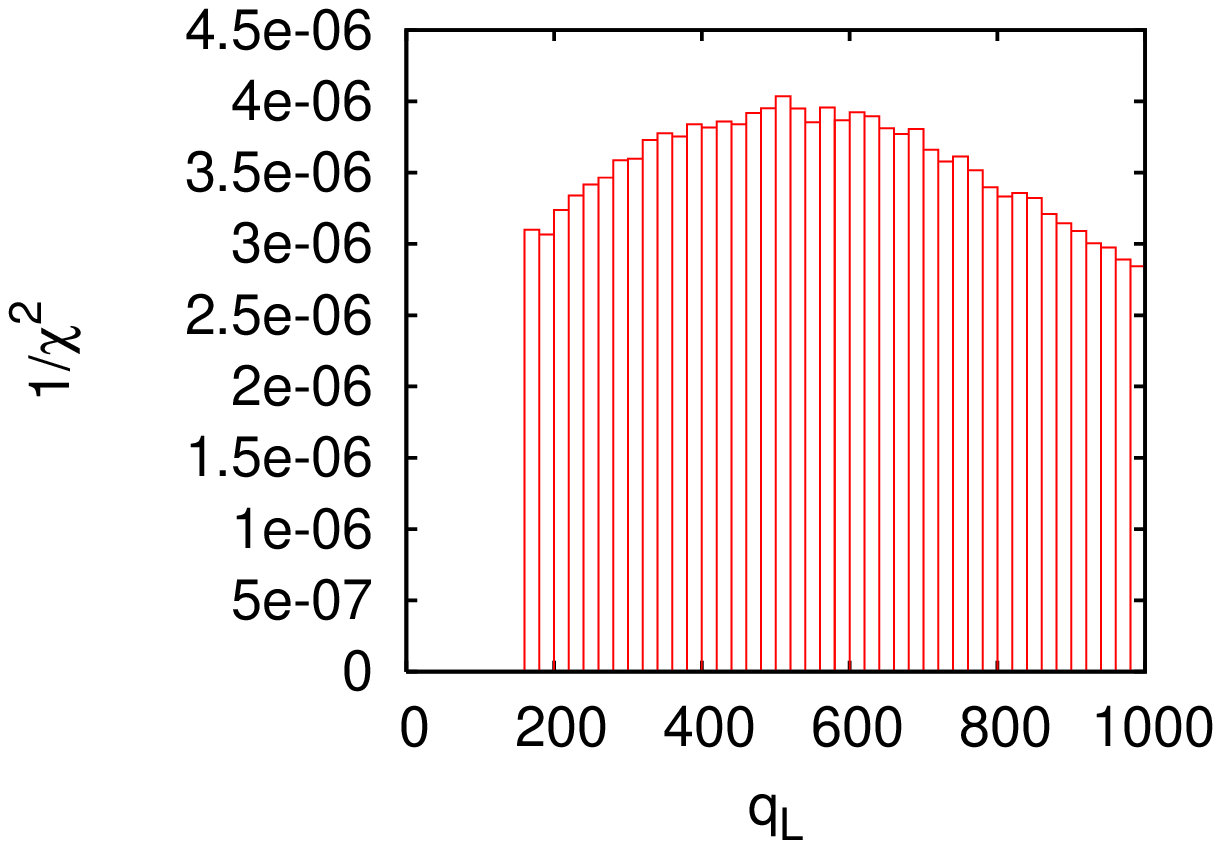} 
\caption[]{Marginalized Bayesian probabilities for the MSSM from SFitter.
 The distributions of the neutralino sector are derived from the
 log-likelihood map of the neutralino sector alone, using the Markov
 chain after step~2 in the SFitter strategy.}
\label{fig:mssm_map_b}
\end{figure}

In addition to the profile likelihoods shown in Fig.~\ref{fig:mssm_map_f},
SFitter also provides Bayesian probability distributions. For the
details of both approaches see section~\ref{sec:sugra}. While the structures 
in the two-dimensional
$M_1-M_2$ plane in Fig~\ref{fig:mssm_map_b} are similar to the profile
likelihood, the one-dimensional histograms show two significant
differences: first, the Bayesian pdf for $M_{1,2}$ shows the same
three physical solutions as the corresponding profile likelihood, namely
one peak around 100~GeV, another one around 200~GeV, separated only by
one bin from the edge of the 100~GeV peak, and a heavy--neutralino peak
above 300~GeV (more visible for $M_1$). However, the peaks in the
Bayesian pdf are much wider, as expected from the discussion of
the MSUGRA case.  The two lower peaks in $M_2$ even appear as one, with
a maximum around 150~GeV, which is a typical Bayesian volume effect.

The second difference between the profile likelihood and the Bayesian
pdf is that the Bayesian pdf can answer the question: which neutralino
is the most likely to be bino--like. Note that only the
neutralino--chargino Markov chain from step~2 is used, so the probabilistic
interpretation has to be taken with a grain of salt. However,
while $M_1$ has a best profile--likelihood entry around 350~GeV,
the Bayesian pdf shows a clear maximum for around 100~GeV. As usual,
SFitter leaves the interpretation of the two different approaches to
the reader.

As expected, the difference between the two signs of
$\mu$ is small, but both of them are driven to small values of $|\mu|$, again
by volume effects. This arises because of the decoupling of one of the
two gaugino masses for a light higgsino, while for two light gauginos
the higgsino mass is still determined by the fourth neutralino.  The
squark mass as a comparably well--measured and less noise--dominated
parameter shows the kind of behavior known from the MSUGRA case: the
profile likelihood is much more strongly peaked than the Bayesian pdf.

\subsection{Precision Analysis}
\label{sec:mssm_prec}

Similarly to the MSUGRA case, one of the most important outcomes of
the SFitter parameter extraction is the proper definition of the
errors of all extracted model parameters. The
flat theory errors are now only weak--scale uncertainties, for example
due to the translation of mass parameters into physical masses or due
to higher--order effects in the observables. Compared to the MSUGRA
case a proper error analysis in the MSSM is even more important: the
errors at the end of the day will determine if and how well we can
extract information on the SUSY breaking mechanism.

\subsubsection{Errors on MSSM parameters}

\begin{table}[t]
\begin{tabular}{|l|r@{$\pm$}rr@{$\pm$}rr@{$\pm$}rr|}
\hline
                     & \multicolumn{2}{c}{LHC}    & \multicolumn{2}{c}{ILC}     & \multicolumn{2}{c}{LHC+ILC} & SPS1a \\
\hline
$\tan\beta$          &       9.8 & 2.3             &      17.6 & 9.6             &      16.4 & 7.0             &     10.0 \\
$M_1$                &     101.5 & 4.6             &     102.8 & 0.72            &     102.7 & 0.53            &    103.1 \\
$M_2$                &     191.7 & 4.8             &     192.3 & 2.6             &     191.7 & 1.7             &    192.9 \\
$M_3$                &     575.7 & 7.7             &\multicolumn{2}{c}{fixed 500}&     578.0 & 6.3             &    577.9 \\
$M_{\tilde{\tau}_L}$ &     196.2 & $\om(10^2)$     &     185.4 & 14.3            &     187.8 & 13.6            &    193.6 \\
$M_{\tilde{\tau}_R}$ &     136.2 & 36.5            &     142.0 & 16.4            &     139.0 & 15.1            &    133.4 \\
$M_{\tilde{\mu}_L}$  &     192.6 & 5.3             &     194.4 & 0.53            &     194.4 & 0.51            &    194.4 \\
$M_{\tilde{\mu}_R}$  &     134.0 & 4.8             &     135.8 & 0.26            &     135.7 & 0.16            &    135.8 \\
$M_{\tilde{e}_L}$    &     192.7 & 5.3             &     194.4 & 0.24            &     194.4 & 0.22            &    194.4 \\
$M_{\tilde{e}_R}$    &     134.0 & 4.8             &     135.8 & 0.17            &     135.7 & 0.12            &    135.8 \\
$M_{\tilde{q}3_L}$   &     478.2 & 9.4             &     509.1 &$\om(2\cdot10^2)$&     489.6 & 10.7            &    480.8 \\
$M_{\tilde{t}_R}$    &     429.5 & $\om(10^2)$     &     427.6 & $\om(10^2)$     &     402.9 & 50.3            &    408.3 \\
$M_{\tilde{b}_R}$    &     501.2 & 10.0            &\multicolumn{2}{c}{fixed 500}&     494.4 & 10.5            &    502.9 \\
$M_{\tilde{q}_L}$    &     523.6 & 8.4             &\multicolumn{2}{c}{fixed 500}&     526.7 & 4.9             &    526.6 \\
$M_{\tilde{q}_R}$    &     506.2 & 11.7            &\multicolumn{2}{c}{fixed 500}&     508.2 & 10.8            &    508.1 \\
$A_\tau$             &\multicolumn{2}{c}{fixed 0}  &    2496.3 & $\om(10^4)$     &    2681.6 & $\om(10^4)$     &   -249.4 \\
$A_t$                &    -500.6 & 58.4            &    -521.8 & 160.1           &    -490.3 & 166.8           &   -490.9 \\
$A_b$                &\multicolumn{2}{c}{fixed 0}  &\multicolumn{2}{c}{fixed 0}  &    3084.9 & $\om(10^4)$     &   -763.4 \\
$A_{l1,2}$           &\multicolumn{2}{c}{fixed 0}  &\multicolumn{2}{c}{fixed 0}  &\multicolumn{2}{c}{fixed 0}  &   -251.1 \\
$A_{u1,2}$           &\multicolumn{2}{c}{fixed 0}  &\multicolumn{2}{c}{fixed 0}  &\multicolumn{2}{c}{fixed 0}  &   -657.2 \\
$A_{d1,2}$           &\multicolumn{2}{c}{fixed 0}  &\multicolumn{2}{c}{fixed 0}  &\multicolumn{2}{c}{fixed 0}  &   -821.8 \\
$m_A$                &     446.1 & $\om(10^3)$     &     393.4 & 1.1             &     393.4 & 1.1             &    394.9 \\
$\mu$                &     350.9 & 7.3             &     355.2 & 2.5             &     355.2 & 2.3             &    353.7 \\
$m_t$                &     171.4 & 1.0             &     171.4 & 0.12            &     171.4 & 0.12            &    171.4 \\
\hline
\end{tabular}
\caption[]{Result for the general MSSM parameter determination in
  SPS1a assuming vanishing theory errors. As experimental measurements
  the kinematic endpoint measurements given in
  Tab.~\ref{tab:edges} are used for the LHC column, and the mass measurements
  given in Tab.~\ref{tab:mass_errors} for the ILC column. In the LHC+ILC column these two
  measurements sets are combined. Shown are the nominal parameter values
  and the result after fits to the different data sets. All masses are
  given in GeV.}
\label{tab:mssm_ilc_noerrors}
\end{table}

\begin{table}[t]
\begin{tabular}{|l|r@{$\pm$}rr@{$\pm$}rr@{$\pm$}rr|}
\hline
                     & \multicolumn{2}{c}{LHC}    & \multicolumn{2}{c}{ILC}     & \multicolumn{2}{c}{LHC+ILC} & SPS1a \\
\hline
$\tan\beta$          &      10.0 & 4.5             &      12.1 & 7.0             &      12.6 & 6.2             &     10.0 \\
$M_1$                &     102.1 & 7.8             &     103.3 & 1.1             &     103.2 & 0.95            &    103.1 \\
$M_2$                &     193.3 & 7.8             &     194.1 & 3.3             &     193.3 & 2.6             &    192.9 \\
$M_3$                &     577.2 & 14.5            &\multicolumn{2}{c}{fixed 500}&     581.0 & 15.1            &    577.9 \\
$M_{\tilde{\tau}_L}$ &     227.8 & $\om(10^3)$     &     190.7 & 9.1             &     190.3 & 9.8             &    193.6 \\
$M_{\tilde{\tau}_R}$ &     164.1 & $\om(10^3)$     &     136.1 & 10.3            &     136.5 & 11.1            &    133.4 \\
$M_{\tilde{\mu}_L}$  &     193.2 & 8.8             &     194.5 & 1.3             &     194.5 & 1.2             &    194.4 \\
$M_{\tilde{\mu}_R}$  &     135.0 & 8.3             &     135.9 & 0.87            &     136.0 & 0.79            &    135.8 \\
$M_{\tilde{e}_L}$    &     193.3 & 8.8             &     194.4 & 0.91            &     194.4 & 0.84            &    194.4 \\
$M_{\tilde{e}_R}$    &     135.0 & 8.3             &     135.8 & 0.82            &     135.9 & 0.73            &    135.8 \\
$M_{\tilde{q}3_L}$   &     481.4 & 22.0            &     499.4 &$\om(10^2)$      &     493.1 & 23.2            &    480.8 \\
$M_{\tilde{t}_R}$    &     415.8 & $\om(10^2)$     &     434.7 &$\om(4\cdot10^2)$&     412.7 & 63.2            &    408.3 \\
$M_{\tilde{b}_R}$    &     501.7 & 17.9            &\multicolumn{2}{c}{fixed 500}&     502.4 & 23.8            &    502.9 \\
$M_{\tilde{q}_L}$    &     524.6 & 14.5            &\multicolumn{2}{c}{fixed 500}&     526.1 & 7.2             &    526.6 \\
$M_{\tilde{q}_R}$    &     507.3 & 17.5            &\multicolumn{2}{c}{fixed 500}&     509.0 & 19.2            &    508.1 \\
$A_\tau$             &\multicolumn{2}{c}{fixed 0}  &     613.4 & $\om(10^4)$     &     764.7 & $\om(10^4)$     &   -249.4 \\
$A_t$                &    -509.1 & 86.7            &    -524.1 & $\om(10^3)$     &    -493.1 & 262.9           &   -490.9 \\
$A_b$                &\multicolumn{2}{c}{fixed 0}  &\multicolumn{2}{c}{fixed 0}  &     199.6 & $\om(10^4)$     &   -763.4 \\
$A_{l1,2}$           &\multicolumn{2}{c}{fixed 0}  &\multicolumn{2}{c}{fixed 0}  &\multicolumn{2}{c}{fixed 0}  &   -251.1 \\
$A_{u1,2}$           &\multicolumn{2}{c}{fixed 0}  &\multicolumn{2}{c}{fixed 0}  &\multicolumn{2}{c}{fixed 0}  &   -657.2 \\
$A_{d1,2}$           &\multicolumn{2}{c}{fixed 0}  &\multicolumn{2}{c}{fixed 0}  &\multicolumn{2}{c}{fixed 0}  &   -821.8 \\
$m_A$                &     406.3 & $\om(10^3)$     &     393.8 & 1.6             &     393.7 & 1.6             &    394.9 \\
$\mu$                &     350.5 & 14.5            &     354.8 & 3.1             &     354.7 & 3.0             &    353.7 \\
$m_t$                &     171.4 & 1.0             &     171.4 & 0.12            &     171.4 & 0.12            &    171.4 \\
\hline
\end{tabular}
\caption[]{Result for the general MSSM parameter determination in
  SPS1a assuming flat theory errors. As experimental measurements the
  kinematic endpoint measurements given in Tab.~\ref{tab:edges} are
  used for the LHC column, and the mass measurements given in
  Tab.~\ref{tab:mass_errors} for the ILC column. In the LHC+ILC column
  these two measurements sets are combined. Shown are the nominal
  parameter values and the result after fits to the different data
  sets. All masses are given in GeV.}
\label{tab:mssm_ilc}
\end{table}

For the best--fit
parameter point, we show the results for the error determination in
Tables~\ref{tab:mssm_ilc} and \ref{tab:mssm_ilc_noerrors}. The general
feature is that the LHC is not sensitive to several parameters.
Some of them, namely the trilinear mixing terms $A_i$ are fixed in the
fit. Others, like the heavier stau--mass and stop--mass parameters or
the pseudoscalar Higgs mass turn out to be unconstrained. In the stau
sector only the lighter of the two mass eigenstates is observed in
Tab.~\ref{tab:mass_errors}. Because of the non--zero mixing between
the two staus, the relative error on the mass parameter is much larger
than the experimental error on the lighter stau mass. Because the
heavy Higgses are for all practical purposes decoupled at the LHC, the
parameters in the Higgs sector are $\tan\beta$ and the lightest stop
mass. Because the sbottom masses are known from the gluino cascade
decay, the stop mass matrix has two remaining free parameters.

As expected in the slepton sector, the ILC improves the precision by an order
of magnitude in the parameters be it with or without theory errors. Again
the ILC alone, where parameters can be measured, dominates the precision. 
It is instructive to compare the effect of theory errors on the parameter determination. 
While the ILC loses a factor~5 in precision, going from a per-mille determination 
to half a percent, the LHC looses roughly less than a factor~2. The naive 
expectation would have called for only the ILC measurement being
affected. However, the
LHC measurements being functions of several sparticle masses, the error 
propagation leads also to a significant theory error (Table~\ref{tab:edges}). In particular 
the $\ell\ell$ mass theory error is larger than the experimental error. 
The strength of the LHC is clearly visible in the sector of sparticles
with color quantum numbers.

While for the LHC and ILC separately not all parameters can be
determined, the combination of the two machines allows to determine
all parameters (with the exception of the first and second generation
trilinear couplings) with good precision. The combination of LHC and
ILC measurements can be particularly useful to determine the link
to dark--matter observables~\cite{masters,leszek,ellis_olive,ilc_dm,with_dan}.

\subsubsection{Testing unification}
\label{sec:uni}

Once the parameters of the weak--scale MSSM--Lagrangian have been
determined, the next step is to extrapolate the parameters all the way
to the Planck scale. Inspired by the apparent unification of the gauge
couplings~\cite{coupling_uni} in the MSSM the question arises if any
other running parameters unify at a higher scale
as shown in the pioneering work in~\cite{msugra,blair}.
Such structures can give hints for example about
supersymmetry--breaking. For two reasons, the prime candidates for
unification in supersymmetry are the gaugino masses: first, in
contrast to the scalar masses, the three gaugino masses can well be
argued to belong to the same sector of physics, being the partners of
gauge bosons of a possibly unified gauge group. Secondly, interactions
between the hidden SUSY--breaking sector and the MSSM particle content
can affect the unification pattern, in particular for scalars. In that
case, scalar mass unification might be replaced by much less obvious
sum rules for scalar masses at some high scale~\cite{martin}.\bigskip

\begin{figure}[t]
 \includegraphics[width=7cm]{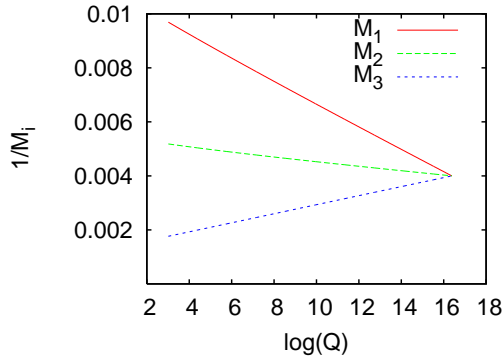}
\caption[]{SFitter/SuSpect output for the upward renormalization
  group running of the three gaugino masses in the MSSM.  The central
  values are shown without error bars, a more detailed study of
  bottom--up running is beyond the scope of this
  paper~\cite{sfitter_uni}.} 
\label{fig:uni_mi}
\end{figure}

Technically, upwards running is considerably more
complicated~\cite{suspect,sfitter_uni} than starting from a
unification--scale and testing the unification hypothesis by comparing
to the weak--scale particle spectrum. For example, it is by no means
guaranteed that the renormalization group running will converge for
weak--scale input values far away from the top--down prediction. In
Figure~\ref{fig:uni_mi} the extrapolation of the central
values of the gaugino mass parameters is shown using SuSpect. As expected in
SPS1a, the mass parameters unify at the GUT scale. This figure is only
a proof of concept for the SFitter approach to testing unification. A
full study of the extrapolation to the high scale including error
estimate is beyond the scope of this paper~\cite{sfitter_uni}.

\section{Outlook}

If the LHC is successful in discovering physics beyond the Standard
Model, the focus of its running will be on the interpretation of this
new physics, identifying the ultraviolet completion of the Standard
Model. The situation would be similar to current fits to electroweak
precision data, flavor--physics data and dark--matter constraints, but
likely considerably more complex. This increase in complexity is a
challenge to the statistical tools employed to study high--dimensional
physics parameter spaces.\bigskip

SFitter translates measurements for example of new particles' masses
into information on the weak--scale Lagrangian. It uses a combination
of (weighted) Markov chains and modified Minuit algorithms. The 
roughly 20-dimensional and highly
correlated weak--scale MSSM parameter space can be controlled by
SFitter. The correct description of all errors is a challenge for any
high--dimensional parameter determination.  However,
especially to distinguish different new--physics models, a proper
error propagation is crucial.  Therefore, SFitter includes the proper
treatment of statistical and systematic experimental errors as well as
(flat) theory errors, including arbitrary correlation.\bigskip

As an example two physics models, the low--dimensional toy model
MSUGRA and the effectively 19-dimensional MSSM, are analyzed in
detail.  SFitter first produces an unintegrated log-likelihood map
using Markov--chain techniques. For both models this likelihood map is
studied and distinct local maxima are identified, which SFitter
resolves using modified Minuit algorithms. For the best--fitting
parameter points the error on the extracted model parameters are
determined, properly including all experimental and theory errors.

Alternative maxima are for example due to the sign of the higgsino
mass parameter, to the structure of the neutralino mass matrix, or to
a correlation between the top Yukawa and the trilinear mixing
parameter. While for MSUGRA these local maxima correspond to different
values of the log-likelihood, they are degenerate in the MSSM and
cannot be resolved using the relative likelihood values \bigskip

Following either a profile likelihood or a Bayesian probability
approach SFitter then computes lower--dimensional
likelihood/probability distributions. For MSUGRA as well for the MSSM
distributions in one and two dimensions are shown, illustrating the
strengths and weaknesses of each of the two approaches. In the MSSM
parameter space the complete log-likelihood map is complemented by
corresponding maps over the approximately orthogonal gaugino--higgsino
and scalar parameter spaces. Such analyses of lower--dimensional
spaces lead to a less noisy likelihood map and can be useful in
addition to the completely exclusive likelihood map.\bigskip

The determination of the parameters of the weak--scale Lagrangian from
the LHC and the ILC and their errors are an essential ingredient to test 
unification. 
The SFitter approach is not limited to studies of the
supersymmetric parameter space. It can and will be used to study any
problem including mapping high--dimensional measurement and parameter
spaces in the LHC era.

\subsection*{Acknowledgments}

We are grateful to Ben Allanach, Chris Lester, Kyle Cranmer, Andreas
H\"ocker and others, who unsuccessfully tried to teach us statistics
and should not be held responsible for our results. We are also
grateful to the GDR Supersym\'etrie (CNRS), the 
Les Houches Workshops as well as the SUPA ultra-mini
workshops, during which many of the ideas were developed.

\appendix

\section{Weighted Markov Chains}
\label{sec:mcmc}

Markov--chain Monte Carlos (MCMCs) have for a long time been a tool to
evaluate functions for systems with a very large number of degrees of
freedom. An example in BSM physics would be the prediction of a
distribution for squark--gluino cross sections at the LHC, given the
currently available data and a supersymmetric
model~\cite{ben}. Computing LHC cross sections involves integrating
over parton densities and is therefore expensive. Similarly, one can
predict distributions of dark--matter detection rates given the
current data, which again is a fairly expensive computation for each
parameter point. The role of the MCMC is to provide us with a
representative sample of parameter points, where in our case
`representative' is defined by the likelihood $p(d|m)$ describing the
probability of a model parameter point being correct given our LHC
data. In general, this can be any normalized probability distribution
$p(m)$.

We produce a sample which with respect to $p(m)$ is a smaller copy of
the complete parameter space using the Metropolis--Hastings
algorithm~\cite{metropolis_hastings}.  This algorithm is nothing but
an iterative chain of decisions if a new point is accepted as part of
the Markov chain. As long as the probability of proposing a point $m'$
while sitting in $m$ is the same as the probability of proposing $m$
sitting in $m'$, the decision if the new point gets accepted depends
solely on the values $p(m)$ and $p(m')$ of the probability we want to
map: if the new $p(m')>p(m)$ then the new point is accepted, otherwise
it gets accepted with the probability $p(m')/p(m)$. Once this decision
is made, the next parameter point $m''$ is proposed, starting from
either $m$ or $m'$.\bigskip

The proposal probability is the probability $q(m \to m')$ with which
we find new points which then get suggested as new entries in the
Markov chain. Its choice is an internal choice in the
Metropolis--Hastings algorithm, but can have a huge impact on the
efficiency of probing the model--parameter space. For example,
dark--matter constraints are notoriously difficult, because they
generate narrow ridges in $p(m)$ which are not aligned with any of the
model parameters~\cite{ben,leszek}.  LHC measurements for example are
less restrictive, but more likely to develop distinct local
maxima. The proposal function must be able to jump back and forth
between these hills efficiently. For example a Gaussian distribution,
which is indeed symmetric between the starting point and the target
point, will have too suppressed tails to cover the MSSM parameter
space. We could instead add a constant to the proposal probability, or
use a Breit--Wigner distribution instead. In the more general case
where the proposal distribution is not symmetric, the decision for a
new point is not based on $p(m')/p(m)$, but on $[p(m') \; q(m \to
m')]/[p(m) \; q(m' \to m)]$. The only two requirements on the choice
of $q(m \to m')$ are that the proposal probability cannot have a
memory of the earlier points in the Markov chain (detailed balance),
and any point must have a non-zero probability of being proposed after
a finite number of steps. The latter ensures coverage of the whole
parameter space.  The proposal function can for example be symmetric
in $m$ and $m'$ or it can be independent of $m'$ altogether. The
efficiency for building a useful Markov chain is of course closely
linked to the efficiency of finding new parameter points which get
accepted with a reasonable probability.  Generally, $25 \%$ is
considered an optimal choice.\bigskip

In comparison to the usual Markov chain, the problem we are tackling
with SFitter is simpler: we are only interested in the likelihood of
some LHC measurement given a parameter point in our model, interpreted
as a map over the model's parameter space: $p(m) \equiv
p(d|m)$. Starting from this likelihood map we can either compute
profile likelihoods of lower--dimensional parameter spaces or a
Bayesian posterior probability distribution $p(m|d)$. This means that
naively we would produce a representative sample with respect to this
probability $p(m)$, then evaluate again the same probability $p(m)$,
add an integration measure or find the profile likelihood, bin it, and
obtain a likelihood or probability distribution in a subspace of the
complete vector $m$.  To save computing time we should obviously
retain the probability of each point in the Markov chain, similar to a
phase--space Monte Carlo where we produce weighted events for
integrated cross sections.\bigskip

To briefly illustrate the possible gain in efficiency consider a
binary system, where each parameter point enters one of two bins and
the probability $p$ of the two bins is divided as $10\% : 90\%$. We
need at least 10 unweighted entries in the Markov chain to get the
correct answer for the first time. Until then the probability
associated with the first bin will be either zero or too large. If we
use weighted events, two entries can already be sufficient, and each
additional entry can improve the error on our extraction of the
relative probability.\bigskip

Obviously, we cannot just keep the weight for each point in the Markov
chain and multiply it into the binning procedure, since this would
double--count this weight. Instead, we use a modified form of
binning~\cite{FerrenbergSwendsen}. We first consider the case that
$p\ne0$ everywhere and then generalize this result to also include
regions with $p=0$.

We define an inverse averaging in each bin as
\begin{equation}
P_{\rm bin}(p\ne0) = \frac{\text{bincount}}{\sum_{i=1}^{\text{bincount}} 1/p}
\end{equation}
where the sum in the denominator is over all points in the Markov
chain which belong into this bin, counted with their correct
multiplicity. It is easy to see that this gives the right answer. The
numerator can be written as 
$\sum_{i=1}^{\text{bincount}} 1$. Now we take the limit of infinitely
many points, so both sums turn into an integral
\begin{equation}
P_{\rm bin}(p\ne0) = \frac{\int \di x w(x) \cdot 1}{\int \di x w(x)/p(x)}
\quad , 
\end{equation}
where $w(x)$ is an arbitrary weight function with $\int \di x w(x) = 1$.
We choose $w(x)=p(x)$ and obtain the desired result
\begin{equation}
P_{\rm bin}(p\ne0) = \frac1{V(p\ne0)} \int \di x p(x) 
\quad ,
\end{equation}
where $V(p\ne0)$ is the volume of the bin in the parameter space.\bigskip

Note that this expression is only defined for $p \neq 0$.  This means
we need to correct for regions where $p=0$, as points in such regions
will never enter the Markov chain. We store all points which we
generate as suggested points during the evaluation of the Markov
chain, and which are rejected because the probability is zero and
compute the correction factor
\begin{equation}
P_{\rm bin} = P_{\rm bin}(p\ne0) \cdot 
  \left (1- \frac{\sum_{i=1}^{\text{zerocount}}
          P(m_i\rightarrow m_i')^{-1}
         }{\text{zerocount} \cdot V_{\rm bin}}
  \right) \quad .
\end{equation}
$P(m\rightarrow m')$ is the probability of suggesting $m'$ from
$m$. For our Weighted-Markov-Chain technique $m$ is the previous
point in the Markov chain and $m'$ is the proposed point with $p=0$.
$V_{\rm bin}$ is the volume of the bin.  We need to show that the
second term in the bracket turns to $V_{\rm bin}(p=0)/V_{\rm bin}$,
the fraction of volume inside the bin where $p$ vanishes.  To do this
we add an additional sum
\begin{equation}
\frac{\sum_{i=1}^{\text{zerocount}}
          P(m_i\rightarrow m_i')^{-1}
         }{\text{zerocount} \cdot V_{\rm bin}}
= \frac{\sum_{i=1}^{\text{zerocount}} \sum_{j=1}^{k}
          P(m_i\rightarrow m_{i,j}')^{-1}
         }{\text{zerocount} \cdot V_{\rm bin}}
\end{equation}
with $k=1$ and $m_{i,1}' = m_i'$. We now take the newly introduced sum
in the numerator as a very crude approximation to the corresponding
Monte Carlo integral, effectively taking the limit of infinite
$k$. This is exactly the probability of hitting the region where $p=0$
times the total volume, which is just $V_{\rm bin}(p=0)$.
$P(m_i\rightarrow m_{i,j}')$ is the weight function of the Monte Carlo
integration. Canceling $\text{zerocount}$ in numerator and
denominator gives the desired form.

So far, we have discussed this weighting technique using a probability
$p$. Markov chains, however, are more general. They allow every
function $f$ which is non-negative everywhere to be used as potential,
and SFitter uses $1/\chi^2$ as potential. It is easy to see that the
expressions given above remain valid, as the normalization constant
drops out in the final results. The resulting $P$ is
then an average of $f$ over the bin. In the special case that $f$ is
constant we would obtain $f$ again.

For details of these Weighted Markov Chains (WMC), including their
features under marginalization see Ref.~\cite{wmc}.

\section{Toy Model}
\label{sec:toy}

To illustrate the SFitter results and output we use a simple toy
model: we evaluate a potential (likelihood) $V(m)$ over a
5-dimensional parameters space $m$. The potential has five
distinct maxima, a small and a large sphere, a cigar and two cuboids,
one of which is tilted. The background consists of a constant term and
a flat parabola centered at the origin:
\begin{align}
V_{\text{small sphere}} = 
       75 & \cdot \Biggl[1-\left(\frac{m_1-650}{100}\right)^2
                          -\left(\frac{m_2-250}{100}\right)^2
                          -\left(\frac{m_3-350}{100}\right)^2
                           \nonumber\\
                        & -\left(\frac{m_4-350}{100}\right)^2
                          -\left(\frac{m_5-350}{100}\right)^2
                \Biggr]_{\! +}^{\,\frac12} \displaybreak[0]\\
V_{\text{large sphere}} = 
       12 & \cdot \Biggl[1-\left(\frac{m_1-350}{300}\right)^2
                          -\left(\frac{m_2-650}{300}\right)^2
                          -\left(\frac{m_3-650}{300}\right)^2
                           \nonumber\\
                       &  -\left(\frac{m_4-650}{300}\right)^2
                          -\left(\frac{m_5-650}{300}\right)^2
                \Biggr]_{\! +}^{\,\frac12} \displaybreak[0]\\
V_{\text{tilted cuboid}} =
       60 & \cdot \biggl( 
             0.8+0.2\left[1-\frac{|m_1+2m_2-1300|}{250}\right]_{\! +} \biggr)   
            \nonumber\\
          & \cdot \biggl( 
             0.8+0.2\left[1-\frac{|2m_1-m_2-1475|}{125}\right]_{\! +} \biggr) 
            \cdot \biggl( 
             0.8+0.2\left[1-\frac{|m_3-650|}{100}\right]_{\! +} \biggr)   
            \nonumber\\
          & \cdot \biggl( 
             0.8+0.2\left[1-\frac{|m_4-650|}{100}\right]_{\! +} \biggr) 
            \cdot \biggl( 
             0.8+0.2\left[1-\frac{|m_5-650|}{100}\right]_{\! +} \biggr)
                                                         \displaybreak[0]\\
V_{\text{cuboid}} =
       25 & \cdot \biggl( 
             0.8+0.2\left[1-\frac{|m_1-750|}{50}\right]_{\! +} \biggr)   
            \cdot \biggl( 
             0.8+0.2\left[1-\frac{|m_2-750|}{50}\right]_{\! +} \biggr) 
            \nonumber\\
          & \cdot \biggl( 
             0.8+0.2\left[1-\frac{|m_3-450|}{150}\right]_{\! +} \biggr)   
            \cdot \biggl( 
             0.8+0.2\left[1-\frac{|m_4-450|}{150}\right]_{\! +} \biggr) 
            \nonumber\\
          & \cdot \biggl( 
             0.8+0.2\left[1-\frac{|m_5-450|}{150}\right]_{\! +} \biggr) 
                                                         \displaybreak[0]\\
V_{\text{cigar}} =
     6 & \cdot \exp\left(-\frac{\left(m_1+m_2-500\right)^2}{2 \cdot 50^2}\right)
         \cdot \exp\left(-\frac{\left(m_1-m_2\right)^2}{2 \cdot 150^2}\right)
         \nonumber\\
       & \cdot \exp\left(-\frac{\left(m_3-550\right)^2}{2 \cdot 150^2}\right)
         \cdot \exp\left(-\frac{\left(m_4-550\right)^2}{2 \cdot 150^2}\right)
         \nonumber\\
       & \cdot \exp\left(-\frac{\left(m_5-550\right)^2}{2 \cdot 150^2}\right)
                                                         \displaybreak[0]\\
V_{\text{background}} = 0.1 & + 4 
  \cdot \left(\frac{m_1}{1000}\right)^2 
  \cdot \left(\frac{m_2}{1000}\right)^2 
  \cdot \left(\frac{m_3}{1000}\right)^2 
  \cdot \left(\frac{m_4}{1000}\right)^2 
  \cdot \left(\frac{m_5}{1000}\right)^2 
\end{align}
The symbol $\left[x\right]_{\! +}$ means $V_{\text{object}} = 0$ for
$x<0$.\bigskip

SFitter analyzes this parameter space using two approaches: first, we
produce a set of Markov chains sampling the entire parameter space as
described in Appendix~\ref{sec:mcmc} corresponding to $p(m)
\equiv V(m)$.  This means we produce a sample of $10^8$ points,
distributed equally over ten individual chains, which form a
likelihood map of the parameter space $m$.\bigskip

In a second step SFitter starts from the maxima in the Markov chain
for $V(m)$ and searches for the local maxima with improved
resolution. For the Bayesian probability functions this step is
strictly speaking not necessary, as long as we are only interested in
marginalized distributions. On the other hand, we always want to have
a good idea what structure $V(m)$ exhibits over the parameter space
and where its maxima are.

\begin{figure}[t]
 \begin{tabular}{rr@{$\pm$}r@{, }r@{$\pm$}r@{, }r@{$\pm$}r@{, }r@{$\pm$}r@{, }r@{$\pm$}r}
     $V=75.1$ &( 650 & 16.3 & 
                 250 & 16.3 &
                 350 & 16.3 &
                 350 & 16.3 &
                 350 & 16.3 ) \\
     $V=60.1$ &( 850 & 4.2  &
                 225 & 8.2  &
                 650 & 8.3  &
                 650 & 8.3  &
                 650 & 8.3 ) \\
     $V=25.1$ &( 750 & 10.0 &
                 750 & 10.0 &
                 450 & 29.9 &
                 450 & 29.9 &
                 450 & 29.9 ) \\
     $V=16.1$ &( 250 & 28.4 &
                 250 & 28.4 &
                 550 & 53.9 &
                 550 & 53.9 &
                 550 & 53.9 ) \\
     $V=12.1$ &( 350 & 120.0 &
                 650 & 119.9 &
                 650 & 119.9 &
                 650 & 119.9 &
                 650 & 119.9 ) \\
      \dots
 \end{tabular} \\
 \includegraphics[width=6cm]{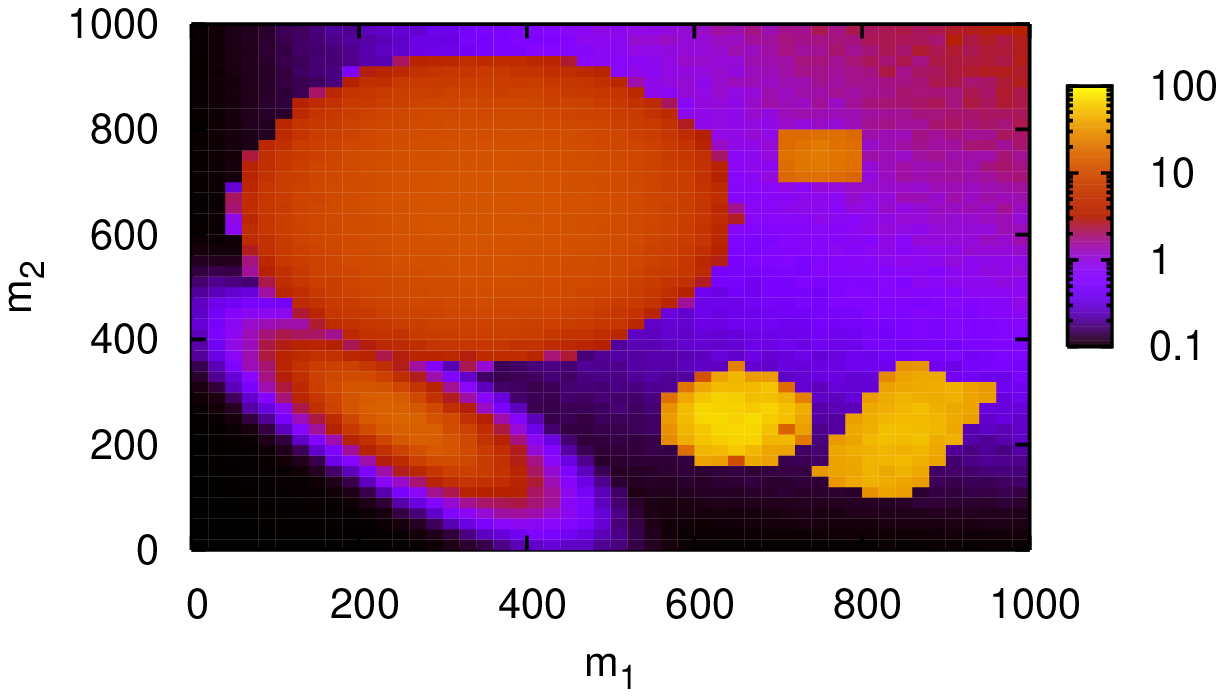} \hspace*{1cm}
 \includegraphics[width=6cm]{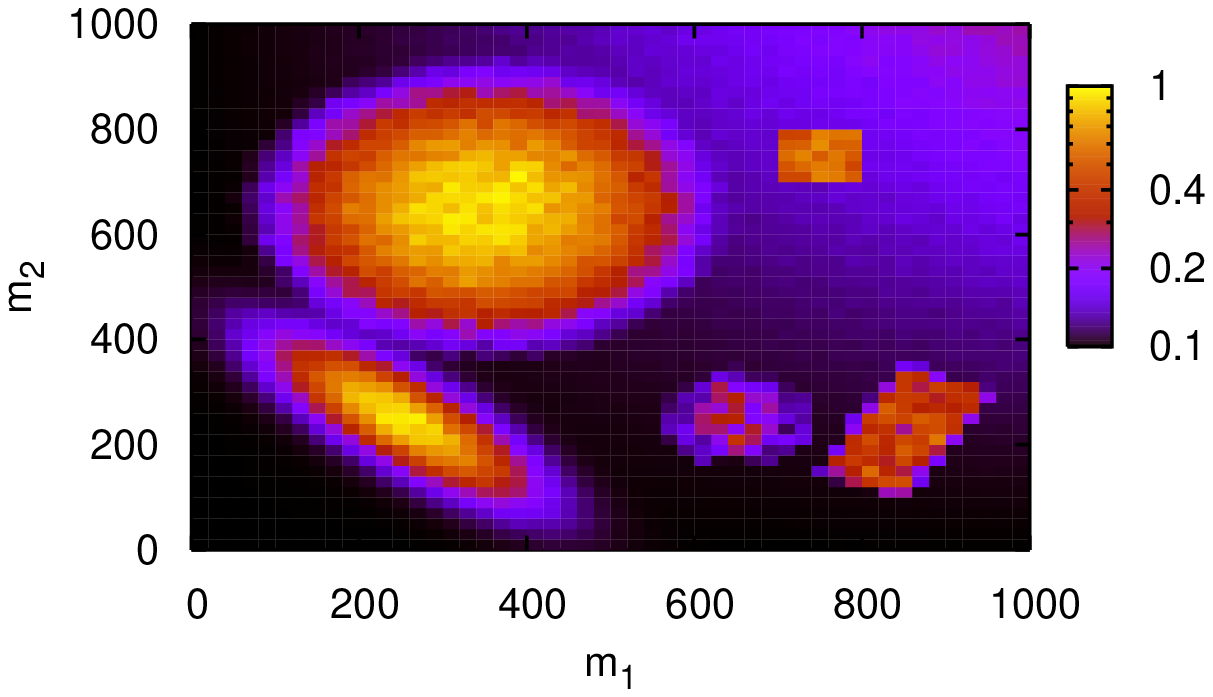} \\
 \includegraphics[width=5cm]{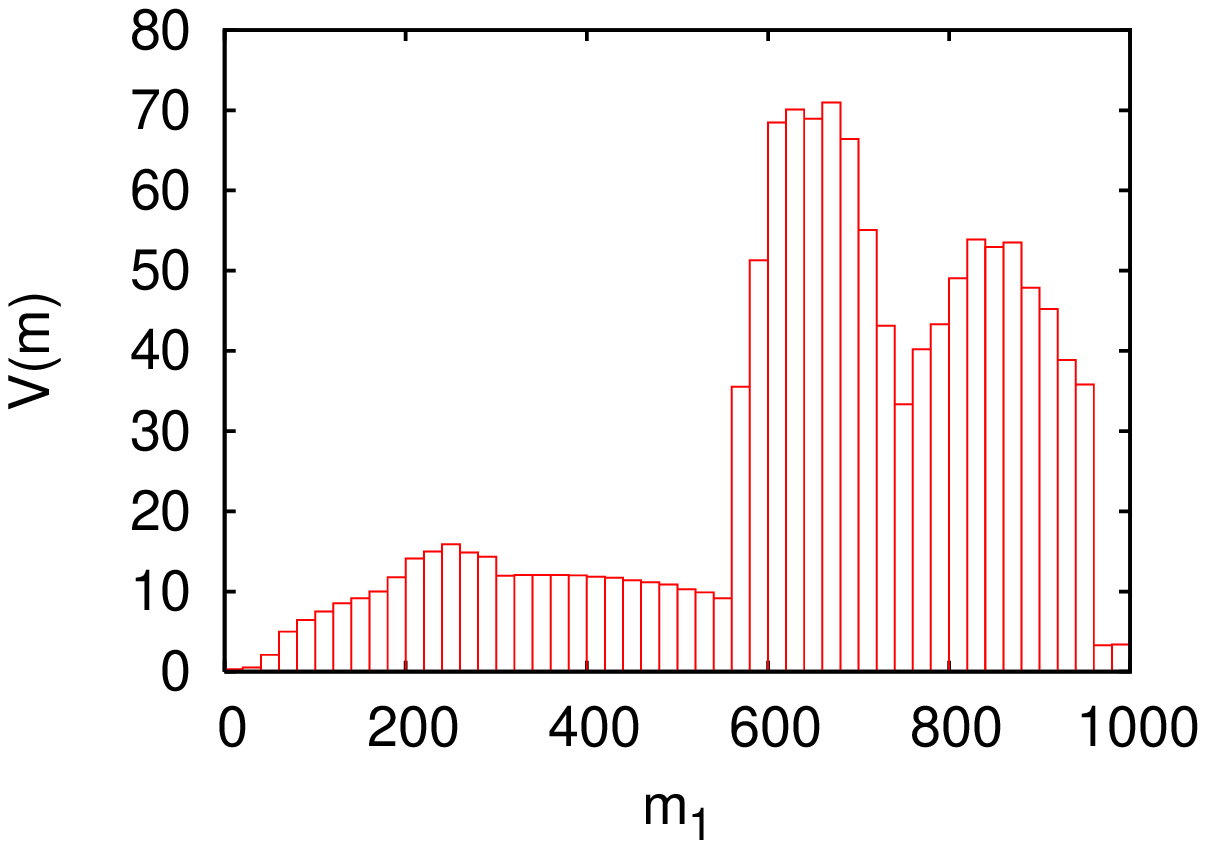} \hspace*{2cm}
 \includegraphics[width=5cm]{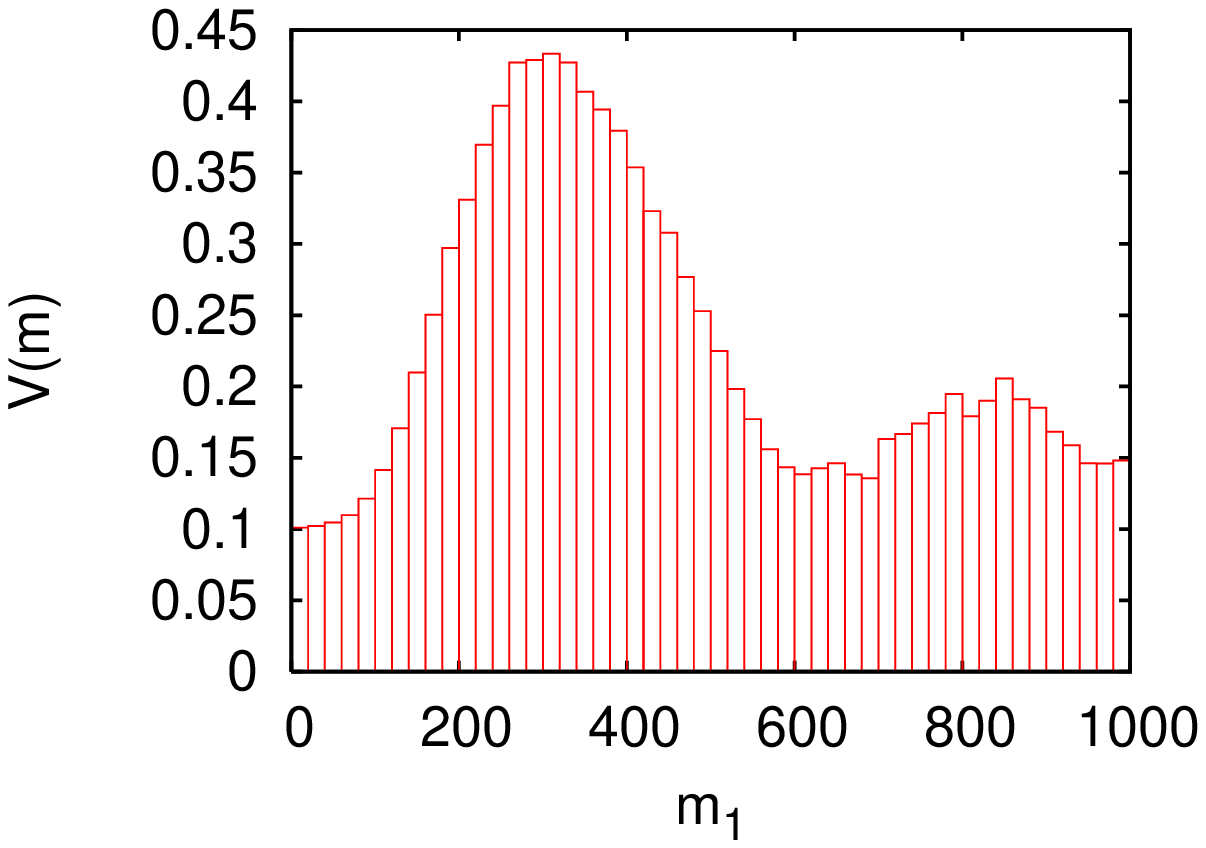} 
 \caption[]{SFitter output for the 5--dimensional potential.  First
 row: list of the largest values of $V(m)$ in the entire
 parameter space.  Second row: logarithmic map of $V(m_1,m_2)$, as a
 profile likelihood (left), or marginalized over $m_{3,4,5}$ (right).
 Third row: distribution for $V(m_1)$, as a profile likelihood (left)
 or marginalized over $m_{2,3,4,5}$ (right).}
\label{fig:toy_model}
\end{figure}

We eliminate local--maxima candidates if they are too close in
parameter space and produce the ranked list of the largest values of
$V(m)$ in the 5--dimensional parameter space, shown in
Fig.~\ref{fig:toy_model}.  We see that as an isolated point the small
sphere has the highest value of $V(m)$.\bigskip

Technically, because the resolution of the Markov chain will in
general be too coarse to match the data errors, we need an additional
hill--climbing algorithm.  We use a modified version of
Minuit~\cite{minuit}. For the gradient and diagonal second
derivatives, we replace the simple three-point formulae in the
standard Minuit version with Ridders' method~\cite{Ridders}. This
algorithm starts with the three-point formulae using a large step
size, then iteratively shrinks the step size (typically by a factor of
two) and computes an estimate using all points calculated so far. The
result of the three-point formula using only the new points is used to
estimate the calculation's uncertainty. The iterations terminate when
the desired accuracy is reached or when numerical uncertainties
dominate for very small step sizes. In this method, not only all
odd--power terms in the Taylor expansion of the derivative cancel, but
also the leading even--power terms, in turn improving the accuracy. In
addition, the step size is dynamically adjusted to its optimal value.

A slight complication arises from our box--shaped theory errors,
because the function has a discontinuous second derivative. The Minos
error estimate is in principle not affected by this, but this
discontinuity breaks Ridders' algorithm: the higher derivative can now
vastly differ between two neighboring points, and the terms listed
above do not cancel any longer. To solve this problem, we replace the
likelihood function by its original shape around the discontinuity:
suppose the parameter point for which we want to compute the
derivatives falls into the central region of eq.(\ref{eq:flat_errors})
where $\log \mathcal{L}=0$. For the derivatives we always assume $\log
\mathcal{L}=0$, no matter if the parameter point probed by Ridders'
algorithm falls inside or outside the flat region. Similarly, in case
the parameter point we are interested in is on the positive branch of
the parabola, for the derivatives we just replace the flat region with
the opposite branch of the parabola. Note that this is only a
technical trick to improve the estimate of the derivative and that the
calculated values of $\log \mathcal{L} $ are not used anywhere
else.\bigskip

To reduce the number of dimensions over which we would like to compute
a probability distribution we have three options: first we can simply
slice the parameter space in $m_{3,4,5}$, which is useful to
illustrate the behavior of $V(m)$ but has no statistical meaning
whatsoever.  Second, we can compute the profile likelihood described
in Sec.~\ref{sec:sugra_likelihood}, just projecting out dimensions by
replacing the reduced--dimensional value of $V$ by its maximum in the
removed dimensions. And finally, we can marginalize over the
dimensions. Note that only marginalization will produce a
mathematically well--defined lower--dimensional probability
distribution. Technically, marginalization means nothing but binning
the pdf and collecting its values in a histogram for the two remaining
dimensions $m_{1,2}$.

In the second row of Fig.~\ref{fig:toy_model} we immediately see that
the small sphere appears more prominent in the profile likelihood
version of $V(m_1,m_2)$ while the large sphere dominates the
two--dimensional Bayesian distribution of $V(m_1,m_2)$. The same
effect we see in the one--dimensional distributions $V(m_1)$, where in
the profile likelihood case one of the cuboids appears prominently, as
expected from the list of best values for $V(m)$. If $V$ were a
pdf we could conclude that the small sphere contains the most likely
parameter points while the large sphere is the most likely physics
configuration.  This dominance of the large sphere over the most
likely single point in the small sphere is an effect of the
marginalization, \ie an example for a volume effect.\bigskip

The question if such volume effects should be considered, if instead
the best single point is preferable, or if actually the third--best
point should be picked out by a theory bias cannot and should not be
answered by SFitter as a tool. Instead, SFitter provides all
information needed by the user to correctly answer each of these
different questions.


\end{document}